\documentclass[sn-nature]{sn-jnl}

\usepackage{graphicx}%
\usepackage{multirow}%
\usepackage{amsmath,amssymb,amsfonts}%
\usepackage{amsthm}%
\usepackage{mathrsfs}%
\usepackage[title]{appendix}%
\usepackage{xcolor}%
\usepackage{textcomp}%
\usepackage{manyfoot}%
\usepackage{booktabs}%
\usepackage{algorithm}%
\usepackage{algorithmicx}%
\usepackage{algpseudocode}%
\usepackage{listings}%
\usepackage{threeparttable}  
\usepackage{url} 
\usepackage{placeins}
\usepackage{float}

\begin{document}

\def \nuprocess{$\nu$-process}
\def \nodata{. . .}
\def \degree{$^{\circ}$}
\def \Msolar{M$_{\odot}$}
\def \alphafe{[$\alpha$/Fe]}
\def \HI{H\ion{I}}
\def \sion{\ion{II}}
\def \vninety{v$_{90}$}
\def \Lbol{L$_{\rm bol}$}
\def \Mstar{M_{\star}}
\def \logMstar{\log(M_{\star}/\mathrm{M}_{\odot})}
\def \logRimp{log(R$_{\rm imp}$/kpc)}
\def \logLbol{log(L$_{\rm AGN}$/erg s$^{-1}$)}
\def \logsSFR{log(\mathrm{sSFR / yr}^{-1})}
\def \kms{km s$^{-1}$}
\def \zabs{z$_{\rm abs}$}
\def \zem{z$_{\rm em}$}
\def \Rimp{$\rho_{\rm imp}$}
\def \Rvir{$\rho_{\rm vir}$}
\def \deltaEW{${\rm \Delta log(EW/m\AA)}$}
\def \RadRat{$f_{\rm AGN}/f_{\rm HM01}$}
\def \mnfe{[Mn/Fe]$_{\rm DC}$}
\def \Mgeqw{W$_{0}^{2796}$}
\def \Feeqw{W$_{0}^{2600}$}
\def \fracMgFe{ ${\rm{W}_{0}^{2796}}$/${\rm{W}_{0}^{2600}}$}
\def \omegaDLA{$\Omega_{\rm H \textsc{i}}$}
\def \ndla{30}
\def \npdla{46}
\def \nlpdla{41}
\def \nxpdla{5}
\def \nmdla{27}
\def \nlmdla{21}
\def \nxmdla{6}
\def \CosmoZ{$\langle Z/Z_{\odot} \rangle$}
\def \fNX{$f(N,X)$}
\def \deltasfms{\Delta \langle \log(\mathrm{SFR})\rangle_{\mathrm{MS}}}
\def \dlogsfms{\Delta \langle \log(\mathrm{SSFR})\rangle_{\mathrm{MS}}}
\def \fgas{f_{\mathrm{gas}}}
\def \mstar{M_{\star}}
\def \logsfr{\log({\mathrm{SFR}})}
\def \dlogsfr{\Delta\log({\mathrm{SFR}})}
\def \logssfr{\log({\mathrm{SSFR}})}


\newcommand{\actaa}{Acta Astron.} 
\newcommand{\araa}{Annu. Rev. Astron. Astrophys.} 
\newcommand{\aar}{Astron. Astrophys. Rev.} 
\newcommand{\ab}{Astrobiol.} 
\newcommand{\aj}{Astron. J.} 
\newcommand{\apj}{Astrophys. J.} 
\newcommand{\apjl}{Astrophys. J. Lett.} 
\newcommand{\apjs}{Astrophys. J. Suppl. Ser.} 
\newcommand{\ao}{Appl. Opt.} 
\newcommand{\apss}{Astrophys. Space Sci.} 
\newcommand{\aap}{Astron. Astrophys.} 
\newcommand{\aapr}{Astron. Astrophys. Rev.} 
\newcommand{\aaps}{Astron. Astrophys. Suppl.} 
\newcommand{\baas}{Bull. Am. Astron. Soc.} 
\newcommand{\caa}{Chinese Astron. Astrophys.} 
\newcommand{\cjaa}{Chinese J. Astron. Astrophys.} 
\newcommand{\cqg}{Class. Quantum Gravity} 
\newcommand{\gal}{Galaxies} 
\newcommand{\gca}{Geochim. Cosmochim. Acta} 
\newcommand{\icarus}{Icarus} 
\newcommand{\jcap}{J. Cosmol. Astropart. Phys.} 
\newcommand{\compastro}{Comput. Astrophys.} 
\newcommand{\jgr}{J. Geophys. Res.} 
\newcommand{\jgrp}{J. Geophys. Res.: Planets} 
\newcommand{\jqsrt}{J. Quant. Spectrosc. Radiat. Transf.} 
\newcommand{\memsai}{Mem. Soc. Astron. Italiana} 
\newcommand{\mnras}{Mon. Not. R. Astron. Soc.} 
\newcommand{\nat}{Nature} 
\newcommand{\nastro}{Nat. Astron.} 
\newcommand{\ncomms}{Nat. Commun.} 
\newcommand{\nphys}{Nat. Phys.} 
\newcommand{\nrevphys}{Nat. Rev. Phys.}
\newcommand{\na}{New Astron.} 
\newcommand{\nar}{New Astron. Rev.} 
\newcommand{\physrep}{Phys. Rep.} 
\newcommand{\pra}{Phys. Rev. A} 
\newcommand{\prb}{Phys. Rev. B} 
\newcommand{\prc}{Phys. Rev. C} 
\newcommand{\prd}{Phys. Rev. D} 
\newcommand{\pre}{Phys. Rev. E} 
\newcommand{\prl}{Phys. Rev. Lett.} 
\newcommand{\psj}{Planet. Sci. J.} 
\newcommand{\planss}{Planet. Space Sci.} 
\newcommand{\pnas}{Proc. Natl Acad. Sci. USA} 
\newcommand{\procspie}{Proc. SPIE} 
\newcommand{\pasa}{Publ. Astron. Soc. Aust.} 
\newcommand{\pasj}{Publ. Astron. Soc. Jpn} 
\newcommand{\pasp}{Publ. Astron. Soc. Pac.} 
\newcommand{\rmxaa}{Rev. Mexicana Astron. Astrofis.} 
\newcommand{\sci}{Science} 
\newcommand{\sciadv}{Sci. Adv.} 
\newcommand{\solphys}{Sol. Phys.} 
\newcommand{\sovast}{Soviet Ast.} 
\newcommand{\ssr}{Space Sci. Rev.} 
\newcommand{\uni}{Universe} 


\newcounter{extfig}
\renewcommand{\theextfig}{\arabic{extfig}}

\setlength{\footskip}{150pt}



\definecolor{violet}{rgb}{0.5, 0, 0.8}
\newcommand{\cb}[1]{\textcolor{violet}{#1}}


\title[]{Galaxy and black hole coevolution in dark matter haloes not captured by cosmological simulations}



\author*[1,2,3]{\fnm{Hassen M.} \sur{Yesuf}}\email{yesufh@shao.ac.cn}

\author[4]{\fnm{Connor} \sur{Bottrell}}

\affil[1]{\orgdiv{Key Laboratory for Research in Galaxies and Cosmology}, \orgname{Shanghai Astronomical Observatory, Chinese Academy of Sciences}, \orgaddress{\street{80 Nandan Road}, \city{Shanghai}, \postcode{200030}, \country{China}}}

\affil[2]{\orgdiv{Kavli Institute for the Physics and Mathematics of the Universe}, \orgname{University of Tokyo}, \orgaddress{\street{5-1-5 Kashiwanoha Campus}, \city{Kashiwa}, \postcode{277-8583}, \state{Chiba}, \country{Japan}}}

\affil[3]{\orgdiv{Kavli Institute for Astronomy and Astrophysics}, \orgname{Peking University}, \orgaddress{\street{5 Yiheyuan Road}, \city{Beijing}, \postcode{100871}, \country{China}}}

\affil[4]{\orgdiv{International Centre for Radio Astronomy Research}, \orgname{University of Western Australia}, \orgaddress{\street{35 Stirling Hwy}, \city{Crawley}, \postcode{6009}, \state{WA}, \country{Australia}}}

\maketitle

Star formation in galaxies is governed by internal and environmental processes, yet their relative roles are not well understood. In particular, uncertainties in measurements of active galactic nuclei (AGN) host galaxies, combined with modeling limitations, obfuscate the impact of supermassive black hole feedback across environments and over time. Here we address this with a comprehensive analysis of $\sim$ 60,000 nearby AGNs ($z < 0.15$) and new environment and halo-mass measurements for $\sim 500,000$ AGN and non-AGN host galaxies. This benchmark enables unified comparisons with three prominent cosmological simulations—SIMBA, TNG, and EAGLE—and reveals major, contrasting shortcomings. Simulations fail to reproduce observed trends linking star formation, quiescence, AGN luminosity, stellar mass, and halo mass. While simulations qualitatively capture that AGNs are more common in low-mass halos than in rich groups or clusters, detailed host demographics diverge strongly from observations. Partial agreement exists in the stellar mass distribution within large-scale structures, yet all simulations overproduce quenched low-mass satellites in massive halos, while misrepresenting quenched fractions of massive central galaxies and those in low-density environments, which are sensitive to feedback implementation. Improved AGN physics and modeling of multi-phase gas cooling and flows are required to capture the observed interplay between black holes, galaxies, and halos.
    
\section*{Main}\label{sec:res}

Star formation is regulated by both external and internal processes. External drivers include gas accretion, galaxy mergers, and large-scale gravitational interactions \citep{PengY+10,Cortese+21}, while internal processes involve feedback from supernovae, stellar winds, and accreting supermassive black holes (SMBHs) \citep{Croton+06,Hopkins+08,KormendyHo13,DonahueVoit22}. These processes regulate the available gas supply within and around galaxies, for example, in the interstellar medium (ISM), where stars form, and in the circumgalactic reservoir that may replenish the ISM and fuel future star formation. To model these effects on galaxy demographics, cosmological hydrodynamical simulations are typically used \citep{CrainvdVoort23}. By calibrating unresolved feedback implementations against key observations, state-of-the-art simulations such as EAGLE \citep{2015MNRAS.450.1937C,2015MNRAS.446..521S}, IllustrisTNG \citep{2018MNRAS.475..648P,2018MNRAS.475..624N}, and SIMBA \citep{simba2019} broadly reproduce distributions of stellar masses, star formation rates, and gas fractions \citep{Dave+20, Ward+22, MaWenlin+22}. These three simulations, also used in this study, implement AGN feedback as a primary mechanism for quenching star formation in massive galaxies and reproducing the observed stellar mass and star formation rate functions \citep{2018MNRAS.475..648P,simba2019,Furlong+15,Katsianis+17}, motivating the feedback prescriptions adopted in each case.

Nevertheless, we find major discrepancies between observed galaxy properties and those predicted by simulations, raising important questions about how baryonic physics are implemented in numerical efforts that adopt the $\Lambda$ cold dark matter ($\Lambda$CDM) galaxy formation framework. Using large SDSS and GAMA samples, we provide a unified comparison of galaxy and AGN properties---stellar mass, star formation, black hole accretion, and environment---between observations and simulations. The following sections present the stellar mass functions of star-forming and quiescent galaxies, the environmental dependence of quiescent fraction, and the demographics of active SMBHs and their environments.

\subsubsection*{Galaxy stellar mass functions}

Fig.~1a presents the stellar mass function (SMF) for nearby galaxies from the GAMA and SDSS surveys \citep{Weigel+16,Driver+22}, compared with the three simulations analyzed in this study. Because these simulations were calibrated to match the observed SMF at $z = 0$, their agreement with the total galaxy population in Fig.~1a is expected \citep{2018MNRAS.475..648P,simba2019,Furlong+15}.

Building on this baseline comparison, Fig.~1b--d decompose the SMF into three subpopulations according to galaxies' offsets from the star-forming main sequence (SFMS) at fixed stellar mass: upper-SFMS ($\Delta\log\mathrm{SFR} > 0$), lower-SFMS ($-1 < \Delta\log\mathrm{SFR} < 0$), and quiescent ($\Delta\log\mathrm{SFR} < -1$). This decomposition has not been examined in previous SMF studies and provides a new, sensitive diagnostic of how feedback and quenching prescriptions regulate galaxy number densities (Extended Data Figs.~1 and 2). By separating galaxies according to their star-formation state---rather than only their mass---we reveal discrepancies that are hidden in the global SMF: all three simulations show pronounced and systematic departures from observations in these subpopulations, with deviations often exceeding $5\sigma$ ( Table~1 and Fig.~1).

The three simulations---particularly TNG and SIMBA---show a qualitative tendency toward the bimodal, or double-Schechter, form of the quenched galaxy SMF (Fig.~1d), consistent with internal and environmental quenching processes \citep{PengY+10}. Next, we examine how quiescent fractions vary with both stellar mass and environment (halo mass). These comparisons also reveal significant tensions with observations.

\subsubsection*{Star formation trends with local environment and halo mass}

As a preliminary comparison, Fig.~2 contrasts the multiscale environments of observations and simulations. On $0.5$--$8\,\mathrm{Mpc}/h$ scales, all three models broadly reproduce the stellar-mass overdensity distributions of GAMA and SDSS galaxies at $z < 0.12$.

Fig.~3 presents the quiescent fraction as a function of stellar mass and environment, where environment is quantified by the stellar mass excess within a $1\,\mathrm{Mpc}/h$ aperture relative to the median \citep{yesuf22}. Inset panels quantify the statistical significance of differences between SDSS and each simulation per bin (see also Table~2).

Fig.~3a reveals the expected trends: quiescence is most common among massive galaxies in dense environments and least common among low-mass galaxies in underdense regions. Mixed populations appear in two key regimes: (1) low-mass galaxies in high-density environments and (2) intermediate-mass galaxies, where the quiescent fraction rises steeply with the mass overdensity, by 30--50\% from the lowest to highest density bins. The remaining panels of Fig.~3 show residuals of the quiescent fraction, defined as simulation minus SDSS. All three simulations show deviations in quiescent fractions exceeding 30\% in some regions, with systematic trends across stellar mass and environmental density; in each case, the combined discrepancy corresponds to an RMSE of $\sim 0.2$, significant at $>5\sigma$.

~3 presents analogous analyses of mean SFRs as a function of $M_\star$ and environment for star-forming galaxies. 

Using the ability of simulations to reproduce multiscale stellar mass distributions (Fig.~2), we develop a halo mass estimator with XGBoost regression. The method links galaxy observables to halo mass with high precision ($R^2 \approx 0.97$, scatter $\sigma \approx 0.15$\,dex), providing a robust, model-informed mapping consistent across simulations. Validating this prediction will require future observational efforts with accurate, independent halo mass measurements. Our estimates also show reasonable agreement with previous group catalog results \citep{YangX+07,Robotham+11,Tempel+17}, though those lack the precision needed for the validation.

As an application of our measurements, Fig.~4 shows the quiescent fraction of satellite galaxies as a function of halo mass at fixed $M_\star$, comparing SDSS with the three simulations analyzed in this study. As expected, the quiescent fraction increases with both $M_h$ and $M_\star$, but more strongly with $M_\star$ (Fig.~4a). All three simulations overpredict the quiescent fraction of low-mass satellites in massive halos ($M_h > 10^{13}\,M_\odot$) by $\sim$30\% or more. SIMBA overpredicts quiescent satellites at intermediate stellar masses ($M_\star \approx 10^{10}\,M_\odot$), even in low-mass halos, but agrees better with observations for more massive galaxies. In contrast, TNG underpredicts the number of intermediate-mass quiescent satellites in low-mass halos and does not match the abundance of quiescent satellites at the highest stellar masses. EAGLE similarly underpredicts quiescent satellites in low-mass halos across intermediate to high stellar masses. The trends presented in Fig.~3 (and Fig.~4) are qualitatively similar when the analysis is repeated using GAMA data as shown in Supplementary Information and are also consistent with previous findings \citep{Donnari+21fq}. ~2 illustrate how different implementations of feedback, or its absence, in the SIMBA and EAGLE variants influence the trends of quiescent fraction as a function of $M_h$ and $M_\star$.

The SMF and environmental discrepancies indicate that quenching in current simulations---whether linked to halos, supermassive black holes, or both---fails to reproduce the long-term evolution of galaxy quiescence at fixed halo mass or environmental density. Next we examine active SMBH demographics to probe their immediate feedback and reveal further tensions with the models.

\subsubsection*{Characteristics of active supermassive black holes and their host galaxies}

Previous studies have examined AGN and host demographics, including AGN luminosity and black hole mass functions \citep[e.g.,][]{Rosas-Guevara+16,Weinberger+17,Habouzit+21,Habouzit+22}, as well as the sSFR distributions of luminous AGN hosts \citep{Ward+22}. While these works provided essential groundwork, they differ from our study in scope, sample definition, methodology, and completeness treatment. Here we analyze a uniformly selected optical AGN sample of $\sim$60{,}000 sources, with completeness corrections tied to X-ray–selected AGNs \citep{Webb+20}.

Fig.~5 compares the number densities of observed and simulated AGNs as functions of (a) AGN luminosity ($L_\mathrm{AGN}$), (b) host-galaxy velocity dispersion ($\sigma_\star$) , (c) stellar mass ($M_\star$), and (d) specific star-formation rate (sSFR $\equiv \mathrm{SFR}/M_\star$). For unobscured AGNs, velocity dispersions are inferred from their black hole masses \citep{Greene+20}.

None of the three simulations reproduces the joint demographics of AGNs and their host galaxies, exhibiting systematic discrepancies across four key observables (with $>5\sigma$ tensions and RMSEs of $\sim 0.4$$-$1.0\,dex; see Table~3, Fig.~5, and Supplementary Information). TNG and SIMBA overpredict low-luminosity AGNs, whereas EAGLE underpredicts them. Both TNG and SIMBA overproduce AGNs in low-mass hosts, while all three underproduce the number of massive AGNs. Notably, simulated AGN populations occupy very different stellar-mass ranges from observations (Fig.~5 and ~4). These stellar-mass offsets correspond to the discrepancies in the sSFR function. SIMBA better matches the sSFRF and SMF of massive AGN hosts ($\log M_\star/M_\odot \gtrsim 10$), though their $\sigma_\star$ distributions remain discrepant. By contrast, EAGLE and TNG more severely underproduce quiescent or low-sSFR AGNs; TNG further overproduces low-$\sigma_\star$ AGNs, while EAGLE and SIMBA underproduce them. These discrepancies are also reflected in the distributions of black hole-to-stellar mass ratios, Eddington ratios, and the AGN black hole mass functions (Extended Data Figs.~4 and 5).

Fig.~6 compares the environments and halo masses of AGNs in three simulations and in SDSS at $z < 0.15$. The simulations qualitatively reproduce the observed trend that most AGNs reside in low-mass halos ($\log M_h/M_\odot < 13$) and in environments that are neither highly overdense nor underdense, although a minority occur in extreme environments. Despite this broad agreement, the simulated halo mass and halo-scale density distributions differ significantly from one another ($p < .001$), partly reflecting variations in host stellar mass. EAGLE, for instance, favors higher halo masses and densities. Moreover, all simulations remain in strong tension with SDSS ($p < .001$). 

\begin{figure*}
\includegraphics[width=\linewidth]{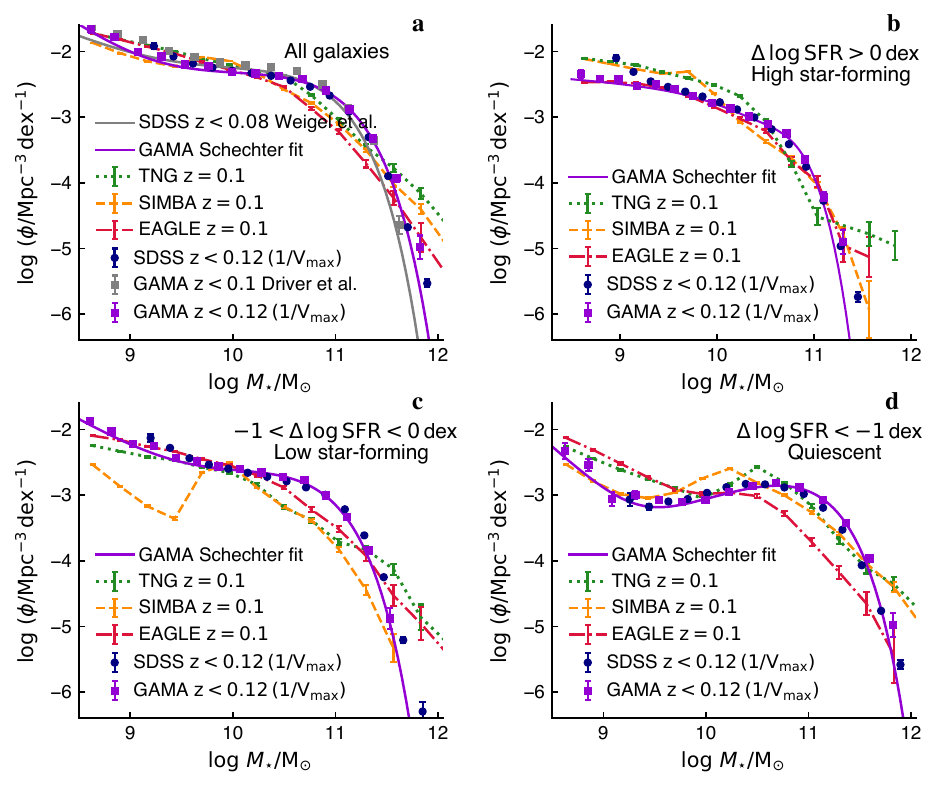}
\caption{\textbf{Galaxy stellar mass functions.} Stellar mass functions (SMFs) of galaxies from the Sloan Digital Sky Survey (SDSS), the Galaxy And Mass Assembly survey (GAMA), and three cosmological simulations (TNG, SIMBA, and EAGLE) at $z \sim 0.1$ are shown. Panel (a) presents the SMF for all galaxies, while panels (b)–(d) show SMFs for galaxy subpopulations ranked by decreasing star-formation rate: high star-forming galaxies (b), low star-forming galaxies (c), and quiescent galaxies (d). Error bars denote standard deviations estimated assuming Poisson statistics. The simulations are calibrated to match the total SMF in panel (a).}
\label{fig:SMF1}
\end{figure*}

\begin{figure*}
\includegraphics[width=0.99\linewidth]{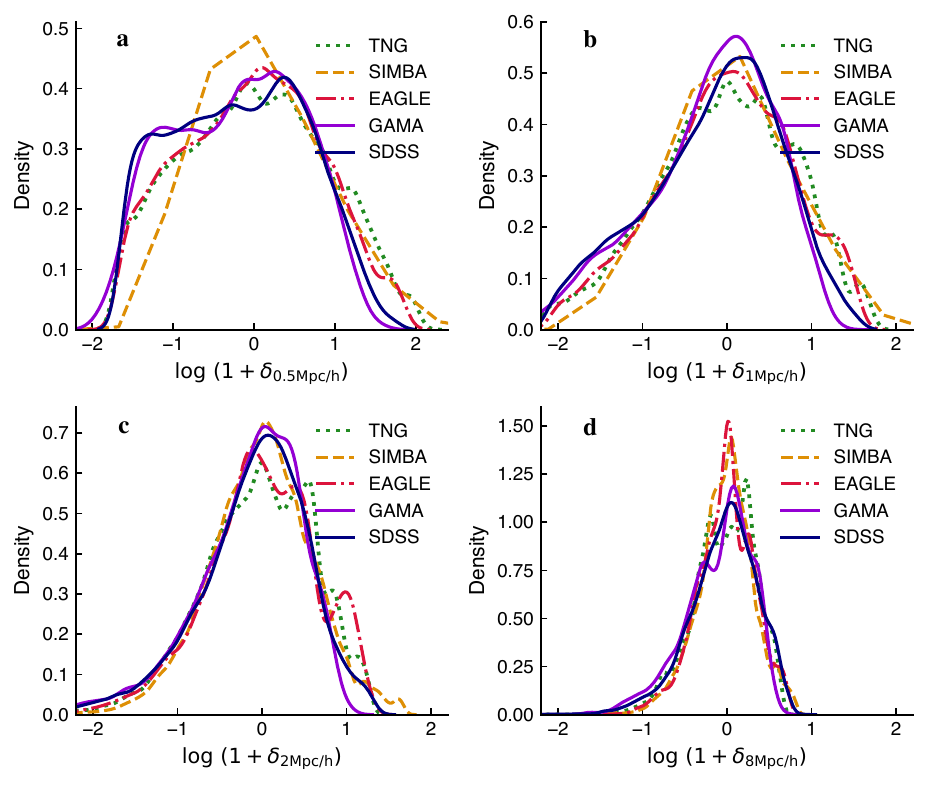}
\caption{\textbf{Comparison of stellar mass overdensity distributions across multiple scales}. Panels (a)–(d) show the stellar mass overdensity distributions measured on scales of 0.5, 1, 2, and 8\,Mpc/$h$, respectively, for galaxies at $z<0.12$ in SDSS and GAMA, compared with predictions from the TNG, SIMBA, and EAGLE simulations.}
\label{fig:multiDelta}
\end{figure*}

\begin{figure*}
\includegraphics[width=\linewidth]{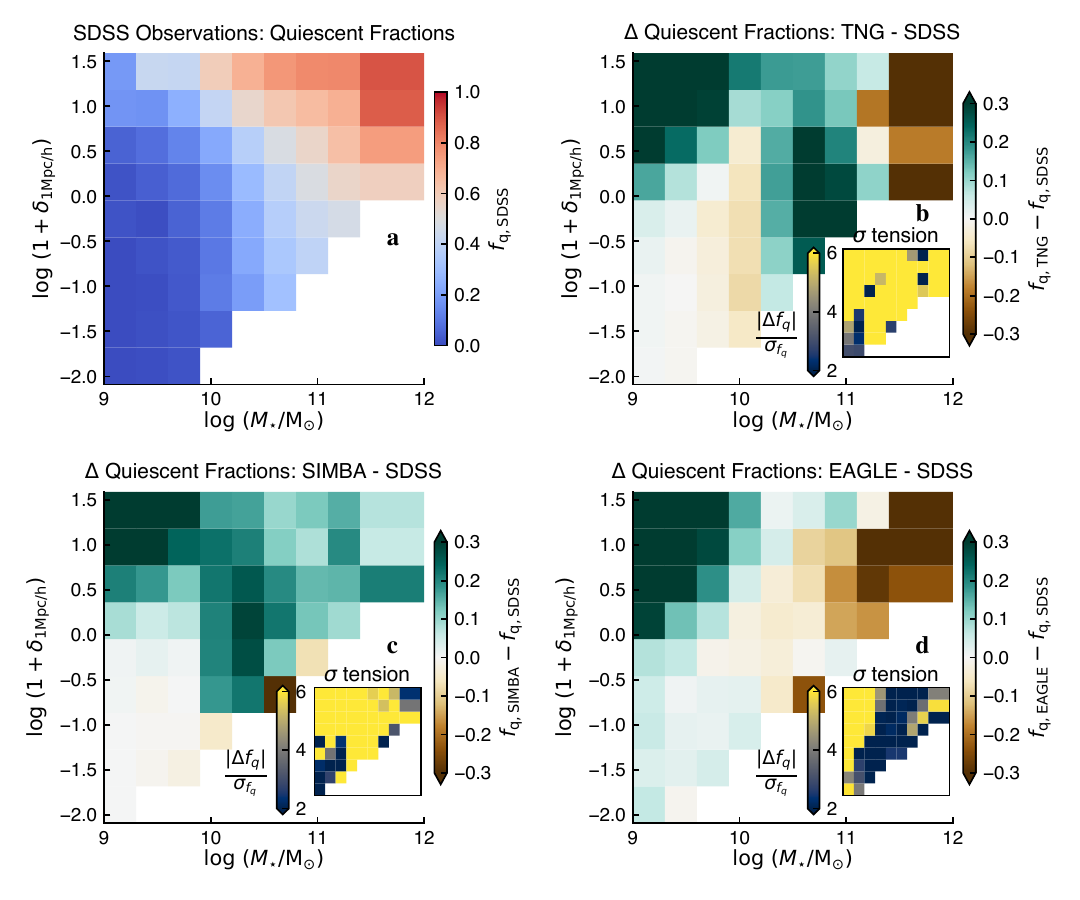}
\caption{\textbf{Trends of quiescent galaxy fractions with stellar mass and environment in SDSS and simulations.} Panel (a) shows SDSS quiescent fractions in the plane of stellar mass and stellar mass excess (overdensity) within 1\,Mpc/$h$ of each galaxy. Panels (b)--(d) display residuals between the quiescent fractions in each simulation and SDSS, where green (brown) indicates higher (lower) quiescent fractions in the simulation relative to SDSS for galaxies of the same mass in the same environment. Insets show $|\Delta f_q| / \sigma_{f_q}$, where $\sigma_{f_q}$ is obtained by adding in quadrature the binomial standard errors of the SDSS and simulation quiescent fractions, providing a quantitative measure of the statistical tension ($\sigma$ level).}
\label{fig:ME_SDSSsim}
\end{figure*}

\begin{figure*}
\includegraphics[width=0.99\linewidth]{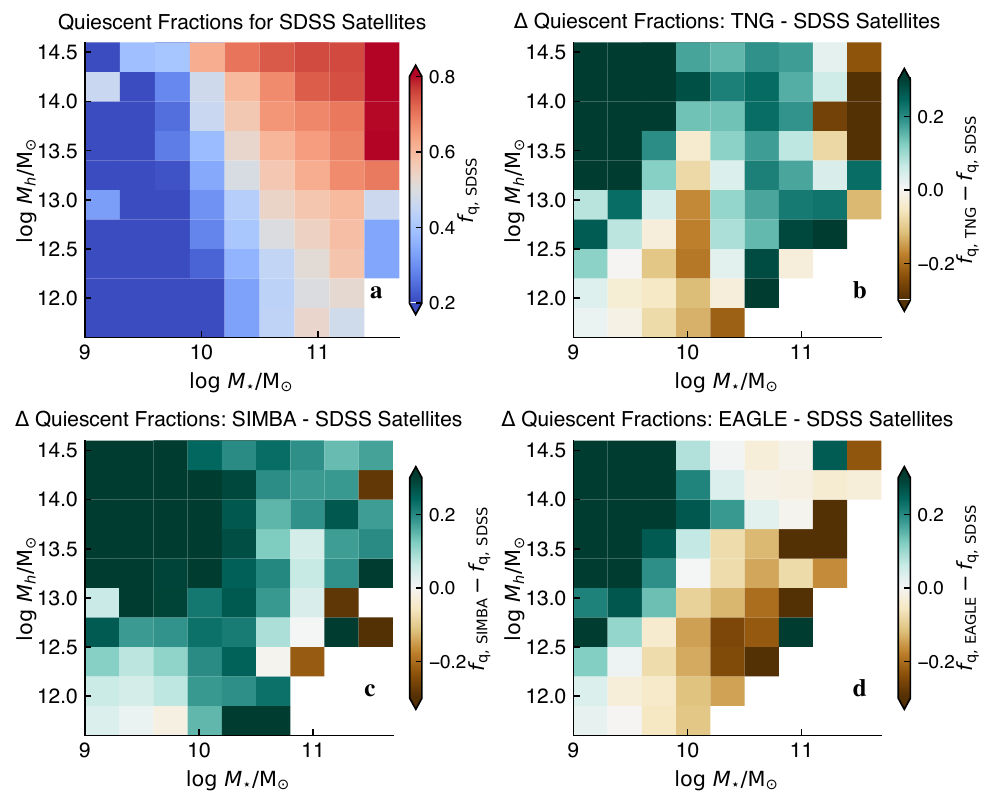}
\caption{\textbf{Quenched fractions of \emph{satellite} galaxies as a function of stellar and halo mass.} Panel (a) shows quiescent fraction trends for satellite galaxies in SDSS. Panels (b--d) show residuals between the quiescent fractions in each simulation and SDSS, where green (brown) indicates higher (lower) quiescent fractions in the simulation relative to SDSS at fixed stellar and halo mass. For SDSS, halo masses are derived from our new measurements combining multi-scale stellar-mass overdensities with additional observational halo-mass tracers. Panel (a) adopts the SIMBA-based halo-mass calibration, while panels (b--d) use the calibrations of the corresponding simulations.}
\label{fig:MsMh_SDSSsim}
\end{figure*}

\begin{figure*}
\includegraphics[width=0.99\linewidth]{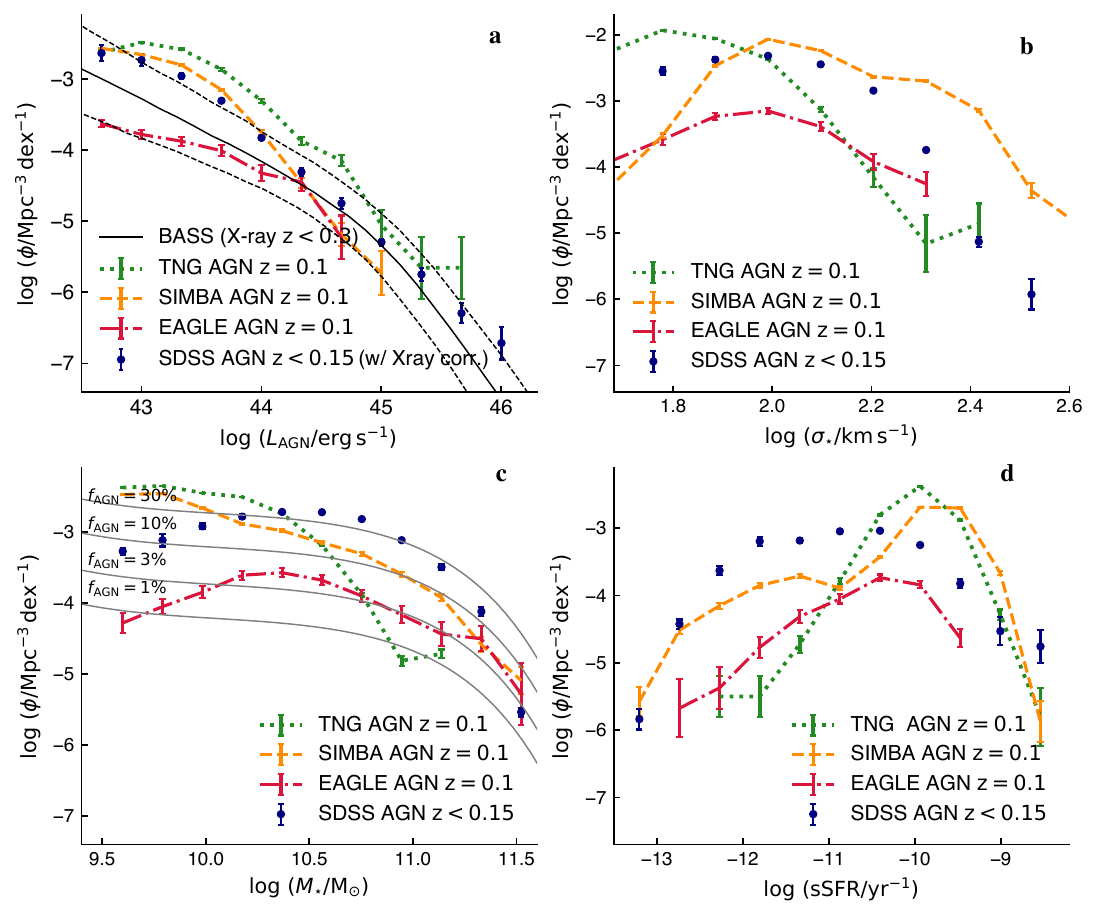}
\caption{\textbf{Demographics of active galactic nuclei (AGN) and their host galaxies} in the SDSS and the simulations. Panels show (a) AGN luminosity functions, (b) host stellar velocity dispersion functions (c) host stellar masses, and (d) host specific star formation rates (sSFR $\equiv \mathrm{SFR}/M_\star$). Error bars denote standard deviations assuming Poisson statistics. The median, 16th, and 84th percentiles of the fitted \textit{Swift}-BAT X-ray AGN luminosity function \citep{Ananna+22} are overplotted in panel (a) for comparison. Scaled-down versions of the global stellar mass function from Fig.~1a is shown in panel (c) to indicate the AGN fraction ranging from 1\% to 30\%.}
\label{fig:AGNfunc}
\end{figure*}

\begin{figure*}
\includegraphics[width=0.99\linewidth]{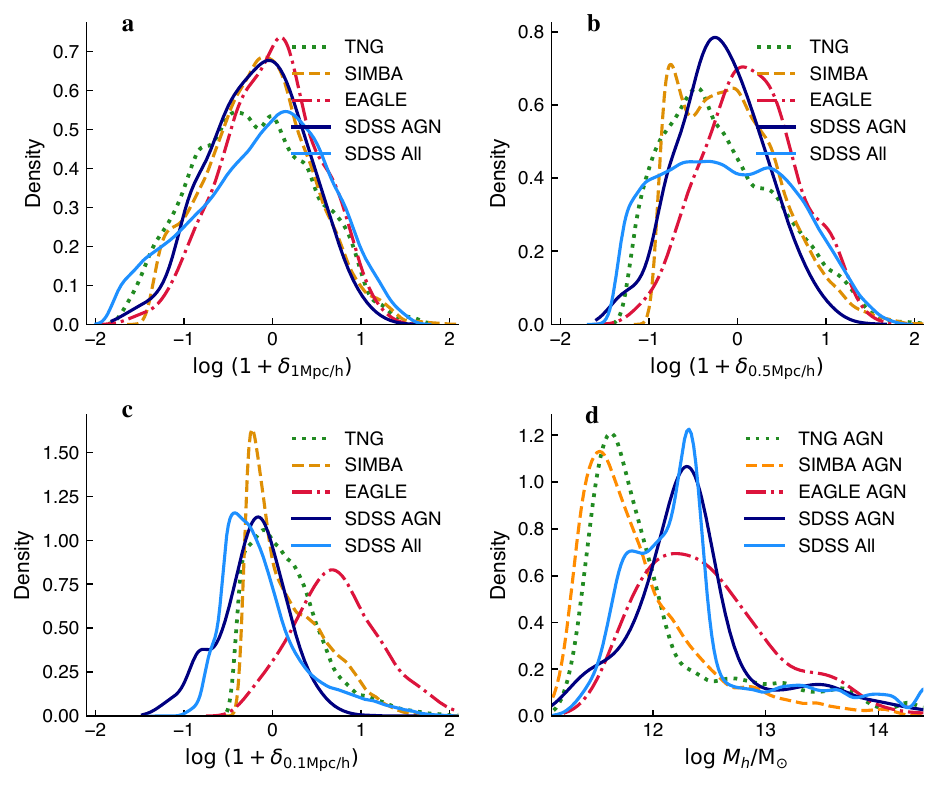}
\caption{\textbf{Distributions of halo-scale stellar mass overdensity and halo mass for simulated and SDSS AGNs and galaxies}. SDSS distributions are $1/V_\mathrm{max}$ weighted to correct for sample incompleteness. Panels (a)--(c) show stellar mass overdensity at scales of 1, 0.5, and 0.1\,Mpc/$h$, respectively. Panel (d) shows halo mass distributions, with SDSS AGNs in navy and the full SDSS sample in blue. \\
}
\label{fig:AGN}
\end{figure*}
\section*{Discussion} 

\subsection*{Stellar mass functions and star formation}

EAGLE, TNG, and SIMBA reproduce many galaxy observables through calibrated subgrid feedback models \citep{CrainvdVoort23, Crain+17}. Yet their simplified treatments of star formation, gas flows, and black hole growth lead to discrepancies when galaxies are divided by star formation and black hole activity or environment (Figs.~1, 3--5); none match stellar mass functions across all subpopulations. Still, these simulations broadly recover the cosmic star formation and stellar mass buildup to $z \sim 4$ \citep{CrainvdVoort23, Scharre+24, Furlong+15, Katsianis+17, 2019MNRAS.486.2827D, 2018MNRAS.473.4077P, Lagos+25}. 

Subdividing the galaxy population into star-forming and quiescent systems provides more physically informative constraints on feedback models. We demonstrate this using a suite of SIMBA and EAGLE feedback variants (~1). For instance, SIMBA matches the number density of low-mass ($M_\star \lesssim 10^{10}\,M_\odot$) quiescent galaxies better, primarily due to its stellar feedback model. However, this model also leads to an overproduction of SFGs above the star formation main sequence (SFMS) and an underproduction of galaxies below the SFMS or in transition. At higher stellar masses, the number density of QGs is already close to observed values even without AGN feedback, suggesting that the main challenge is controlling the overabundance of massive SFGs. SIMBA's AGN feedback---especially in its jet mode---is essential to suppress this excess, but it often does so at the expense of overproducing QGs or underproducing SFGs at intermediate masses ($M_\star \sim 10^{10} - 10^{11}\,M_\odot$), highlighting the difficulty in simultaneously matching all galaxy subpopulations.

Although SIMBA better reproduces the quiescent stellar mass function (Fig.~1d), many low-mass quiescent galaxies are up to three times larger than their SDSS counterparts \citep{2019MNRAS.486.2827D,Appleby+20} and are disproportionately found in dense environments (Fig.~3), a trend also seen in TNG and EAGLE. These discrepancies demonstrate that calibrating subgrid models solely to the global stellar mass function is insufficient.
 
\subsection*{Quenching in the right place at the right time}

EAGLE, TNG, and SIMBA broadly reproduce the global SMF and the stellar mass distribution on $\sim0.5$–$8\,\mathrm{Mpc}/h$ scales (Fig.~2), and predict a tight correlation between stellar mass within 2\,Mpc and halo mass. While all three qualitatively capture the environmental dependence of quiescent and star-forming galaxies, their quiescent fractions and SFR–density relations systematically deviate from observations.

\subsubsection*{Satellites}  

Our results, together with prior studies, highlight the critical roles of both AGN feedback and environmental mechanisms in quenching satellite galaxies within massive halos \citep{Hirschmann+14, Donnari+21}. Although current AGN activity is typically weak in such halos at $z = 0$, the strong correlation between black hole mass and halo mass suggests that AGN feedback at earlier epochs has influenced halo thermodynamics \citep{Zinger+20}. For massive satellites, AGN-driven heating may be the dominant quenching mechanism \citep{Donnari+21}.

In contrast, lower-mass satellites ($M_\star \sim 10^9$–$10^{10}\,M_\odot$) are more likely quenched through prolonged environmental effects, including ram pressure stripping, tidal interactions, and indirect AGN feedback from the central galaxy \citep{KangX+08,Dashyan+19}. Gas stripping generally operates on gigayear timescales \citep{Fillingham+15,XieL+20}, whereas AGN feedback can rapidly suppress star formation by heating or depleting halo gas. However, SIMBA and EAGLE variants without AGN feedback show that most low-mass satellites in massive halos still quench, indicating that environmental processes or internal gas exhaustion dominate (Extended Data Fig.~2).

Although TNG, EAGLE, and SIMBA simulations reproduce the basic signatures of gas stripping \citep{YunK+19}, they do not fully resolve the multiphase structure of the CGM and ISM. For numerical reasons, these simulations impose a $\sim 10^4\,\mathrm{K}$ temperature floor, preventing the formation of cold and molecular phases that respond differently to gas stripping than the warm diffuse component. They also do not resolve chaotic cold accretion (CCA)---a potentially important channel for black hole fueling and for regulating the heating–cooling balance in group and cluster halos \citep{Gaspari+17, Lovisari+21}---likely contributing to discrepancies with observed hot-halo thermodynamic profiles \citep{Chadayammuri+22, ZhangY+24}.

\subsubsection*{Centrals}

Quiescent fractions among central galaxies provide a useful constraint on AGN feedback efficacy in simulations, particularly for massive halos or dense environments (Extended Data Fig.~6). SIMBA predicts higher quiescent fractions for massive centrals ($M_\star \gtrsim 10^{11}\,M_\odot$) than TNG100 and EAGLE, most likely due to the strength of its jet-mode AGN feedback \citep{2019MNRAS.486.2827D}. However, this feedback scheme may also over-quench massive satellites. At intermediate masses ($M_\star = 10^{10}$–$10^{11}\,M_\odot$), SIMBA and TNG both overpredict quiescent fractions, EAGLE underpredicts them beyond $M_\star \sim 5 \times 10^{10}$–$10^{11}\,M_\odot$ by about 30\%, while TNG100 and TNG300 give inconsistent fractions (by $\sim$20--30\%) in the same mass range. These differences broadly mirror previously reported trends in cold gas \citep{Crain+17, Diemer+19, Dave+20, MaWenlin+22}: SIMBA retains excess cold gas at low to intermediate masses, TNG maintains too much cold gas in the outskirts of massive centrals, and EAGLE’s thermal AGN feedback depletes cold gas in low-SFR systems while leaving substantial hot gas in massive halos \citep{ZhangY+24}. Notably, SIMBA shows higher gas masses than observed at $M_\star \approx 1$--$3 \times 10^{10}\,M_\odot$ \citep{MaWenlin+22}, a range with persistent discrepancies across our analysis. 

\subsection*{AGN properties and their implications for feeding and feedback}

Differences in black hole accretion and AGN feedback implementations strongly influence both present-day AGN and host galaxy properties, as well as the cumulative co-evolution of galaxies and black holes \citep{Habouzit+21,Habouzit+22,Zinger+20,Terrazas+20,Appleby+20}. Although EAGLE, SIMBA, and TNG all include AGN feedback, they differ in black hole growth models and energy deposition, producing divergent predictions for AGN prevalence, quenching efficiency, and observable feedback signatures.

Comparing SDSS AGN and host properties with the three simulations at $z \lesssim 0.15$ reveals systematic mismatches (Fig.~5, Extended Data Figs.~4 and 5, and Supplementary Information). TNG overproduces luminous AGNs, primarily in lower-mass, star-forming hosts, and exhibits narrower Eddington ratio distributions than the other simulations or SDSS. EAGLE underpredicts luminous AGNs at high $M_\star$, while SIMBA overproduces them in low-mass star-forming or quiescent hosts but more closely matches the abundance in massive galaxies. All simulations underrepresent massive, low-luminosity AGNs in quiescent or green-valley hosts and fail to reproduce host galaxy velocity dispersion functions. Black hole mass distributions also differ: SIMBA overproduces low-mass BHs, TNG overproduces intermediate-mass BHs ($\log M_\mathrm{BH}/M_\odot \approx 7$–8), and EAGLE underpredicts across the same range.

These discrepancies reflect the influence of black hole seeding, accretion prescriptions, and feedback modes on the simulated AGN and host populations. In TNG, AGN feedback only transitions to the efficient kinetic mode once black holes exceed $\log (M_\mathrm{BH}/M_\odot) \approx 8$, corresponding to $\log (M_\star/M_\odot) \approx 10.5$ or $\sigma_\star \approx 100\,\mathrm{km\,s^{-1}}$ \citep{Weinberger+17,Terrazas+20}. This explains the sharp transition and the paucity of low-SFR AGNs in intermediate-mass hosts (Figs.~3, 4, and 5). By contrast, SIMBA overproduces AGNs in low-mass hosts ($M_\star \lesssim 10^{10}\,M_\odot$) due to early efficient black hole growth via cold gas accretion and the delayed onset of kinetic jet feedback \citep{Habouzit+21}.

All three simulations underpredict low-luminosity AGNs in low-SFR or gas-poor galaxies, as weak AGNs are quenched by strong feedback or unresolved low-level accretion. Conversely, in star-forming, gas-rich hosts, SMBHs in TNG and SIMBA may accrete too efficiently, overproducing AGNs at $M_\star \lesssim 5 \times 10^{10}\,M_\odot$ and highlighting broader limitations in modeling AGN fueling and quenching \citep{Katsianis+21}.

These low-redshift discrepancies likely reflect the cumulative impact of each model’s black hole accretion and feedback history. At $z \sim 1$–4, EAGLE reproduces the number densities of low-luminosity AGNs most accurately, whereas SIMBA and TNG tend to overpredict them \citep{Habouzit+22}. All three underpredict the abundance of luminous AGNs ($L_\mathrm{bol} \gtrsim 10^{45}\,\mathrm{erg\,s^{-1}}$) across cosmic time, with EAGLE showing the largest deficit. These rare, high-luminosity AGNs correspond to brief, high-Eddington accretion episodes that dominate black hole growth and quasar-mode feedback \citep{Weinberger+17,McAlpine+17}. The bright end of the AGN luminosity function is sensitive to short-timescale variability \citep{Habouzit+22} and to simulation limitations in box size, temporal sampling, and spatial resolution, which reduce the likelihood of capturing rare, luminous AGNs. Accurately modeling AGN and host galaxy properties, therefore, requires resolving the timing and physical triggers of accretion mode transitions \citep[][]{Weinberger+17}. In SIMBA and TNG, kinetically decoupled supernova feedback is less effective at removing or heating cold ISM gas, enabling excessive star formation and rapid black hole growth in low-mass galaxies, leading to AGN overabundance and excess molecular gas \citep{2019MNRAS.486.2827D,Dave+20,Habouzit+21}. By contrast, EAGLE’s thermal feedback more effectively regulates star formation in low-mass galaxies and delays black hole growth until hosts become more massive \citep{Bower+17,McAlpine+17}.

The contrasting AGN host populations in the three simulations highlight complementary strengths but also persistent weaknesses in current feedback prescriptions. Discrepancies in the quenched fractions of massive galaxies ($\log M_\star/M_\odot = 10$--11) in low-density environments (Fig.~3) and among massive centrals (Extended Data Fig.~6) underscore mismatches in AGN feedback implementation. In principle, combining features from different models could yield more realistic AGN populations, for example, integrating SIMBA’s gradual jet activation with EAGLE’s efficient thermal stellar feedback. However, because AGN feedback is coupled to black hole accretion, star formation, and stellar feedback, hybrid implementations might be challenging. Encouragingly, recent work has begun incorporating jet-mode feedback into the EAGLE framework \citep{Husko+24}.

Observationally, radio-mode AGNs are linked to massive, quiescent hosts \citep{Best+05,Jin+25}. Disabling jet-mode feedback in SIMBA significantly reduces the quenched fraction (Extended Data Figs.~1 and 2). By contrast, TNG300 overpredicts the quenched fraction at the massive end while TNG100 underpredicts it (Extended Data Fig.~6) \citep{Donnari+21fq}. In EAGLE, the primary issue is the underproduction of both massive weak AGNs and quenched galaxies. 

AGN prevalence in low-mass halos and low-density environments reflects both the availability of cold gas and the regulatory effect of stellar feedback, which limits black hole growth \citep{Bower+17,McAlpine+17}. Cold gas accretion is efficient in these regions because processes that dominate in dense environments---shock heating, strangulation, and ram-pressure stripping---have little impact \citep{DekelBirnboim09,Catinella+13,Stark+16}. The halo mass distribution of AGNs in Fig.~6 broadly aligns with rapid black hole growth and high accretion rates, peaking around $M_h \approx 10^{12}\,M_\odot$ in EAGLE \citep{Bower+17,McAlpine+17}.

Beyond gas availability, AGN fueling in low-density environments can be enhanced by mergers and secular processes that drive gas inward. Structural asymmetries---including bars, nuclear spirals, and lopsided features---redistribute angular momentum and promote inflow \citep{Storchi-Bergmann+19}. These asymmetries are common in star-forming galaxies \citep{Yesuf+21,Bottrell+24}, although additional mechanisms, such as gravitational torques and angular momentum dissipation, are needed for gas to reach the nucleus \citep{HopkinsQuataert+11,Angles-Alcazar+17acc}. 

Over long timescales, AGNs inject turbulence and thermal energy into their halos \citep{Zinger+20}, shaping the circumgalactic medium and regulating star formation. Although strongly accreting AGNs are rare in massive halos, simulations emphasize the growing role of low-accretion, maintenance-mode feedback at late stages \citep{Weinberger+17,2019MNRAS.486.2827D}. Observationally, matched samples show no significant SFR differences between AGN and inactive galaxies, suggesting AGN feedback is weak, delayed, or masked by variability \citep{Harrison17,Ward+22}.

SIMBA produces some massive, quiescent AGN hosts (Fig.~5d), but their abundance remains insufficient and may be artificially boosted by overly efficient jet-mode feedback \citep{Zheng+22}. SIMBA also underproduces starbursts relative to observations \citep{Zheng+22}, indicating that its rapid-quenching channel may not represent the typical starburst-to-post-starburst (PSB) transition \citep[see also][]{Akins+22}.

Variable AGNs capable of inducing rapid quenching would produce a substantial population of recently quenched AGN-off galaxies. In contrast, nearby PSBs are rare, comprising less than 1\% of the local population \citep{Yesuf+14,Wild+16}, and purely AGN-off PSBs are even rarer. Many PSBs exhibit weak AGN signatures, emphasizing that AGN-induced rapid quenching is uncommon. The typical AGN variability timescale is an order of magnitude shorter than the PSB visibility time, so the two phases rarely overlap. If rapid, AGN-driven quenching were common, AGN-off PSBs would be more frequent; their observed scarcity therefore implies that AGN feedback seldom drives rapid quenching, despite overall AGN fractions of $\sim$10--30\% (Fig.~5c).

Strongly accreting AGNs at low redshift preferentially reside in low-density environments (Fig.~6), unlike quiescent galaxies that typically occupy massive halos \citep{yesuf22}. Their relationship is not one of direct or immediate causality: radiative-mode AGNs such as nearby Seyferts and quasars are not the direct progenitors of today’s massive quenched galaxies, whose stellar populations also show they quenched long ago \citep{Thomas+05}. Most luminous optical AGNs are gas-rich and actively star-forming \citep{YesufHo20,Ward+22}, and rapid AGN-driven quenching appears uncommon. Together, these trends imply that radiative AGN feedback at low redshift is generally inefficient at suppressing star formation for the majority of AGNs. Nonetheless, quiescent or post-starburst galaxies in low-density regions, and AGNs with strong outflows also serve as important laboratories for diagnosing recent feedback episodes. Ultimately, halo-scale environment regulates the cold-gas supply, star formation, and AGN fueling---whether through secular processes or mergers \citep{DekelBirnboim09,KormendyHo13}---and can itself be influenced by long-term cumulative AGN activity \citep{Zinger+20}, highlighting the interconnected roles of internal processes and environment in shaping galaxy and black hole evolution.

\subsection*{Summary}

Building on prior successes and discrepancies in cosmological simulations \citep[e.g.,][]{Donnari+21fq,Habouzit+21,Habouzit+22,Katsianis+21,Ward+22}, we use a large, completeness-corrected dataset to place stringent observational constraints on the coevolution of nearby galaxies, their environments, and active SMBHs in the EAGLE, TNG, and SIMBA simulations. A unified comparison across multiple galaxy and AGN observables shows that none of the simulations reproduce the stellar mass functions (SMFs) of galaxy subpopulations at $z \sim 0.1$ (Fig.~1). Each displays systematic inconsistencies in quiescent fractions and star formation rates across stellar mass and environment (Fig.~3). Accounting for multiscale environment, we find that low-mass satellites are overquenched, largely independent of feedback model, while the abundances of massive centrals and satellites remain inaccurate and feedback-sensitive (Fig.~4, Extended Data Figs.~2, and 6). AGN demographics reveal further discrepancies, with each simulation diverging from observations to varying degrees (Fig.~4, 5, and Extended data Fig.~4).

These tensions show that current feedback models and the simulation choices and parameters that influence environmental and gas-dynamical outcomes cannot accurately reproduce the observed quiescent fractions or AGN demographic distributions. Limitations in both resolution and volume further restrict the simulations’ ability to capture long-term galaxy assembly, black hole growth, and the interplay between feedback and environment. We also identify regimes where simulations are in strong mutual tension—differences that are directly testable with observations. Our new constraints and environmental benchmarks provide practical diagnostics for calibrating or testing future models (Extended Data Figs.~1 and 2). Nonetheless, all three simulations broadly reproduce the multiscale stellar mass distribution of the general galaxy population (Fig.~2) and the tendency for AGNs to inhabit low- or average-density environments and low-mass halos (Fig.~6).

Correctly assigning galaxies and AGNs their quiescence status within the cosmic web requires improved treatments of subgrid SMBH and stellar feedback, along with more accurate modeling of multi-phase ISM and CGM, gas cooling and flows, and environmental processes such as ram-pressure stripping and tidal interactions. Further progress depends on advances in unresolved physics, larger, higher-resolution simulations, and hybrid approaches that integrate complementary frameworks guided by observational constraints. Deep, highly complete spectroscopic surveys with robust identification of weak AGNs are equally crucial, providing the improved empirical constraints needed to guide the next generation of cosmological models.

\section*{Methods}\label{sec:method}

\subsection*{The sample selection}

Our nearby AGN candidate sample ($N=191{,}654$), drawn from the Sloan Digital Sky Survey \citep[SDSS;][]{Abazajian+09}, includes both narrow-line and broad-line systems. To ensure coverage of H$\alpha$, we first restrict the parent sample to $z<0.35$. For most of our analysis, however, we adopt a lower redshift cut, $z<0.15$ ($N=149{,}988$), which minimizes incompleteness while preserving the qualitative trends observed for the whole sample. Additional cuts are applied to define subpopulations---$M_\star > 3 \times 10^{9}\,M_\odot$, $L_\mathrm{AGN} > 3 \times 10^{42}$\,erg\,s$^{-1}$, and data-quality/completeness thresholds---yielding subsamples of typically $5\times 10^4$--$8\times 10^4$ objects. 

Narrow-line AGNs (NLAGNs) are identified using the standard BPT diagnostic diagram, based on [N\,II]$\lambda6584$/H$\alpha$ and [O\,III]$\lambda5007$/H$\beta$ ratios \citep{Kauffmann+03}, requiring emission-line signal-to-noise ratios (S/N) $>2$ and continuum S/N $>5$. This selection captures pure AGNs ($N=76{,}656$ Seyferts + Low-Ionization Nuclear Emission-line Regions (LINERs)) and composite objects ($N=109{,}249$). Our results remain qualitatively unchanged when adopting a stricter H$\alpha$ equivalent width cut of $>3$\,{\AA} for pure AGNs to reduce contamination from retired galaxies and by removing composite objects altogether. However, X-ray and SED-fitting comparisons provide no strong evidence that LINERs or composites are non-AGNs (a point for future study). We therefore find no compelling justification for excluding weak AGNs, but the resulting luminosity function may be treated as an upper limit in this regime. Even under this conservative selection, both TNG and SIMBA still overpredict AGN abundances, whereas EAGLE shows comparatively better agreement when composites and low-EW LINERs are excluded. This underscores the need for future observations to better constrain the weak AGN population.

The broad-line AGNs (BLAGNs) were selected from a previous catalog \citep{Liu+19}, which identifies 13{,}603 BLAGNs among objects classified as galaxies or quasars at redshifts $z < 0.35$ in SDSS data release (DR7). This catalog provides detailed spectral properties, including line and continuum fluxes, line widths, and equivalent widths. To checks the reliability of this catalog for our study, we independently performed spectral analysis and remeasured key line fluxes. As the catalog does not include [O\,II] $\lambda\lambda3727$\,{\AA} fluxes, we use our own measurements for this line. 

We also analyze $\sim 9,000$ galaxies at $z<0.35$ from the 4XMM-DR14 catalog \citep{Webb+20} within the SDSS footprint, which helps correct for incompleteness and selection biases inherent in optical AGN identification (see a later section for details). From this sample, we select X-ray AGNs (approximately 2,500) based on detection significance, source quality, and hard-band flux signal-to-noise ratios. Specifically, we retain only point sources with a total detection likelihood $\mathtt{SC\_DET\_ML} > 6$ and a summary quality flag $\mathtt{SUM\_FLAG} \leq 1$. We also require a minimum flux SNR of $\mathtt{sc\_ep\_4\_flux} / \mathtt{sc\_ep\_4\_flux\_err} > 1$ and $\mathtt{sc\_ep\_5\_flux} / \mathtt{sc\_ep\_5\_flux\_err} > 1$ in Bands 4 (2.0--4.5~keV) and 5 (4.5--12~keV), respectively.

Although our optical selection primarily captures moderate- to high-luminosity AGNs ($L_\mathrm{bol} > 10^{42}\,\mathrm{erg\,s^{-1}}$), we note that spatially resolved diagnostics can reveal additional low-luminosity AGNs, particularly in dwarf galaxies, that are missed by integrated spectra \citep{Mezcua+24}. These faint AGNs, often undetected in X-rays and falling below our luminosity threshold, are not expected to impact our results, which focus on the more luminous AGN population.

The comparison sample of over 500,000 non-AGN galaxies is selected from the GALEX-SDSS-WISE Legacy Catalog (GSWLC-X2) \citep{Salim+18}. This catalog integrates data from the Galaxy Evolution Explorer (GALEX), SDSS DR10, and the Wide-field Infrared Survey Explorer (WISE).  The catalog provides measurements such as stellar mass and SFR derived from SED fitting of UV-optical photometry, supplemented by IR luminosity constraints using the CIGALE code \citep{Boquien+19}. The SED fitting incorporates two exponential star formation histories: a younger population (100\,Myr--5\,Gyr) and an older population (formed 10\,Gyr ago), with varying young mass fractions (0–50\%). Stellar population models \citep{BruzualCharlot03} were computed for four metallicities ($0.2–2.5 Z_\odot$) and a Chabrier \citep{Chabrier03} initial mass function (IMF).

Furthermore, the GAMA survey provides valuable data for our study, complementing SDSS. GAMA DR4 includes over 330,000 galaxies with $r < 19.7$\,mag across 286\,deg$^2$, reaching a median redshift of 0.2 (15th--85th percentile range: 0.1--0.3) \citep{Driver+22}. With 95\% redshift completeness, it is well-suited for precise studies of low-redshift galaxy populations and groups.

For our analysis, we utilize the GAMA Galaxy Group Catalog \citep{Robotham+11} and stellar population parameters derived using the energy-balance MAGPHYS code \citep{daCunha+08}. Similar to CIGALE, MAGPHYS employs the Bruzual \& Charlot \citep{BruzualCharlot03} stellar population model, assuming a Chabrier \citep{Chabrier03} IMF, exponential star formation histories, and a two-component dust attenuation model \citep{CharlotFall00}. The code accounts for energy absorbed by dust in UV–optical wavelengths and re-emitted in the IR by various dust components. For this work, we use (median) $M_\star$ and SFR values averaged over the past 0.1\,Gyr, adjusting them by 0.18\,dex and 0.25\,dex, respectively, to align with SDSS measurements \citep{Salim+18}, ensuring consistency between the datasets.

\subsection*{Measuring properties of observed AGNs and their host galaxies}

This subsection details the methodologies employed to measure key properties of observed AGNs and their host galaxies, including stellar mass, star formation rate, AGN luminosity, and environmental characteristics. It also describes the strategies used to correct these measurements for observational biases and incompleteness.

We assume a $\Lambda$ Cold Dark Matter ($\Lambda\mathrm{CDM}$) cosmological model with a Hubble constant $H_0 = 67.74 \, \mathrm{km \, s^{-1} \, Mpc^{-1}}$ ($h = \frac{H_0}{100} = 0.6774$), and a matter density parameter $\Omega_m = 0.3089$.

\subsubsection*{Stellar mass, SFR, and AGN luminosity from SED fitting}

The GSWL catalog does not measure BLAGN properties and corrects NLAGN IR luminosities using [O\,III] $\lambda5007$\,\AA\ equivalent width (EW) prior to SED fitting, without AGN templates. In contrast, we adopt a different SED modeling approach to: (1) measure star formation rates (SFRs) and AGN luminosities for unobscured AGNs---unprecedented for this sample---and (2) estimate these properties for all AGNs using a unified methodology and updated multiwavelength data. These improvements reduce selection biases, improve measurement reliability, and broaden the range of AGN luminosities and host properties explored.

We use the CIGALE code to fit multiwavelength UV-to-IR photometry. Briefly, our model includes stellar emission, ionized gas emission, dust emission by stellar light, AGN contributions, and dust attenuation using a modified starburst attenuation law. We adopt a delayed star formation history (SFH) model, $\mathrm{SFR(t)} \propto t \exp(-t/\tau_\mathrm{main}$), where $\tau_\mathrm{main}$ denotes the e-folding time of the main stellar population. 
 To account for recent star formation, this is supplemented by an exponentially declining burst component characterized by durations $\tau_\mathrm{burst}$ of 50, 100, and 200 Myr, and with the burst mass fraction spanning 0\%–30\%. The stellar component is represented using stellar population models from Bruzual \& Charlot \citep{BruzualCharlot03}, with a Chabrier \citep{Chabrier03} IMF and solar metallicity. Fixing the metallicity to solar is a reasonable approximation, given that AGNs are primarily hosted by massive galaxies and the limited data's inability to distinguish metallicity effects. The reprocessed IR dust emission of UV/optical stellar radiation is modeled with semi-empirical templates \citep{Dale+14}, without AGN contribution. The AGN emission was modeled using the \texttt{SKIRTOR} module \citep{Stalevski+16}, a two-phase torus model incorporating polar dust emission with a single modified blackbody. When available, our fitting includes photometric data from GALEX \citep{Martin+05}, SDSS \citep{Abazajian+09}, PanSTARRS \citep{Chambers+16}, 2MASS \citep{Skrutskie+06}, UKIDSS \citep{Hewett+06,Lawrence+07}, WISE \citep{Wright+10,Eisenhardt+20}, and IRAS 60$\mu$m \citep{Neugebauer+84}. 
 
We implement simple aperture corrections based on calibrations using aperture-matched photometry of nearby galaxies in the GAMA survey. Additionally, the photometry is corrected for Galactic extinction \citep{Blanton+05,SchlaflyFinkbeiner11}. SED modeling is complex with many parameters, and limited data may not constrain them all.  However, key parameters $M_\star$, SFR, $A_V$, and AGN accretion power ($L_\mathrm{AGN}$)--are well-constrained by the data, as also supported by good agreement with previous estimates.

We assess goodness-of-fit using the reduced chi-square ($\chi^2_\nu$) and the consistency of stellar mass estimates, comparing the best-fit mass ($M_{\star,\mathrm{best}}$) with the likelihood-weighted mean ($M_{\star,\mathrm{bayes}}$) from all \textsc{CIGALE} templates \citep{Boquien+19,ZhuangHo23}. We exclude 4\% of AGNs with $\chi^2_\nu > 5$ or $|\log(M_{\star,\mathrm{best}} / M_{\star,\mathrm{bayes}})| > 0.5$\,dex. The median $\chi^2_\nu$ is 0.7, with $\sim 90$\% of fits below 2. Median uncertainties are 0.1\,dex for $M_\star$ and 0.2\,dex for SFR, indicating well-constrained parameters.

AGN luminosities have median uncertainties of 0.6\,dex for $L_\mathrm{AGN} > 3 \times 10^{42}$\,erg\,s$^{-1}$ and 0.2\,dex for $L_\mathrm{AGN} > 3 \times 10^{43}$\,erg\,s$^{-1}$. For weak AGNs, where dust-corrected [O\,III]$\lambda5007$\,\AA\ luminosities are reliable (error $<0.3$\,dex) but SED-derived $L_\mathrm{AGN}$ is poorly constrained (error $>0.5$\,dex), we estimate $L_\mathrm{AGN}$ using the median relation with [O\,III] luminosity and SFR: $\log L_\mathrm{AGN} = 0.274 \, \log L_\mathrm{[O\,III],dc} + 0.461 \, \log \mathrm{SFR} + 32.4.$

The velocity dispersion measurements of NLAGNs from the SDSS database are aperture-corrected to the half-light radius or to $r=2.5$\,kpc using a power-law profile, $\sigma(r) \propto r^\alpha$, with $\alpha = -0.033$ \citep{deGraaff+21}. For simulations, we measure velocity dispersion within $r=2.5$\,kpc. When fiber velocity dispersions are poorly measured ($S/N < 2$, 3.5\% of cases), we substitute them using the empirical relation  $\log \sigma_f = 0.5\log M_\star - 0.5\log R_{50} - 2.88$,  where $R_{50}$ is the $i$-band half-light Petrosian radius.  

For BLAGNs, black hole masses ($M_\mathrm{BH}$) are estimated from H$\alpha$ luminosity and FWHM \citep{Mejia-Restrepo+22}. Their velocity dispersions are then derived from the $M_\mathrm{BH}$--$\sigma$ relation \citep{Greene+20}, with aperture corrections to the half-light radius when available \citep{ZhuangHo23}, and otherwise (in $\sim$20\% of cases) from their median mass--radius relation. Conversely, for NLAGNs in the Supplementary Information, we adopt the same procedure in reverse to estimate $M_\mathrm{BH}$ from their velocity dispersions.  

Star-forming galaxies exhibit a tight correlation between stellar mass ($M_\star$) and SFR, commonly referred to as the star formation main sequence (SFMS). Previous studies have found that the slope of this relation is approximately $\alpha \approx 0.8$ \citep{Donnari+19}. Adopting this value, which we have confirmed to be appropriate for our sample, yields a median linear intercept of $-8.1$. We use this slope and intercept to define the offset from the SFMS, and classify galaxies as quiescent if their SFR lies more than one dex below the SFMS, i.e., $\Delta\log\mathrm{SFR} < -1$.

\subsubsection*{Environmental indicators}

The environmental indicators are based on spectroscopic measurements from SDSS DR17 \cite{yesuf22}. We compute stellar mass overdensities, $\delta_{x\mathrm{Mpc}}$, within apertures of radii $x \in \{0.1, 0.5, 1, 2, 4, 8\}\; h^{-1}$Mpc, using neighbors with rest-frame velocity offsets $|\Delta v| < 1000$ km s$^{-1}$ to exclude line-of-sight contaminants. Stellar masses are derived from SDSS spectra using PCA modeling \citep{ChenY+12}, or from $i$-band mass-to-light ratios and $g-i$ colors when unavailable, and agree with GSWLC estimates for DR7 \citep{Salim+18}. Overdensities are normalized within narrow redshift bins to reduce spectroscopic incompleteness.

We identify central and satellite galaxies using group catalogs from SDSS \citep{YangX+07} and GAMA \citep{Robotham+11}. In the SDSS catalog, galaxies are grouped using an iterative halo-based method, with the central galaxy defined as the most massive member based on stellar mass, and the remaining members classified as satellites. To ensure consistency with the stellar mass estimates used throughout this study, we re-identified centrals in SDSS groups using the same $M_\star$ measurements. This reclassification alters the central-satellite designation for only about 3\% of group members.

In the GAMA catalog, groups are identified using a Friends-of-Friends algorithm with linking lengths optimized using realistic mock catalogs. Among the available options for defining the central galaxy, we adopt the iterative center as our fiducial choice. This method selects the brightest galaxy after iteratively rejecting outliers from the r-band center of light. It yields results comparable to the traditional brightest cluster galaxy (BCG) definition \citep{Davis+19} and is more robust in recovering group centers in mock catalogs \citep{Robotham+11}. Despite differences in central definitions between the SDSS and GAMA catalogs, we find consistent results across the two datasets, and therefore do not reassign centrals in GAMA based on stellar mass. For both surveys, group velocity dispersions are estimated using the gapper estimator \citep{Beers+90}.

\subsubsection*{Correcting for observation bias and incompleteness: $1/V_\mathrm{max}$ method}
Our analysis carefully corrects for observational limits and leverages two complementary surveys (GAMA and SDSS). We correct for Malmquist bias using the $1/V_\mathrm{max}$ method \citep{Pozzetti+10, Baldry+12, Weigel+16}, estimating number densities for galaxies and AGNs as a function of stellar mass, sSFR, black hole mass, and AGN luminosity. The maximum volume, \( V_\mathrm{max} \), is calculated based on the survey flux limits: $r_\mathrm{lim} < 17.77$ for SDSS main galaxies, $i_\mathrm{lim} < 19.1$ for QSOs, and $r_\mathrm{lim} < 19.7$ for GAMA galaxies. We apply k-corrections using color-redshift relations \citep{Chilingarian+10} and dust corrections based on the Calzetti extinction curve \citep{Calzetti+00} to compute the luminosity limits. Weighting each object by \( 1/V_\mathrm{max} \) corrects for the over-representation of luminous objects and yields less biased number densities.

To calculate number densities as a function of stellar mass, we account for variations in the mass-to-light ratio ($M/L$). For each galaxy, we estimate the stellar mass limit from its luminosity limit, assuming a constant $M/L$ over the low-$z$ interval, following previous work. The stellar mass completeness limit as a function of redshift is then derived using 95\% quantile regression.

The incompleteness correction for AGNs accounts for both the survey magnitude limit and the signal-to-noise (S/N) requirements of emission-line measurements. For the continuum limit, we determine the luminosity corresponding to the survey flux threshold in the $r$- or $i$-band (depending on the selection method), and fit the relation between $\log L_{r}$ or $\log L_{i}$ and $\log z$ to define the 95\% completeness boundary.

Spectroscopic incompleteness is treated separately. For BLAGNs we adopt broad H$\alpha$ as the limiting feature, while for NLAGNs we use narrow H$\beta$, the weakest of the BPT lines and thus most restrictive. Line ratios show no systematic redshift trends for composite galaxies, suggesting host dilution is not dominant, and the FWHM of broad H$\alpha$ shows no significant redshift dependence. The accessible volume $V_{\max}$ for each AGN is defined by the more restrictive of the continuum and emission-line limits. For the velocity dispersion function (VDF) and black hole mass function (BHMF) we further restrict to NLAGNs with median continuum S/N $>$ 10 (results are similar for S/N $>$ 5). For the AGN SMF, we apply the stellar mass completeness limit in the same manner as for inactive galaxies.

For the AGN luminosity function, both continuum and emission-line limits are adopted. This is justified by the strong correlation between AGN luminosity and continuum or broad H$\alpha$ luminosity in BLAGNs, and between AGN luminosity and [O\,III] luminosity in NLAGNs, whose detectability is already constrained by H$\beta$. Refining the limits using the faintest 20\% of AGNs or those with $f_{\mathrm{AGN}} < 20\%$ yields similar results.

For number densities as a function of black hole mass, velocity dispersion, or sSFR, we adopt $V_{\max}$ from the minimum of the stellar mass, line, and continuum limits. This approach is supported by correlations between BLAGN luminosities and $M_{\mathrm{BH}}$ ($\rho \approx 0.6$) and between NLAGNs and the $M_{\star}$--$\sigma_{\star}$ relation, although scatter in these scaling relations are significant. As an alternative check, we assume a constant $M_{\mathrm{BH}}/L_{\mathrm{AGN}}$ ratio and scale the continuum limits accordingly, finding consistent results. The distribution of Eddington ratios also shows no significant redshift dependence, further supporting this approximation. Finally, to account for AGNs missed by optical selection, we apply an empirical X-ray completeness correction, as described at the end of this section. The combined BPT and broad-line selections recover $>90$\% of X-ray AGNs when emission lines are measurable; the differences are mainly due to incomplete spectroscopic coverage or the absence of detectable lines.

The $V_\mathrm{max}$ method assumes a uniform galaxy number density, but deviations from large-scale structure (LSS) and galaxy clustering can skew results \citep{Baldry+12}. To correct for this, we adjust $V_\mathrm{max}$ by multiplying it by the ratio of the galaxy number density between $z_\mathrm{min}$ and $z_\mathrm{max}$ to the average density in the reference redshift range, $z=0.02-0.12$. To reduce the impact of incompleteness, the LSS density-defining population is set at $\log M_\star > 10$ for SDSS and $\log M_\star > 9.5$ for GAMA, with the results showing little sensitivity to these mass thresholds. No density correction is applied to AGNs. For inactive galaxies, the SMF accounts for position-dependent spectroscopic completeness in SDSS \citep{YangX+07} and the 83\% sky coverage of the SDSS DR7 survey area (7,748 deg$^2$) using the GSWLC.

In addition to the non-parametric $1/V_\mathrm{max}$ method, we use the \texttt{dftools} code to fit a double Schechter function to model the stellar mass function \citep{Obreschkow+18}. Using the Modified Maximum Likelihood (MML) estimation, \texttt{dftools} provides a Bayesian framework for directly fitting distribution functions without binning the data and includes corrections for measurement errors (Eddington bias). Both approaches yield similar results.

To correct for incompleteness and selection bias in the optical AGN sample, we derive correction factors for AGN and host galaxy properties using a control sample of X-ray-selected AGNs. X-ray AGNs are defined as sources whose hard X-ray (2–10\,keV) luminosities deviate by more than $3\sigma$ (0.5\,dex) above the mean $L_\mathrm{X}$–$M_\star$–SFR relation \citep{Lehmer+16}. The 2–12\,keV fluxes from the \textit{XMM-Newton} catalog \citep{Webb+20} are converted to luminosities, which are then scaled by a factor of 0.9 to obtain 2–10 keV luminosities. The optical AGN sample misses X-ray AGNs because they either do not have SDSS spectra, lack strong emission lines, or have emission line ratios dominated by star formation.

To ensure consistency, we perform SED fitting for the X-ray AGNs using the same procedure as for the optically selected AGNs. We then compare the SED-derived properties of X-ray AGNs with those in our parent optical sample. Specifically, in each bin of ($L_\mathrm{AGN}$), (M$_\star$), and sSFR, we compute the fraction of X-ray AGNs that are also classified as optical AGNs, weighting each object by its optical $V_\mathrm{max}$. The inverse of this fraction is then applied as a completeness correction to the $V_\mathrm{max}$-corrected optical AGN counts. X-ray and optical sources are matched using a $5^{\prime\prime}$ search radius. The velocity dispersion correction is based on the subset of type 2 AGNs with available SDSS measurements in both samples.

To roughly estimate the mean correction and its uncertainty for a given property, we perform 500 bootstrap resamplings, drawing perturbed values for X-ray only AGNs and AGNs in both samples from truncated normal distributions (cut at $2\sigma$), binning the corresponding property arrays, and recomputing the correction factors in each iteration. The total uncertainty on the corrected AGN number densities accounts for both the counting error on the observed optical AGN sample and the uncertainty in the completeness correction factor. These two uncertainties are combined in quadrature to produce a reliable error estimate. In sparsely populated bins where the correction factor is poorly constrained, we replace it with the median value from well-sampled bins and assign an uncertainty equal to $\sqrt{2}$ times the median fractional error. We adopt the X-ray correction factor estimated from the subsample of AGNs at $z < 0.15$, where both optical and X-ray selections are more complete. The corrections typically remain below a factor of two (i.e., less than 50\% incompleteness) in most bins.

\subsection*{Description of the simulations}

We compare our results with three prominent cosmological simulations: EAGLE \citep{2015MNRAS.450.1937C, 2015MNRAS.446..521S}, IllustrisTNG \citep{2018MNRAS.475..676S, 2018MNRAS.475..648P, 2018MNRAS.475..624N, 2018MNRAS.477.1206N, 2018MNRAS.480.5113M}, and SIMBA \citep{2019MNRAS.486.2827D}. Each simulation provides a unique model for the co-evolution of gas, stars, black holes, and dark matter, with distinct physical models and numerical methods for solving gravity and hydrodynamics. The simulation data are publicly available. This section presents the salient features of the simulations, the data used, and our approach to calculating environmental overdensities of simulated galaxies in a manner consistent with the observational sample.

\subsubsection*{IllustrisTNG}
The IllustrisTNG project \citep{2018MNRAS.475..676S,2018MNRAS.475..648P,2018MNRAS.473.4077P,2018MNRAS.475..624N,2018MNRAS.477.1206N,2018MNRAS.480.5113M,2019ComAC...6....2N} is a suite of cosmological magneto-hydrodynamical simulations run using the moving-mesh code AREPO \citep{2010MNRAS.401..791S}. We primarily use the TNG100 simulation from the IllustrisTNG suite, which features a periodic box of side length $L = 75\,h^{-1}\,\mathrm{Mpc}$ and mass resolutions of $m_{\mathrm{dm}} = 7.5\times10^6\,\mathrm{M}_\odot$ for dark matter and $m_{\mathrm{b}} = 1.4\times10^6\,\mathrm{M}_\odot$ for baryons. Supplementary analyses using TNG50 and TNG300 explore the effects of resolution and volume.

Star formation in dense gas ($n_\mathrm{H} > 0.1\,\mathrm{cm}^{-3}$) follows a Kennicutt--Schmidt-like relation \citep{2005MNRAS.364.1105S}. Stellar feedback is implemented via kinetic galactic winds whose launch velocity scales with the square of dark matter velocity dispersion, with a minimum floor of $350\, \mathrm{km\,s^{-1}}$. The mass-loading factor is metallicity-dependent, such that winds are weaker in higher-metallicity gas, and is tied to the chosen wind velocity so that the injected energy from core-collapse supernovae per unit stellar mass formed matches the model assumption. In addition, the wind velocity scales with redshift through the evolution of the Hubble parameter, following a $(H_{0} / H(z))^{1/3}$ dependence \citep{2018MNRAS.473.4077P}. 

Black holes of mass $8 \times 10^{5}\,\mathrm{M}_\odot/h$ are seeded in halos above $5 \times 10^{10}\,\mathrm{M}_\odot/h$ and grow via mergers and Bondi--Hoyle gas accretion, capped at the Eddington limit. Black holes are dynamically repositioned to the local minimum of the gravitational potential, identified within a sphere enclosing a fixed number of region containing equivalent of $n=1000$ mass resolution elements. They also merge if one enters the accretion/feedback region of the other. AGN feedback transitions between thermal and kinetic modes depending on the Eddington ratio compared to a mass-dependent threshold \citep{Weinberger+17}. At high accretion rates, energy is injected continuously and isotropically as thermal energy, with a coupling efficiency of $\epsilon_f \epsilon_r = 0.02$. At low accretion rates, energy is accumulated and then released as directed momentum kicks; while individual kicks are collimated, their orientations vary randomly, resulting in isotropic feedback over time. The minimum energy required to trigger a kinetic feedback event scales with the square of the local dark matter velocity dispersion and the gas mass in the feedback region, ensuring that more massive halos store more energy before launching outflows, in proportion to their gravitational binding energy.

The subgrid feedback physics in the TNG simulation were primarily calibrated using the following observational benchmarks: (1) the global SFR density as a function of time; (2) the $z=0$ galaxy stellar mass function; and (3) the $z=0$ stellar-to-halo mass relation. Secondarily, the following relations were kept under consideration during calibration: (1) the $z=0$ black hole mass vs stellar mass relation; (2) halo gas fractions within $R_{500c}$; and (3) the $z=0$ galaxy size versus mass relation.

\subsubsection*{EAGLE}

The EAGLE project (Evolution and Assembly of GaLaxies and their Environments) \citep{2015MNRAS.450.1937C,2015MNRAS.446..521S} is a suite of cosmological hydrodynamical simulations run using a pressure-entropy formulation of smoothed particle hydrodynamics \citep{2013MNRAS.428.2840H}, implemented in a substantially modified version of the \textsc{Gadget-3} code \citep{2005MNRAS.364.1105S}. Our main analysis uses the Ref-L100N1504 simulation, which evolves $1504^3$ dark matter and baryonic particles in a periodic box of $L = 68\ h^{-1}$\,Mpc. The corresponding particle masses are $m_{\mathrm{dm}} = 9.7 \times 10^6\,\mathrm{M}_\odot$ for dark matter and $m_{\mathrm{b}} = 1.8 \times 10^6\,\mathrm{M}_\odot$ for baryons.

Star formation is modeled stochastically in gas particles that exceed a metallicity-dependent density threshold, with star formation rates tied to the local gas pressure via a prescription motivated by the Kennicutt--Schmidt relation. Stellar feedback is implemented through stochastic thermal energy injection: when a stellar particle reaches an age of 30\,Myr---the typical lifetime of stars that end their lives as core-collapse supernovae---a small number of neighboring gas particles are heated by a fixed temperature increment of $\Delta T = 10^{7.5}$\,K. The strength of the feedback varies with local gas density and metallicity. This model is calibrated primarily to reproduce the $z=0$ stellar mass function and size--mass relation, without explicit tuning to earlier epochs.

Black hole seeds of mass $10^{5}\,\mathrm{M}_\odot/h$ are placed in halos with $M_\mathrm{halo} > 10^{10}\,\mathrm{M}_\odot/h$ if no black hole is already present. They grow via mergers and gas accretion, with the latter modeled using a modified Bondi--Hoyle prescription that accounts for angular momentum suppression and is capped at the Eddington rate \citep{Rosas-Guevara+16}. AGN feedback operates through a single thermal mode active at all accretion rates. Accreted energy is stored in a reservoir and released stochastically: once sufficient energy accumulates to heat at least one neighboring gas particle, it is injected thermally with a temperature increment of $\Delta T = 10^{8.5}$\,K. This temperature is chosen to ensure that feedback energy overcomes radiative losses in dense gas and drives significant outflows, while accounting for the simulation's resolution limitations. The AGN feedback efficiency is calibrated to reproduce the $z=0$ black hole--stellar mass relation, with a net coupling factor of $\epsilon_f \epsilon_r = 0.015$.

\subsubsection*{SIMBA}

The SIMBA project \citep{simba2019} is a suite of cosmological hydrodynamical simulations run using the meshless finite mass code GIZMO \citep{gizmo2015}.
Our primary analysis uses the fiducial large-volume, full-physics simulation from the SIMBA suite, which evolves a periodic box of side length $L = 100\,h^{-1}\mathrm{Mpc}$ with dark matter and baryonic particle masses of $m_{\mathrm{dm}} = 9.6\times10^7\,\mathrm{M}_\odot$ and $m_{\mathrm{b}} = 1.8\times10^7\,\mathrm{M}_\odot$, respectively. For comparison, we also examine feedback variant runs at higher resolution using the $L = 50\,h^{-1}\mathrm{Mpc}$ boxes.

Star formation in SIMBA follows a volumetric Kennicutt--Schmidt relation in molecular gas above a metallicity-dependent density threshold. Stellar feedback is implemented via two-phase kinetic winds that carry metals and dust, with wind velocities scaling with the galaxy circular velocity according to the baryonic Tully--Fisher relation. About 30\% of wind particles are ejected in a hot phase at $T = 10^{5.5}$\,K. The metal-enriched wind mass-loading factor scales with the galaxy stellar mass \citep{simba2019}.

In SIMBA, black hole seeds of mass $M_\mathrm{seed} = 10^4\,\mathrm{M}_\odot/h$ are placed in galaxies once they reach $M_\star > 3 \times 10^5\,M_\mathrm{seed}$ and do not already host a black hole, reflecting the suppression of early BH growth by stellar feedback. Accretion depends on the thermal state of surrounding gas: cold gas ($T<10^5$\,K) follows a torque-limited model capped at three times the Eddington rate, while hot gas undergoes Bondi--Hoyle accretion capped at the Eddington rate. AGN feedback operates via kinetic wind and jet modes, with the dominant mode determined by the black hole's Eddington ratio and mass \citep{simba2019}. At high accretion rates, the wind mode launches bipolar outflows with velocities $v_w = 500 + 500\, (\log M_\mathrm{BH}-6)/3\,$\kms. At low Eddington ratios ($f_{\rm Edd}<0.2$) and for $M_\mathrm{BH} > 10^{7.5}\,M_\odot$, a high-velocity, bipolar jet mode is triggered with $v_\mathrm{jet} = v_w + 7000\,\log \left(0.2/f_\mathrm{Edd}\right)\,$\kms injecting momentum and heating gas to the halo virial temperature. Following jet activation, an X-ray heating mode becomes active in low-gas systems ($f_\mathrm{gas}<0.2$ and $v_\mathrm{jet} \gtrsim 7000$\,km\,s$^{-1}$), contributing secondarily to quenching by suppressing residual gas accretion and star formation in massive galaxies \citep{Appleby+20}.

The subgrid feedback physics in SIMBA are primarily calibrated to the $z=0$ galaxy stellar mass function. Though not explicitly calibrated for SIMBA, the model used previous calibrations for the efficiency with which gas is transported from the inner galactic disk onto a black hole accretion disk, and subsequent transport of accreting gas onto the black hole \citep{2017MNRAS.464.2840A}. These efficiencies were calibrated to reproduce the $z=0$ black hole mass vs galaxy stellar mass relation.

The subgrid physics in each simulation are calibrated to match certain galaxy properties, with the $z=0$ stellar mass function being a common calibration target. Consequently, the mass function is not a true prediction. In contrast, the spatial distribution of stellar mass, most AGN properties, star-forming and quiescent fractions, and their environmental dependence are genuine model predictions. We note that we analyze versions of SIMBA and EAGLE simulations which were re-processed by \citep[][]{Ayromlou+23}.

\subsubsection*{Selecting simulated AGNs}

We extract galaxy properties from the simulation catalogs, including black hole masses, accretion rates, stellar masses, and star formation rates. Accretion luminosities are computed using a variable radiative efficiency, $\epsilon_r$, which depends on the Eddington ratio, $f_\mathrm{Edd}$. We adopt a physically motivated model for $\epsilon_r$ based on axisymmetric 2D hydrodynamical simulations incorporating radiative cooling and black hole feedback \citep{Inayoshi+19}, in particular, the equation given below. While TNG, EAGLE, and SIMBA originally use fixed efficiencies of $\epsilon_r = 0.2, 0.1, 0.1$, respectively, we allow $\epsilon_r$ to vary with accretion rate, capturing the observed transition from low- to high-efficiency regimes. For comparison, we also implement the simpler post-processing method used in \citep{Habouzit+21,Habouzit+22}, which sets $L_\mathrm{AGN} = \epsilon_r \dot{M}_\mathrm{BH} c^2$ for $f_\mathrm{Edd} > 0.1$ and $L_\mathrm{AGN} = 10f_\mathrm{Edd}\epsilon_r \dot{M}_\mathrm{BH} c^2$ for $f_\mathrm{Edd} \le 0.1$. Despite these variations in post-processing, the simulated AGN and host galaxy properties remain in significant tension with observations.

\begin{equation}
\log \epsilon_r =
\begin{cases} 
-1.0 - \left(\dfrac{0.0162}{\dot{m}}\right)^4, & 0.023 \leq \dot{m}, \\[1ex]
a_0 + a_1 \log \dot{m}, & 10^{-4} < \dot{m} < 0.023, \\[1ex]
b_0 + b_1 \log \dot{m} + b_2 (\log \dot{m})^2, & 10^{-8} \leq \dot{m} \leq 10^{-4},
\end{cases}
\end{equation}

\noindent where $\dot{m} \equiv \frac{\dot{M}_\mathrm{BH}}{\dot{M}_\mathrm{Edd}},a_0 = -0.807,  a_1 = 0.27$, and $a_n = 0 \text{ for } n \geq 2,b_0 = -1.749,  b_1 = -0.267,  b_2 = -0.07492,  b_n = 0 \text{ for } n \geq 3.$

In the three simulations, we define the black hole mass as the sum of the masses of all BH particles within a given galaxy. BHs are automatically merged when their host galaxies merge, making it rare for a galaxy to contain multiple BHs \citep{Habouzit+21}. Simulated AGNs are identified as systems with $L_\mathrm{AGN} > 10^{42}\,\mathrm{erg\,s^{-1}}$ and $\lambda_\mathrm{Edd} > 10^{-4}$, consistent with the selection applied to our observed AGN sample. Our results are not sensitive to the specific AGN selection thresholds; discrepancies persist even when the luminosity and Eddington ratio cuts are increased by factors of ten or one hundred.

We compute stellar velocity dispersions in 1D and 3D in spheres extending from each simulated galaxy's gravitational potential minimum. In both cases, the galaxy’s peculiar velocity is subtracted from all member star particles. In 1D, line-of-sight velocity dispersions are computed along the z-axis of the simulation volume. In 3D, velocity dispersions are computed as root mean-square speeds as described in Binney \& Tremaine 1987 (section 4.8.3). Both mass- and light-weighted velocity dispersions are calculated. Light-weighted calculations use each star particle's unattenuated SDSS $i$-band brightness. We adopt the light-weighted 1D velocity dispersions measured within $r=2.5$\,kpc for the fiducial comparison.

\subsubsection*{Simulated galaxy environments}
We characterize the multi-scale environments of galaxies in each simulation by computing stellar mass overdensities, emulating the method of \cite{yesuf22}. First, we project the 3D coordinates of galaxies onto the $z$-axis of and select galaxies in a series of apertures with radii $x \in \{ 0.1, 0.5, 1, 2, 4, 8\}$ $h^{-1}$Mpc. As a proxy for a line-of-sight velocity cut of $|\Delta v|=1000$ km s$^{-1}$, we remove galaxies whose $z$-coordinate offsets with respect to the target galaxy are $\Delta z > |\Delta v| / H(z)$, where $H(z)$ is the Hubble parameter. A peculiar velocity cut of $|\Delta v| < 1000$ km s$^{-1}$ is further applied in the $z$-axis with respect to the target galaxy. Overdensities in each cylindrical volume are then computed using the total stellar masses of the retained simulated galaxies. Only galaxies with total stellar masses $\logMstar>9$ are considered in these calculations. Increasing or decreasing this mass threshold by a factor of 10 does not change the main conclusions. 

Our environmental measurements use projection along the $z$-axis. Using the TNG100 simulation at $z\sim 0.1$ (snapshot 91), we assessed projection effects in Fig.~3 by comparing the variance of quenched fractions obtained from the $x$, $y$, and $z$ projections. The standard deviation of $f_q$ varies by $\sim 2$--$7$\% (median $\sim 2.5$\%) in the highest-density environments or for massive galaxies with $M_\star > 3 \times 10^{10}\,M_\odot$, while the effects are negligible for low-density and low-mass galaxies. Although projection introduces a non-negligible source of uncertainty, our assessment indicates that it does not account for the substantial discrepancy between simulations and observations. Moreover, its impact on the uncertainty of halo mass mapping is also limited.

The observed distributions in Fig.~2 are weighted by $1/V_\mathrm{max}$, and are largely consistent across the two surveys, with small but significant differences at the density extremes. SDSS, benefiting from its wider survey area, contains more galaxies in high-density regions, aligning well with simulation predictions. However, on smaller scales ($<2\,\mathrm{Mpc}/h$), the simulations diverge in their predicted stellar mass overdensity distributions---particularly when broken down by $M_\star$ and $\Delta \log\,\mathrm{SFR}$---with statistically significant discrepancies from the observations ($p \lesssim .001$; see Supplementary Figs.~3--5).

\subsubsection*{Halo mass estimation using simulations}

Our new method provides a scalable alternative for estimating dark matter halo masses from observables by leveraging the multiscale distribution of stellar mass around galaxies, their distance from the group center, and/or their line-of-sight group velocity. This approach exploits the strong correlation between halo mass and galaxy spatial clustering: massive halos preferentially reside in high-density regions, where gravitational collapse is more efficient and galaxy clustering is enhanced. By incorporating multiple scales ($0.1$--$10$\,Mpc), our method delivers accurate halo mass estimates for both central and satellite galaxies across large samples. 

In this work, we train centrals and satellites separately, since the projected group velocity dispersion is poorly defined for isolated centrals or galaxies in low-richness groups, and the stellar mass of a central can be particularly sensitive to feedback prescriptions, introducing potential mismatches with observations. We also tested a unified scheme in which a single model is trained with central/satellite flags, explicit measures of central mass dominance, and a theoretically motivated scaling (imputation) of group velocity with stellar mass for isolated or low-richness systems. Both strategies give consistent results, but here we adopt the separate training. A more optimal framework tailored for individual galaxy–halo predictions will be developed in future work, validated using independent halo mass estimates from alternative tracers.

For satellites in groups with $N > 2$, trained separately, we find that the line-of-sight group velocity dispersion and the mass distribution within $0.5$--$2$\,Mpc/$h$ are strongly correlated with halo mass in the simulations, with correlation coefficients $\rho \gtrsim 0.8$. Nonetheless, substantial scatter ($0.5$--$0.6$\,dex) and heteroscedasticity remain, motivating the use of non-linear machine learning techniques to optimally combine multiple features. To address this, we employ Gradient Boosting Regression, which aggregates an ensemble of decision trees to predict halo mass from the input variables. The training results are robust against variations in hyperparameters. The variable importance rankings are presented in Supplementary Fig.~14.

We train the model independently on each of the cosmological simulations (EAGLE, TNG, and SIMBA) and apply it to observational data from GAMA and SDSS. Our halo mass predictions show good agreement with the GAMA group catalog \citep{Robotham+11}, which has been validated through weak lensing (see Extended Data Fig.~7, Supplementary Figs.~15 and 16), providing observational support for our approach. The model achieves high accuracy, with $R^2 \approx 0.97$ and mean squared error $\sigma \approx 0.1$--$0.15$ on test samples of both satellites and centrals from the simulations. 

We emphasize that the tight coupling between dark matter halos and observables is a robust prediction of the simulations, but it requires validation with future, more accurate and independent observational methods. We are less confident in the accuracy of the small-scale mass distribution ($\sim 0.1$\,Mpc/$h$) in the simulations, and note that spectroscopic incompleteness in current observations remains a limitation. Therefore, we do not place strong weight on the accuracy of the central halo mass predictions here. We will revisit the small-scale mass distribution in a forthcoming paper using improved data.

On Fig.~6, a qualitatively similar result of preferentially low halo masses is also found when using an alternative halo mass estimate for type 2 AGNs \citep{YangX+07, yesuf22}, for which it is more reliable.

\medskip
\textbf{Data availability} 
All observational and simulation data used in this study are publicly available via the links below. Our newly derived measurements will be made publicly accessible in the future to facilitate comparisons and reproducibility. Requests for early access to these data can be accommodated upon reasonable inquiry.
\begin{itemize}
    \item SDSS AGN catalog: \cite{Liu+19} and \\ \href{https://cdsarc.cds.unistra.fr/viz-bin/cat/J/ApJS/243/21}{https://cdsarc.cds.unistra.fr/viz-bin/cat/J/ApJS/243/21}
    \item SDSS galaxy catalog: \cite{Salim+18} and \href{https://salims.pages.iu.edu/gswlc/}{https://salims.pages.iu.edu/gswlc/}
    \item SDSS group catalog: \cite{YangX+07} and \href{https://gax.sjtu.edu.cn/data/Group.html}{https://gax.sjtu.edu.cn/data/Group.html}
    \item Galaxy and Mass Assembly (GAMA): \cite{Driver+22} and \href{https://www.gama-survey.org/dr4/}{https://www.gama-survey.org/dr4/}
    \item GAMA galaxy group catalog: \cite{Robotham+11} and \\ \href{https://www.gama-survey.org/dr4/schema/dmu.php?id=115}{https://www.gama-survey.org/dr4/schema/dmu.php?id=115}
    \item IRAS 60 micron data (median coadds): \href{https://irsa.ipac.caltech.edu/applications/Scanpi/}{https://irsa.ipac.caltech.edu/applications/Scanpi/}
    \item Pan-STARRS: \href{https://catalogs.mast.stsci.edu/panstarrs/}{https://catalogs.mast.stsci.edu/panstarrs/}
    \item 2MASS: \href{https://irsa.ipac.caltech.edu/Missions/2mass.html}{https://irsa.ipac.caltech.edu/Missions/2mass.html}
    \item UKIDSS: \href{http://wsa.roe.ac.uk/}{http://wsa.roe.ac.uk/}
    \item WISE: \href{https://irsa.ipac.caltech.edu/Missions/wise.html}{https://irsa.ipac.caltech.edu/Missions/wise.html}
    \item XMM-DR14\,catalog: \cite{Webb+20} and \href{http://xmmssc.irap.omp.eu/Catalogue/4XMM-DR14/4XMM\_DR14.html}{http://xmmssc.irap.omp.eu/Catalogue/4XMM-DR14/4XMM\_DR14.html}
    \item EAGLE simulations: \href{https://icc.dur.ac.uk/Eagle/database.php}{http://icc.dur.ac.uk/Eagle/database.php}
    \item IllustrisTNG simulations: \href{https://www.tng-project.org/data/}{https://www.tng-project.org/data/}
    \item SIMBA simulations: \href{http://simba.roe.ac.uk}{http://simba.roe.ac.uk}
\end{itemize}

We note that herein we have analyzed versions of SIMBA and EAGLE which were re-processed by \citep{Ayromlou+23} to enable an apples to apples comparison with TNG, not the original versions. 

\medskip

\textbf{Acknowledgments}
We deeply appreciate the valuable consultations and discussions on various aspects of the simulations with Aaron Ludlow, Chris Power, Dylan Nelson, Kate Harborne, and M. Reza Ayromlou. We are also grateful to Lei Hao and her research group for their useful discussions and support, which greatly facilitated the progress and ease of this research. CB gratefully acknowledges support from the Forrest Research Foundation. HY was partially supported by the Research Fund for International Young Scientists of NSFC (11950410492) and JSPS KAKENHI Grant Number JP22K14072. This work was supported by resources provided by the Pawsey Supercomputing Research Centre’s Setonix Supercomputer (https://doi.org/10.48569/18sb-8s43) and Acacia Object Storage (https://doi.org/10.48569/nfe9-a426), with funding from the Australian Government and the Government of Western Australia.

\medskip
\textbf{Author contributions}
HY developed the idea for the project, performed all observational measurements, including environmental measurements, and conducted the data analysis. CB compiled the simulation data and conducted the environmental measurements for the simulations. Both authors contributed to the interpretation of the results and writing the text of the manuscript.

\medskip
\textbf{Competing interests}
The authors declare no competing interests.

\clearpage
\newenvironment{extendedfigure}
  {\refstepcounter{extfig}%
   \renewcommand{\thefigure}{\theextfig}%
   \renewcommand{\figurename}{Extended Data Fig.}
   \begin{figure}}
  {\end{figure}}

\setcounter{figure}{0}

\begin{table}
\centering
\label{tab:SMF_scores}
\caption{\textbf{Comparison between simulation predictions and GAMA stellar mass functions for galaxy subpopulations.} For each subpopulation, we report the root-mean-square error (RMSE), mean absolute error (MAE), and reduced chi-square ($\chi^2_\nu$) to quantify the level of agreement. All values are given in dex. Entries marked with an asterisk ($^\ast$) and dagger ($^\dagger$) correspond to tensions of $1.8\sigma$ and $4.3\sigma$, respectively; all remaining entries exceed $5\sigma$.}
\begin{tabular}{lcc cc cc}
\toprule
\multirow{2}{*}{\textbf{Sample}} & 
\multicolumn{2}{c}{\textbf{SIMBA}} & 
\multicolumn{2}{c}{\textbf{EAGLE}} & 
\multicolumn{2}{c}{\textbf{TNG}} \\
\cmidrule(lr){2-3} \cmidrule(lr){4-5} \cmidrule(lr){6-7}
 & RMSE/MAE & $\chi^2_\nu$ & RMSE/MAE & $\chi^2_\nu$ & RMSE/MAE & $\chi^2_\nu$ \\
\midrule
All Fig.1a & \texttt{0.20/0.16} & \texttt{63} 
                                        & \texttt{0.27/0.22} & \texttt{58}
                                        & \texttt{0.19/0.17} & \texttt{26} \\ 
                                        
Upper SFMS Fig.~1b & \texttt{0.37/0.29} & \texttt{34} 
                                        & \texttt{0.27/0.15} & \texttt{2.5}\tnote{*}
                                        & \texttt{0.31/0.29} & \texttt{55} \\    
                                    
Lower SFMS Fig.~1c & \texttt{0.55/0.47} & \texttt{80}
                                           & \texttt{0.25/0.18} & \texttt{17}
                                           & \texttt{0.42/0.28} & \texttt{51} \\
                                       
GVG Supp. Fig.~2b & \texttt{0.57/0.50} & \texttt{31} 
                                                & \texttt{0.24/0.18} & \texttt{9} 
                                                & \texttt{0.35/0.28} & \texttt{20} \\  
QG Fig.~1d & \texttt{0.21/0.17} & \texttt{61}
                                       & \texttt{0.49/0.42} & \texttt{60}
                                       & \texttt{0.28/0.24} & \texttt{40}\\
                 
SFG Supp. Fig.~2a & \texttt{0.31/0.25} & \texttt{19}
                                       & \texttt{0.40/0.21} & \texttt{5}\tnote{$\dagger$}
                                    & \texttt{0.59/0.36} & \texttt{16} \\
\bottomrule
\end{tabular}
\end{table}

\medskip

\begin{table}
\centering
\label{tab:env_scores}
\caption{\textbf{Aggregate discrepancies in quenched fraction and star formation rate across mass–environment space.} The root-mean-square error (RMSE) and mean absolute error (MAE) are reported for the comparison between simulations and observations. All values exceed the $5\sigma$ level.}
\begin{tabular}{lc c c}
\toprule
Sample & SIMBA RMSE/MAE & EAGLE RMSE/MAE & TNG RMSE/MAE\\
\midrule
GAMA $f_q$ & \texttt{0.17/0.13} & \texttt{0.23/0.15} & \texttt{0.21/0.16}\\
SDSS $f_q$ & \texttt{0.19/0.15} & \texttt{0.23/0.15} & \texttt{0.22/0.16}\\ 
GAMA SFR & \texttt{0.31/0.24} & \texttt{0.12/0.10} & \texttt{0.19/0.16} \\ 
SDSS SFR & \texttt{0.30/0.24} & \texttt{0.19/0.17} & \texttt{0.15/0.10} \\  
\bottomrule
\end{tabular}
\end{table}

\medskip

\begin{table}
\centering
\label{tab:AGN_scores}
\caption{\textbf{Discrepancies between SDSS and simulations in AGN.} This table quantifies the differences in number density functions between SDSS and the simulations. From top to bottom, the distributions correspond to AGN }
\begin{tabular}{lc c c}
\toprule
Sample & SIMBA RMSE & EAGLE RMSE & TNG RMSE\\
\midrule
L$_\mathrm{AGN}$F & \texttt{0.2} & \texttt{0.7} & \texttt{0.4}\\
VDF & \texttt{1.1} & \texttt{1.1} & \texttt{0.8}\\ 
SMF & \texttt{0.5} & \texttt{0.9} & \texttt{0.9}\\
sSFRF & \texttt{0.9} & \texttt{1.3} & \texttt{1.4} \\ 
\bottomrule
\end{tabular}
\end{table}

\begin{extendedfigure}
    \includegraphics[width=0.95\linewidth]{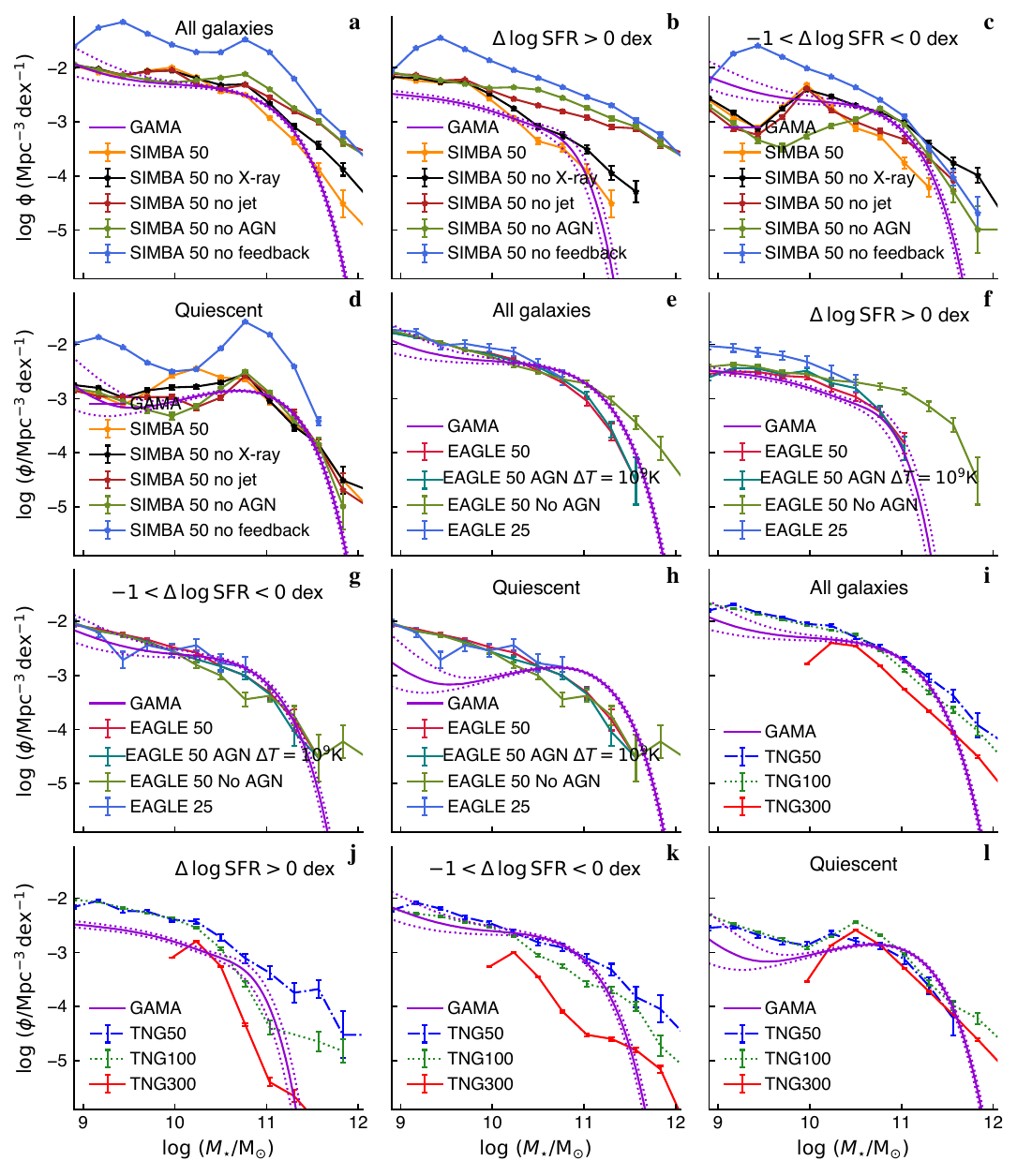}
    \caption{\textbf{Stellar mass functions of the SIMBA, EAGLE, and TNG simulation variants compared with GAMA.} Panels are grouped by simulation (snapshots at $z \approx 0.1$) and, for each simulation, show the stellar mass functions of the global, high-SFR, low-SFR, and quiescent populations, defined by $\deltasfms$ relative to the star-forming main sequence (SFMS). In all panels, the violet curves show the median stellar mass function and the 16th–84th percentile uncertainty range derived from fits to GAMA galaxies at $z < 0.12$. Error bars on the simulation curves indicate standard deviations assuming Poisson statistics. SIMBA (a–d): Black, red, green, and blue curves correspond to SIMBA variants with X-ray feedback disabled; both jet-mode and X-ray AGN feedback disabled; all AGN feedback disabled; and all AGN and supernova feedback processes disabled, respectively. EAGLE (e–h): Red, teal, green, and blue curves correspond to the fiducial EAGLE-50 simulation; the EAGLE-50 variant with increased AGN heating temperature (TAGN $= 10^{9}$ K); the EAGLE-50 simulation with AGN feedback disabled; and the higher-resolution EAGLE-25 simulation, respectively. TNG (i–l): Blue, green, and red curves show the three volume variants of the TNG simulations.
}
\end{extendedfigure}

\begin{extendedfigure}
    \includegraphics[width=0.95\linewidth]{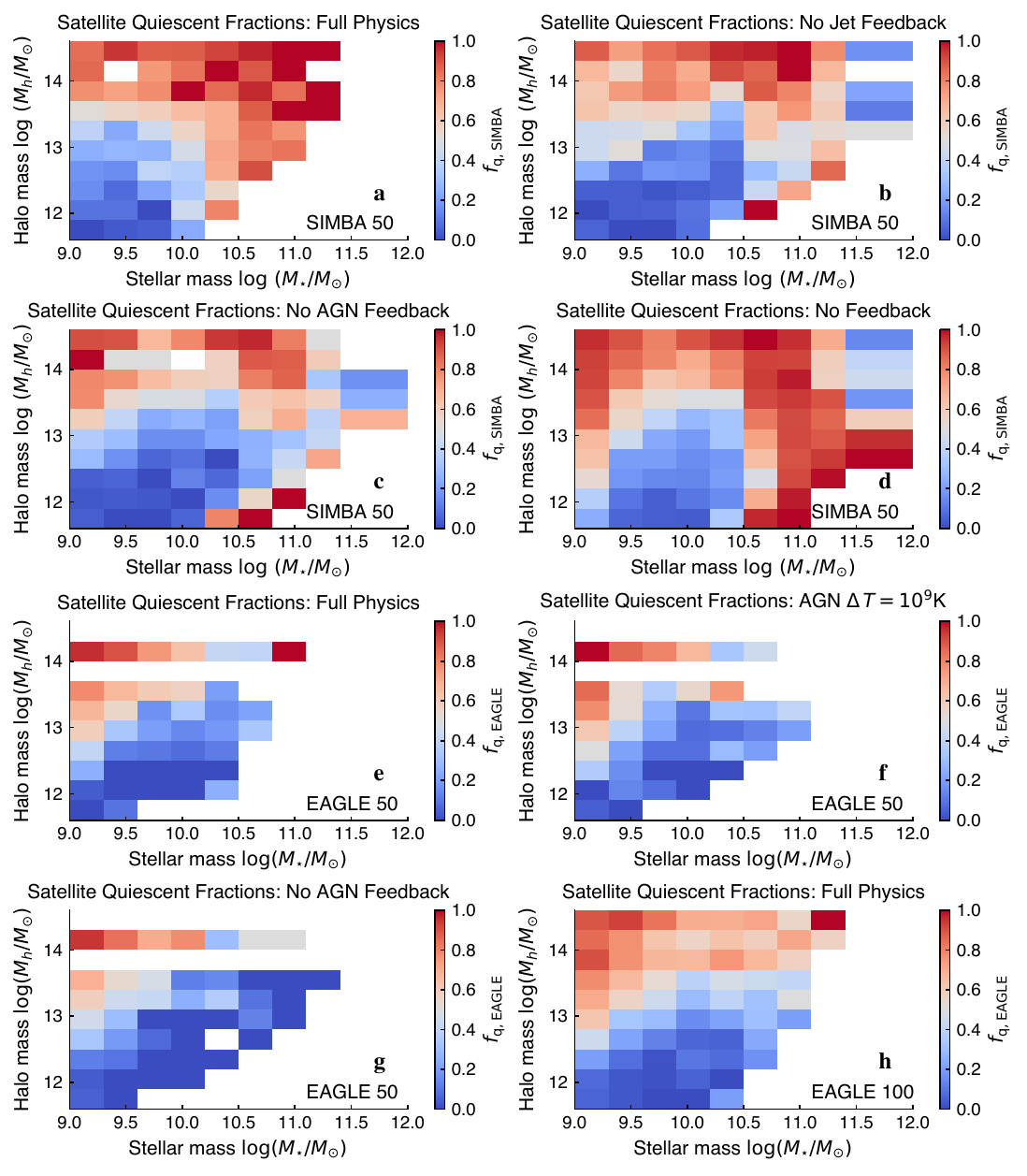}
    \caption{\textbf{Quiescent satellite fractions in SIMBA and EAGLE feedback variants.} Panels (a)–(d) show the SIMBA50 variants: full physics with all feedback enabled; both jet-mode and X-ray AGN feedback disabled; all AGN feedback disabled; and all AGN and supernova feedback disabled, respectively. Panels (e)–(h) show the EAGLE50 variants: full physics; increased AGN heating temperature (TAGN = $10^{9}$\,K); AGN feedback disabled; and the larger-volume EAGLE100 simulation, respectively.}
\end{extendedfigure}

\begin{extendedfigure}
    \includegraphics[width=0.95\linewidth]{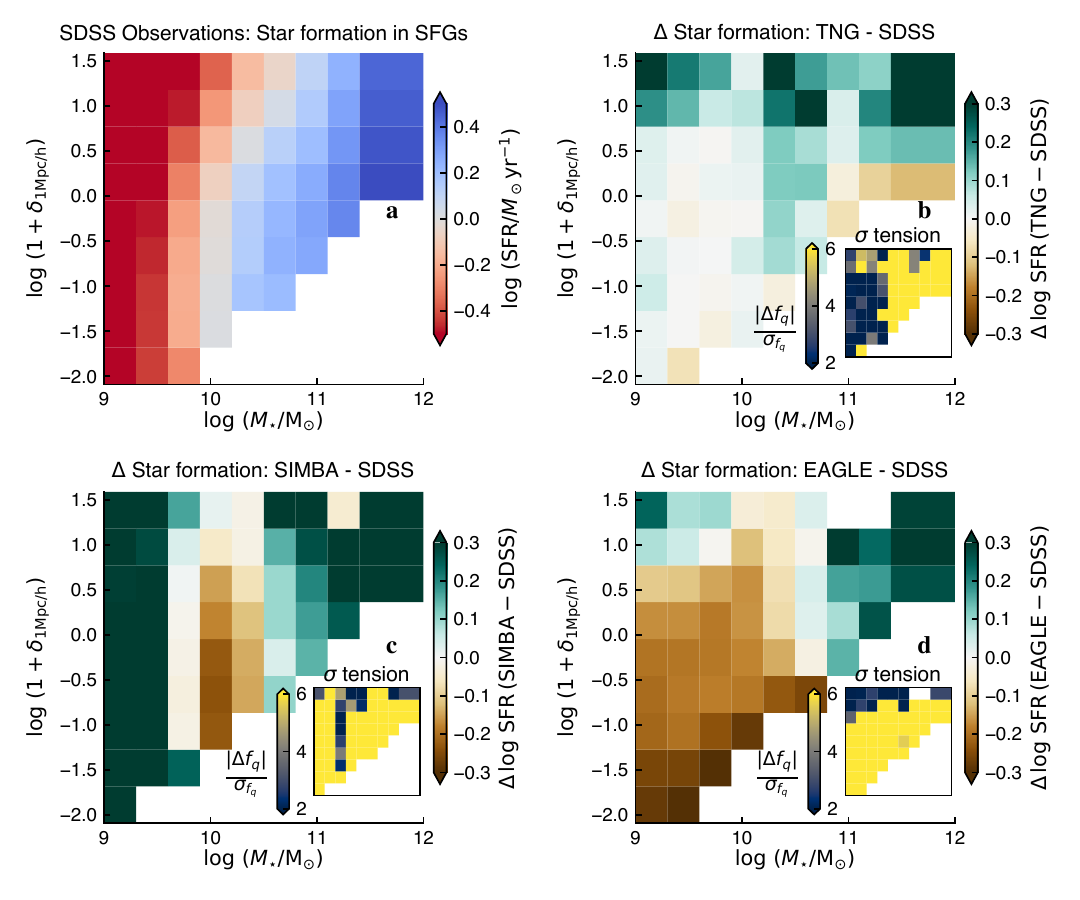}
    \caption{\textbf{Trends of star formation rates (SFRs) with stellar mass and environment in SDSS and simulations.} Only star-forming galaxies (SFGs) are considered. Panel (a) shows the mean SFR trends for SFGs in SDSS, illustrating the expected positive correlation with stellar mass. Panels (b)–(d) show the residuals of SFR between each simulation and SDSS. Green (brown) residuals indicate higher (lower) SFRs in the simulations for galaxies of the same stellar mass in the same environment. Insets show the ratio of the absolute difference to SDSS uncertainties, $|\Delta f_q| / \sigma_{f_q}$, providing a measure of the tension ($\sigma$-level).}
\end{extendedfigure}

\begin{extendedfigure}
    \includegraphics[width=0.95\linewidth]{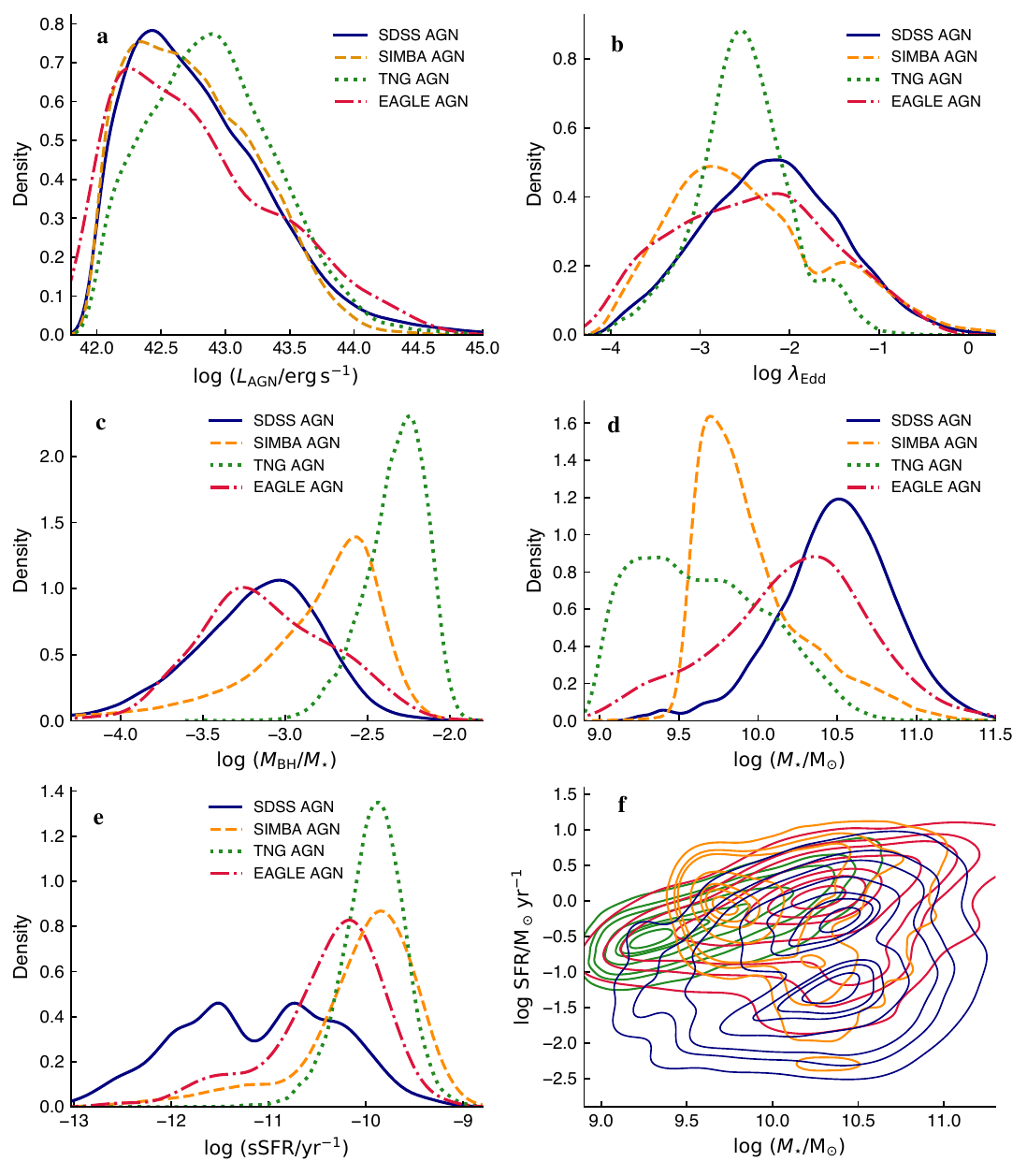}
    \caption{\textbf{Comparison of simulated AGNs and their host galaxies with SDSS observations.} Panels (a)–(e) show the distributions of (a) AGN luminosity, (b) Eddington ratio, (c) black hole–to–stellar mass ratio, (d) stellar mass, and (e) specific star formation rate, while panel (f) shows stellar mass versus star formation rate. The sample is restricted to galaxies with $L_{\rm AGN} > 10^{42}$ erg s$^{-1}$, $M_\star  > 10^9$\,M$_\odot$ and $z < 0.15$. SDSS distributions are weighted by 1/Vmax to correct for sample incompleteness.}
\end{extendedfigure}

\begin{extendedfigure}
    \includegraphics[width=0.95\linewidth]{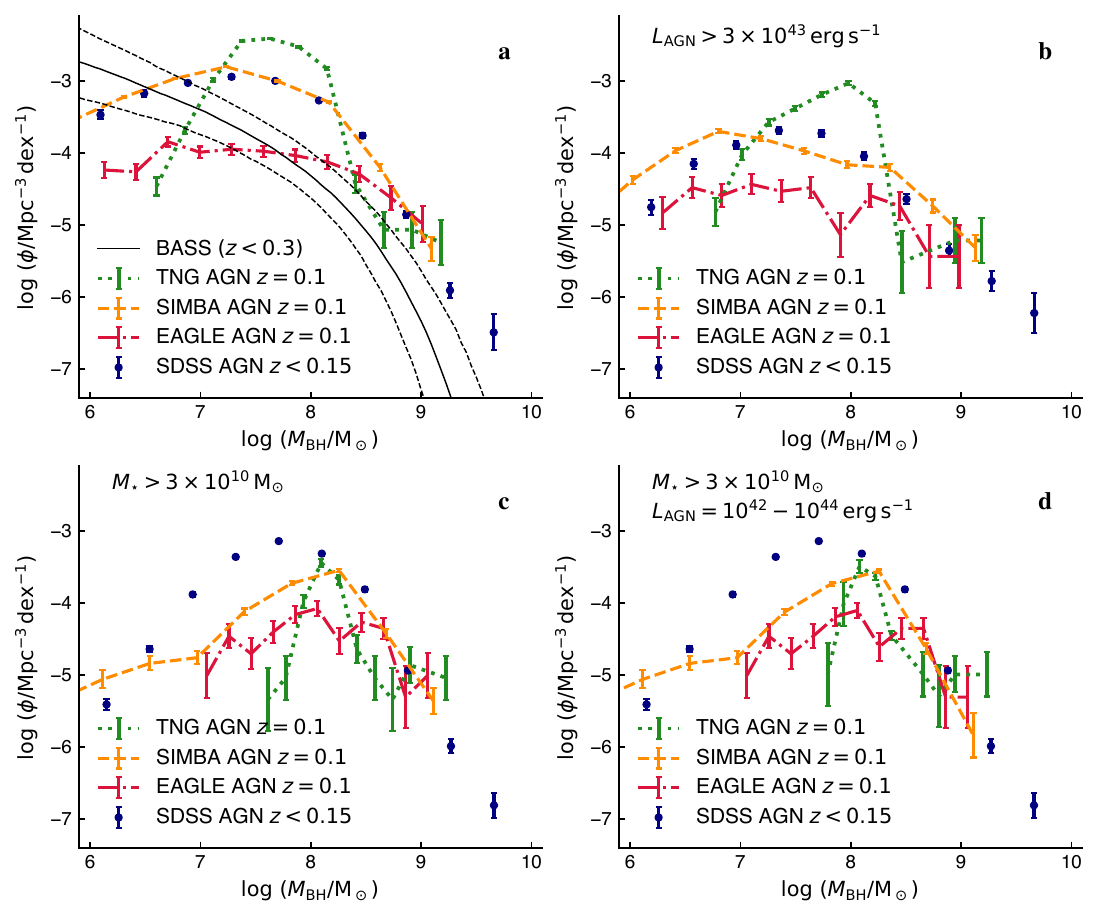}
    \caption{\textbf{Black hole mass functions (BHMFs) from simulations compared with SDSS observations.} Panel (a) shows the full fiducial subsample and includes BHMF fits for Swift-BAT AGNs \citep{Ananna+22}, plotted using their two fitting methods. The remaining panels divide the samples by stellar mass or AGN luminosity. Error bars denote standard deviations assuming Poisson statistics.}
\end{extendedfigure}

\begin{extendedfigure}
    \includegraphics[width=0.95\linewidth]{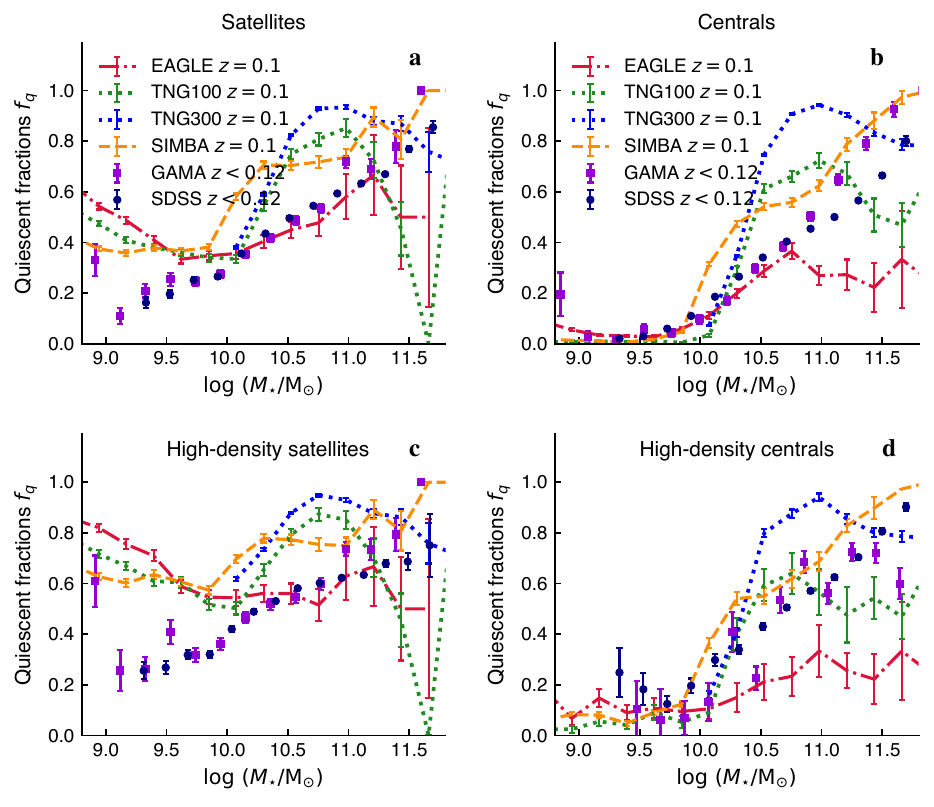}
    \caption{\textbf{Relations between quiescence and galaxy stellar mass for satellites and centrals in SDSS and the three simulations.} Each row of panels compares the quiescent fractions of satellites (left panels) and centrals (right panels). The bottom panels show the subpopulations of satellites and centrals in regions with overdensity greater than three. Error bars on the quiescent fractions represent the standard error of a proportion.}
\end{extendedfigure}

\setlength{\footskip}{150pt}
\clearpage

\renewcommand{\tablename}{Supplementary Table}
\setcounter{table}{0}

\section*{Supplementary Information}
\renewcommand{\thefigure}{\arabic{figure}}  
\renewcommand{\figurename}{Supplementary Fig.}  

\setcounter{figure}{0}

\subsection*{Details on number density of observed and simulated galaxies}

\begin{figure}[h]
\includegraphics[width=0.99\linewidth]{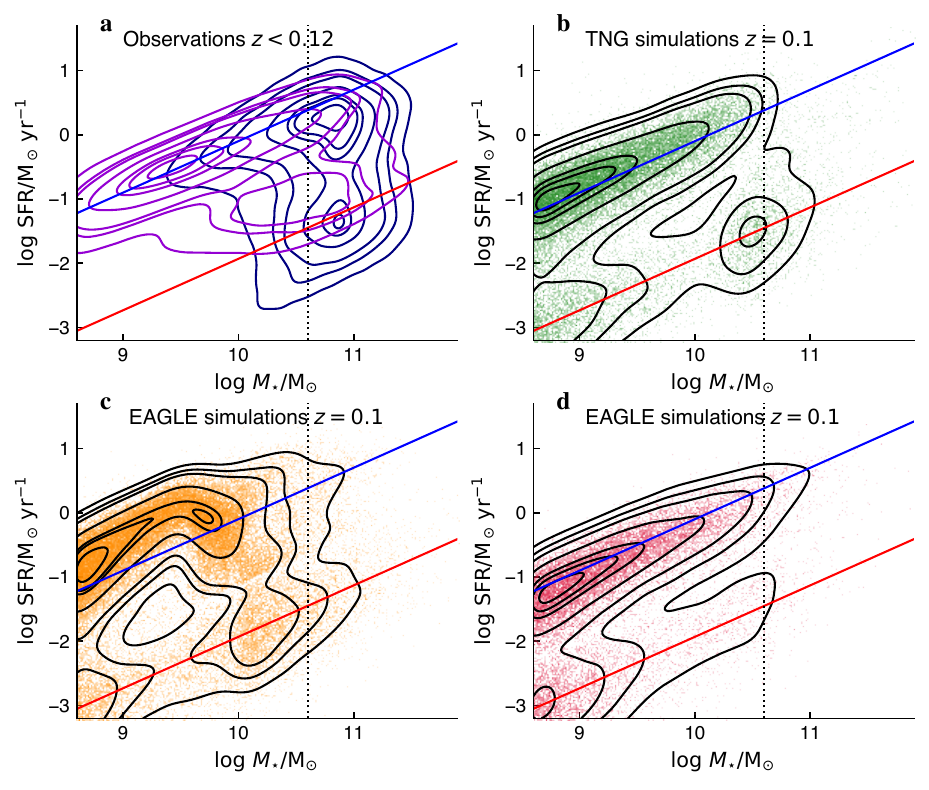}
\caption{\textbf{Star formation rates (SFRs) and stellar masses of observed and simulated galaxies}. Contour levels correspond to 5, 15, 25, 50, 75, 85, and 95 percent of the data. Panel (a) shows the completeness-corrected SDSS galaxy distribution (navy) above $M_\star \gtrsim 10^9\, M_\odot$, alongside the GAMA distribution (violet). Panels (b-d): Distributions from the TNG, SIMBA, and EAGLE simulations, respectively. The solid blue and red lines in each panel indicate the mean loci of SDSS star-forming and quiescent galaxies to facilitate comparison. To approximate observational uncertainties, SFR values for quiescent galaxies (QGs) in the simulations have been randomly jittered. The dotted black lines mark the peak location of QGs in SDSS. Note the horizontal offset for SFGs and the vertical shift for massive QGs in the simulations. SIMBA's SFRs for SFGs (main sequence) are significantly offset from the observations, clustering around $M_\star \approx 10^{10}\, M_\odot$ where many galaxies are transitioning from SFGs to QGs. Additionally, SIMBA’s massive QGs peak at lower $M_\star$ than SDSS and TNG, potentially due to differences in black hole mass thresholds for kinetic feedback activation. EAGLE underpredicts the number of massive QGs. These trends are further quantified with the mass function (Fig.~1 and Supplementary Fig.~\ref{fig:SMF_SFG_GV}). \label{fig:compSFMS}}
\end{figure}

Supplementary Fig.~\ref{fig:compSFMS} shows the distribution of galaxies in $M_\star$ versus SFR space for the three simulations and SDSS observations, with the observed distributions weighted by $1/V_{\mathrm{max}}$. The SFMS of EAGLE and TNG broadly agree with the observations. In contrast, SIMBA’s SFMS is offset by $\sim 0.4$\,dex and exhibits a bend around $M_\star \approx 10^{10}\,M_\odot$, marking the transition from star-forming to quiescent galaxies. This bend is absent in SDSS, GAMA, and the other two simulations, as well as in the SIMBA variant with AGN feedback switched off. The average $M_\star$ of massive quiescent galaxies in SIMBA is also significantly lower than in SDSS, while EAGLE shows a sparse population of massive quiescent galaxies, in apparent tension with SDSS. All three simulations display a concentration of low-mass quiescent galaxies around $M_\star \sim 10^9\,M_\odot$. These qualitative trends are quantified more precisely using the stellar mass function (SMF; Fig.~1).

\begin{figure}
\includegraphics[width=0.99\linewidth]{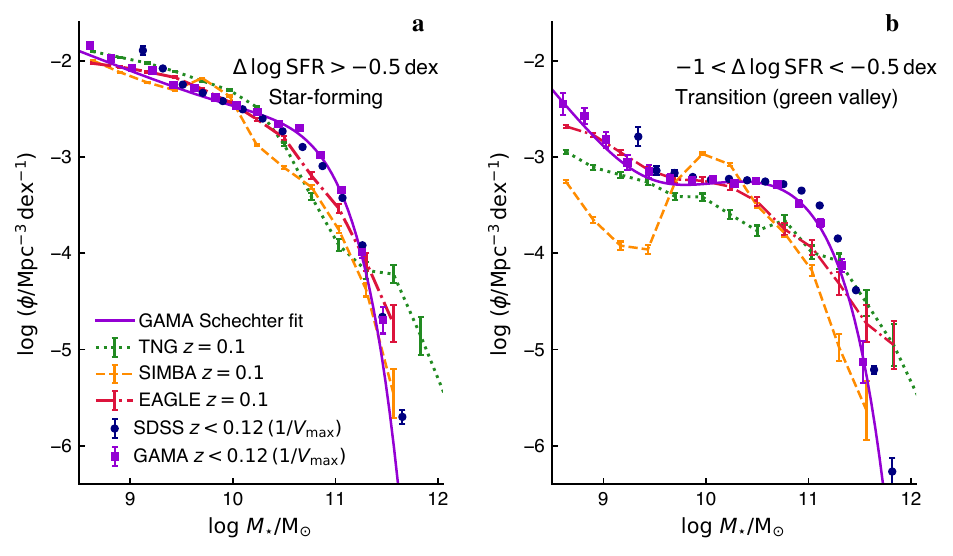}
\caption{\textbf{Galaxy stellar mass functions of star-forming and green-valley galaxies.} Stellar mass functions (SMFs) measured from the Sloan Digital Sky Survey (SDSS) and the Galaxy And Mass Assembly survey (GAMA) are compared with predictions from three cosmological simulations: TNG, SIMBA, and EAGLE. Panel (a) shows the SMFs of all star-forming galaxies, while panel (b) presents those of transition (green-valley) galaxies. Error bars denote one standard deviation, estimated assuming Poisson statistics.}
\label{fig:SMF_SFG_GV}
\end{figure}

Supplementary Fig.~\ref{fig:SMF_SFG_GV}a compares the SMF of SFGs in the three simulations with SDSS and GAMA observations. While the simulations reproduce the number densities of low-mass SFGs reasonably well, they underpredict the number of SFGs in the mass range $M_\star \approx 10^{10}-10^{11}\,M_\odot$ by a factor of two to three. As noted in the Main section, SIMBA significantly underpredicts the number of low-mass ($<10^{10}\,M_\odot$) SFGs with below-average star formation rates (i.e., lower SFMS galaxies) by a several factor, while EAGLE and TNG produce results broadly consistent with observations for these populations. Similarly, SIMBA underproduces the number of low-mass green-valley galaxies ($-1 < \Delta \log\,\mathrm{SFR} < -0.5$), whereas EAGLE and TNG, particularly EAGLE, show good agreement with GAMA data (Supplementary Fig.~\ref{fig:SMF_SFG_GV}b). Additionally, both SIMBA and TNG overpredict the number of upper SFMS galaxies by about a factor of two (Fig.~1b). Adjusting SIMBA’s SFMS for its +0.4 dex offset does not resolve its number density mismatch with observations; instead, this adjustment results in five times fewer massive upper SFMS galaxies ($M_\star > 10^{10}\,M_\odot$) and overproduces QGs with $M_\star \approx 1-3 \times 10^{10}\,M_\odot$ by about a factor of three. Overall, EAGLE provides the closest match to observed number densities for SFGs and green-valley galaxies among the three simulations. SIMBA, on the other hand, shows better agreement with the number densities of QGs. However, each simulation exhibits distinct strengths and limitations when compared to observational data of galaxy subpopulations.

\subsection*{Comparing the environments of observed and simulated galaxies}

\begin{figure}
\includegraphics[width=0.48\linewidth]
{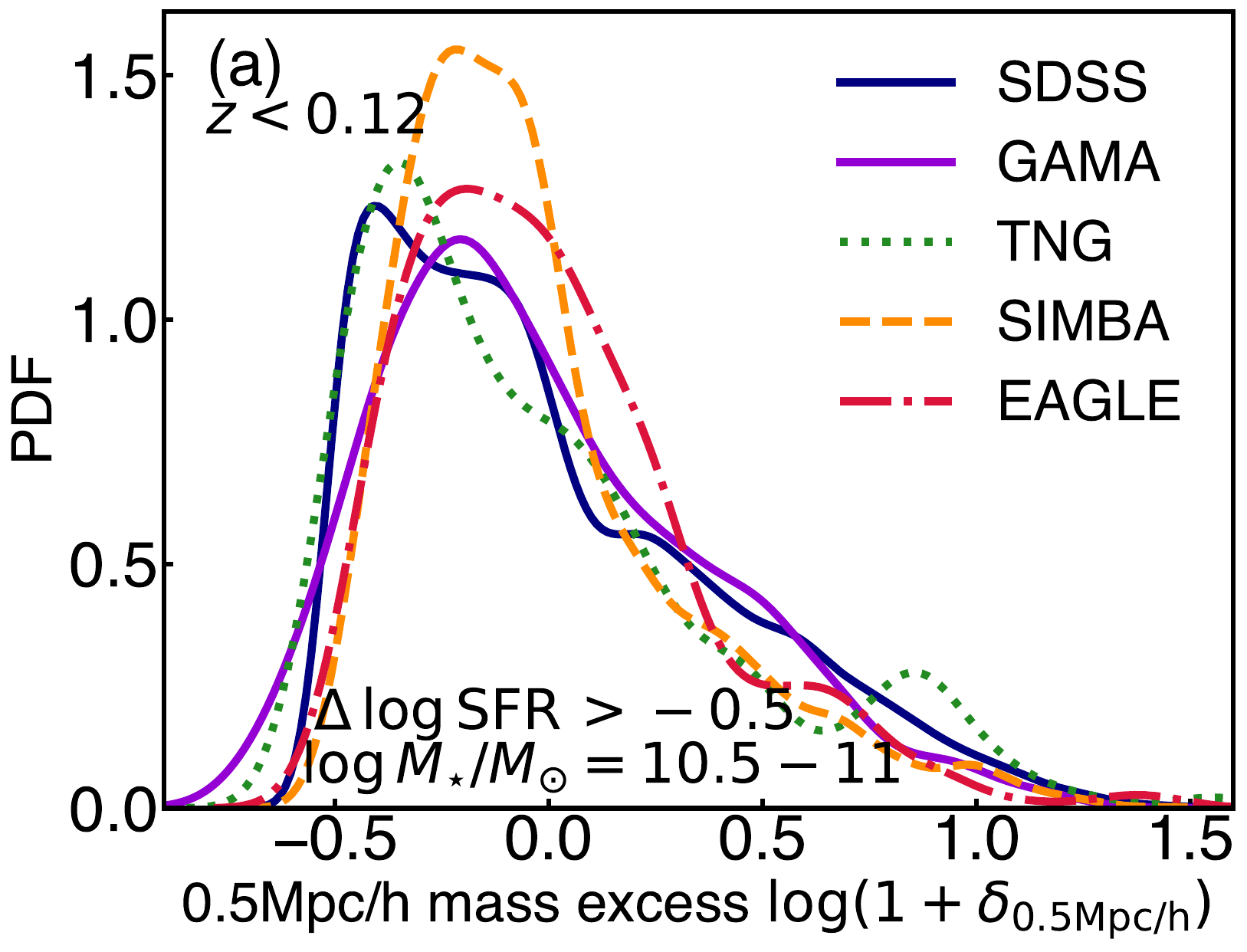} 
\includegraphics[width=0.48\linewidth]
{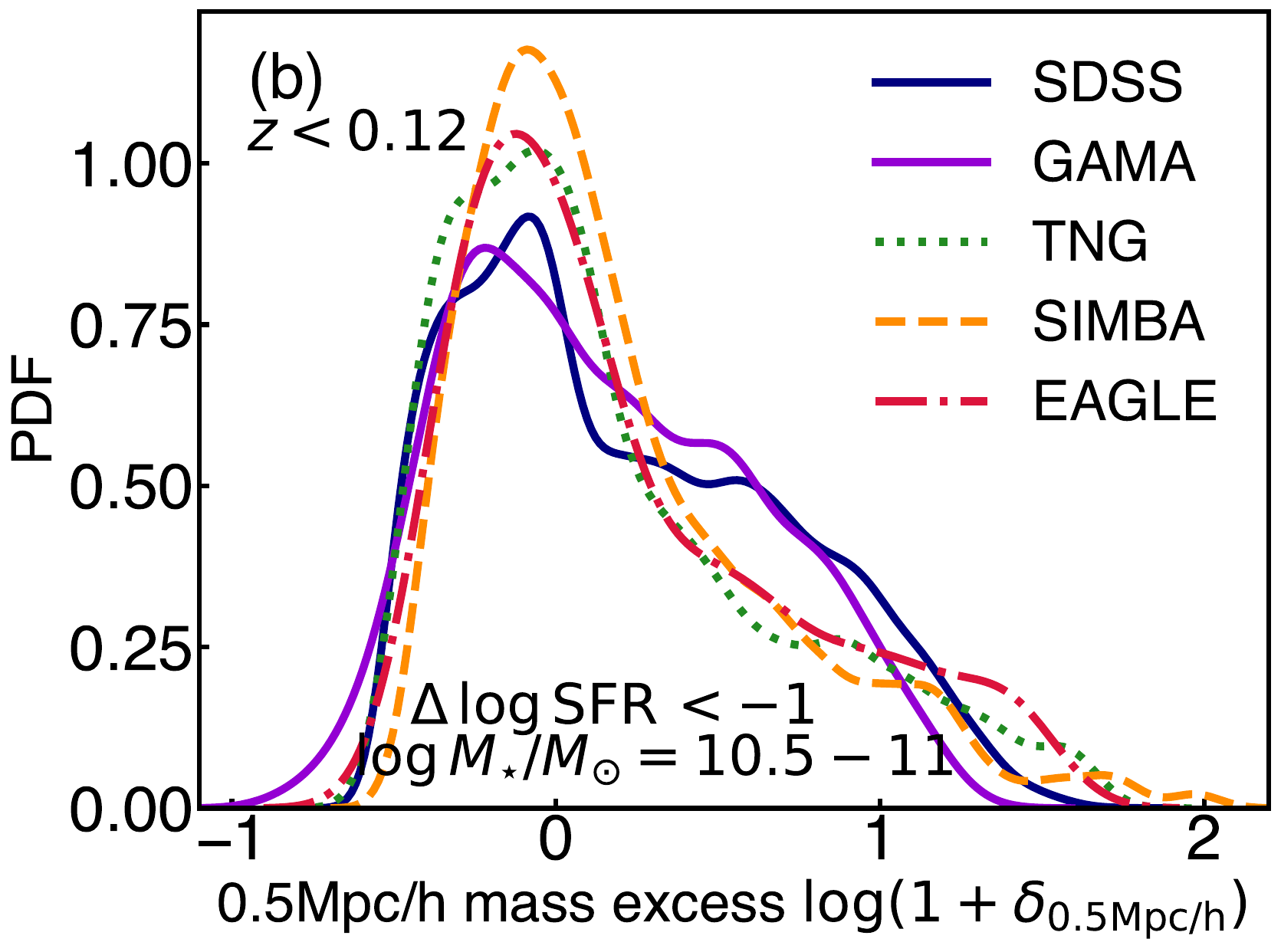}
\includegraphics[width=0.48\linewidth]
{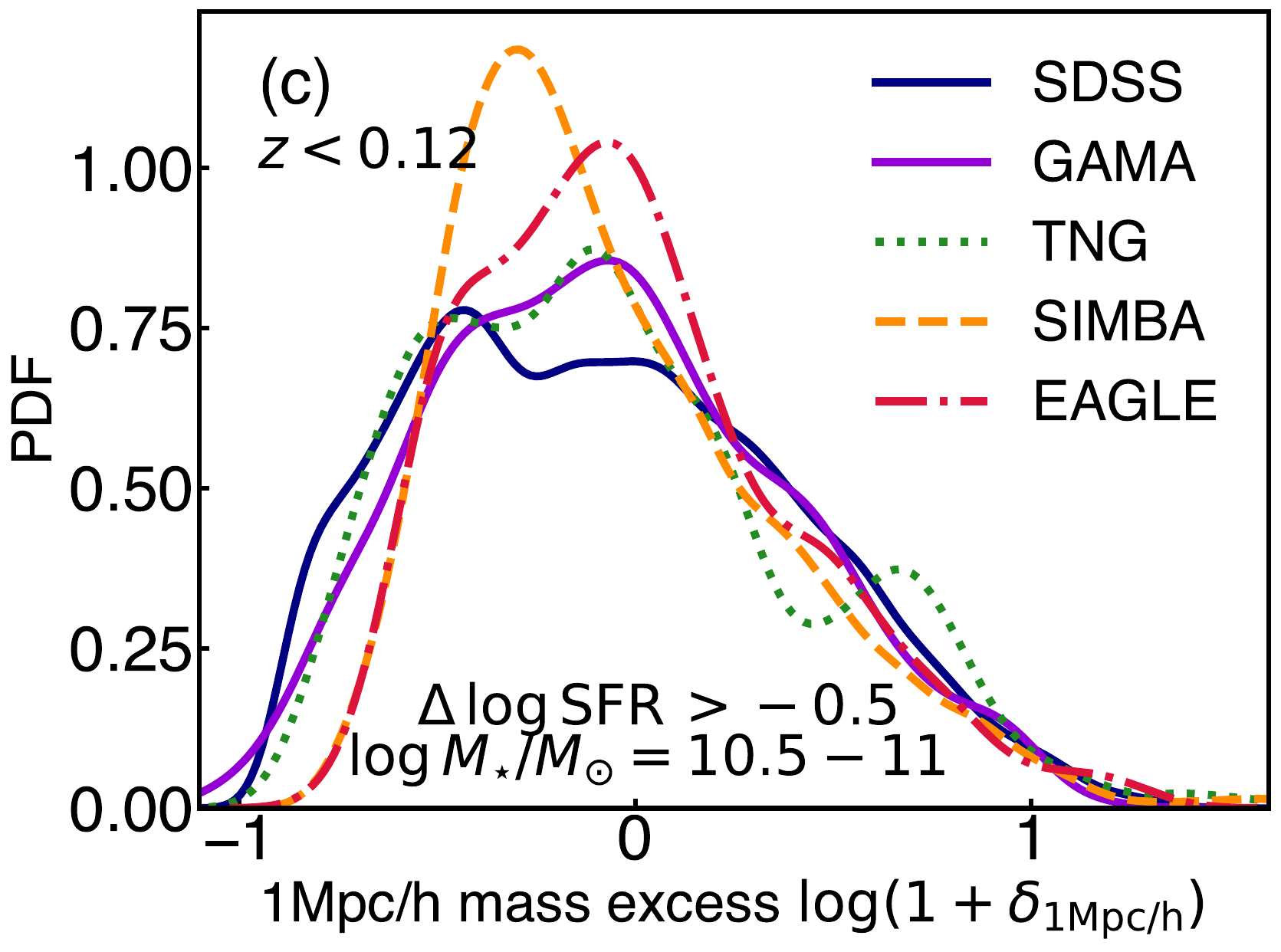}
\includegraphics[width=0.48\linewidth]
{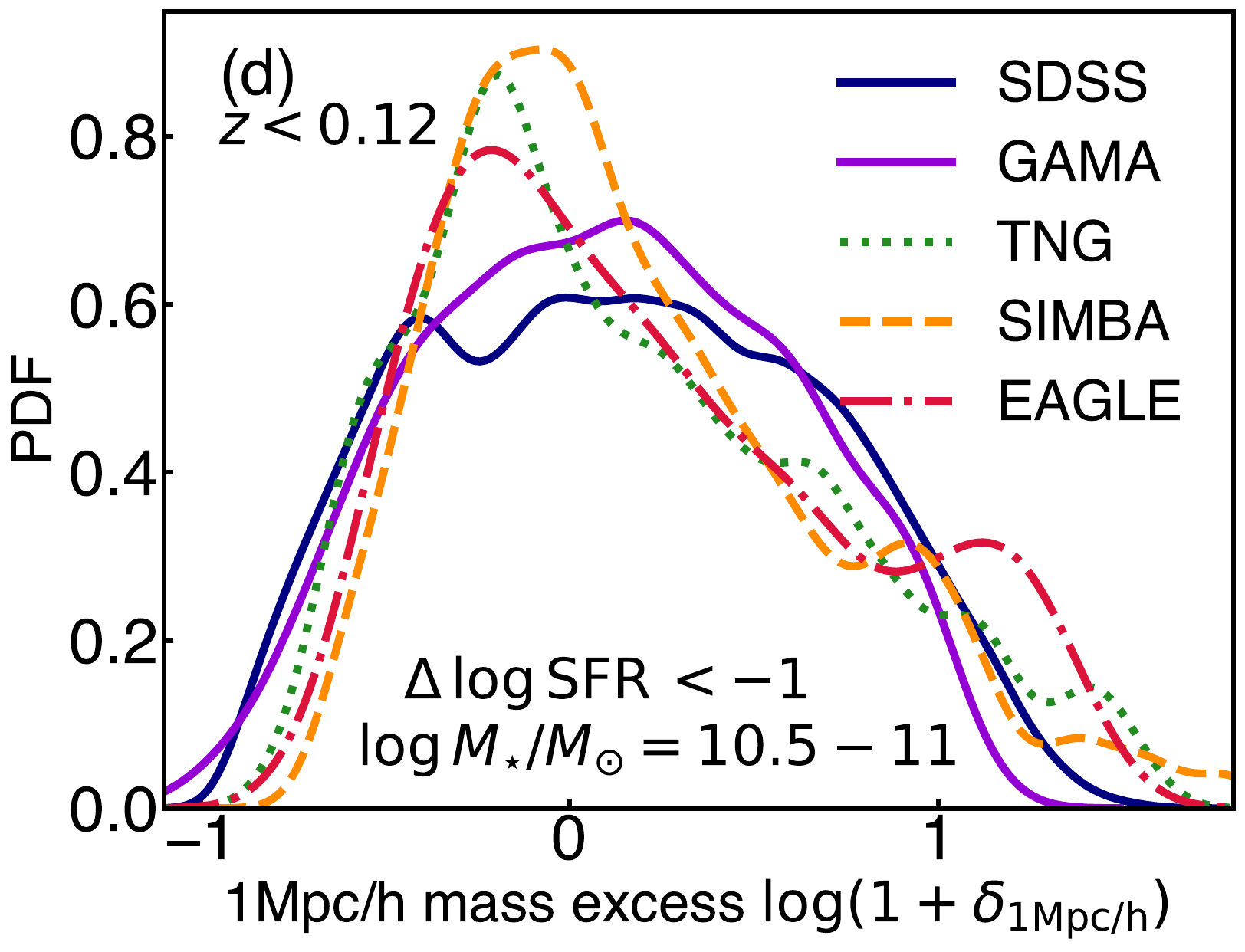}
\includegraphics[width=0.48\linewidth]
{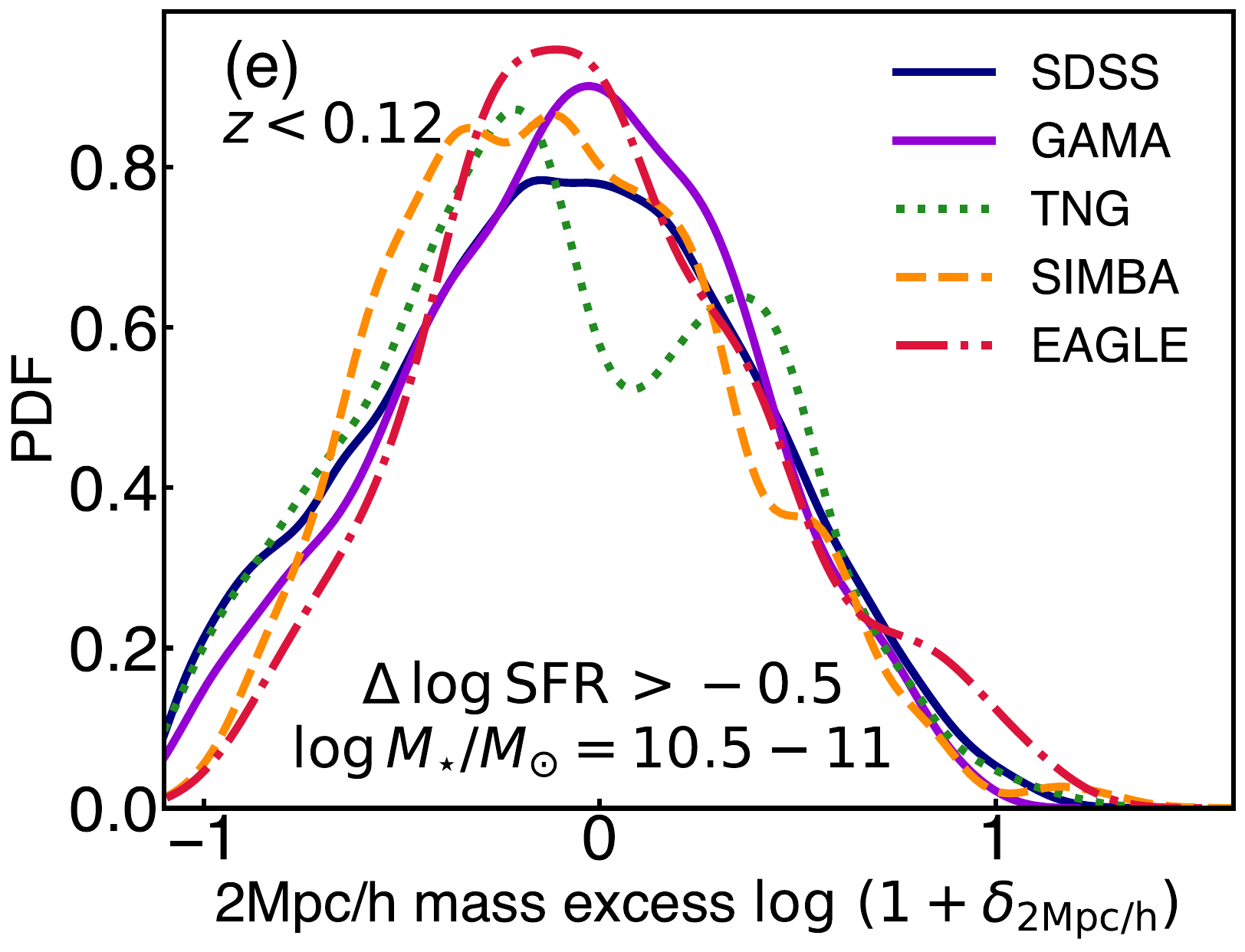}
\includegraphics[width=0.48\linewidth]
{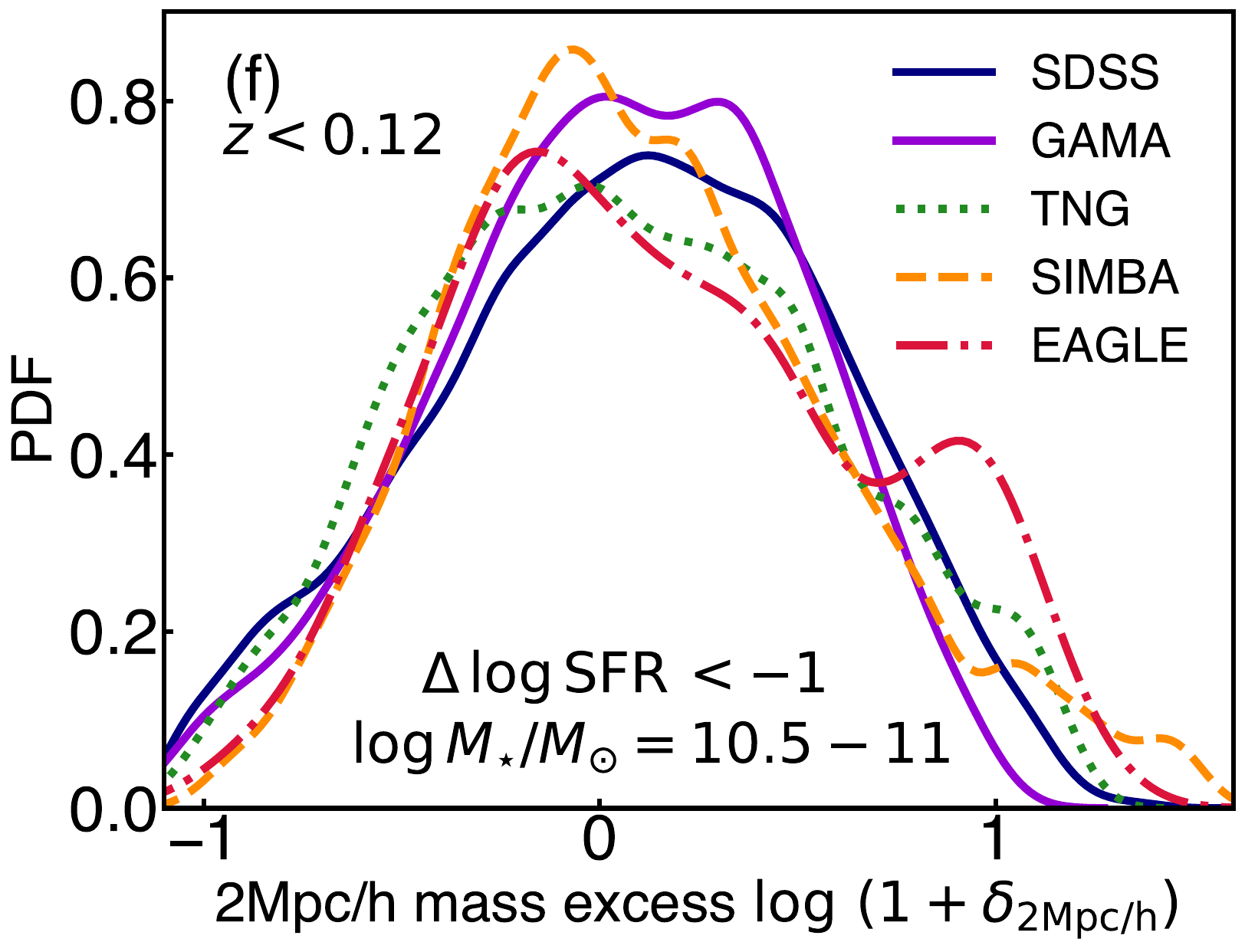}
\caption{\textbf{Comparison of stellar mass overdensity distributions for massive galaxies.}
Galaxies in the stellar mass range $\log\,(M_\star/M_\odot)=10.5$–11 and at redshift $z<0.12$ in SDSS and GAMA are compared with three cosmological simulations---TNG, SIMBA, and EAGLE---at spatial scales of 0.5, 1, and 2\,Mpc\,$h^{-1}$, shown in the first, second, and third rows, respectively. The left panels present the overdensity distributions for star-forming galaxies, while the right panels present those for quiescent galaxies.
\label{fig:M105del}}
\end{figure}

\begin{figure}
\includegraphics[width=0.48\linewidth]
{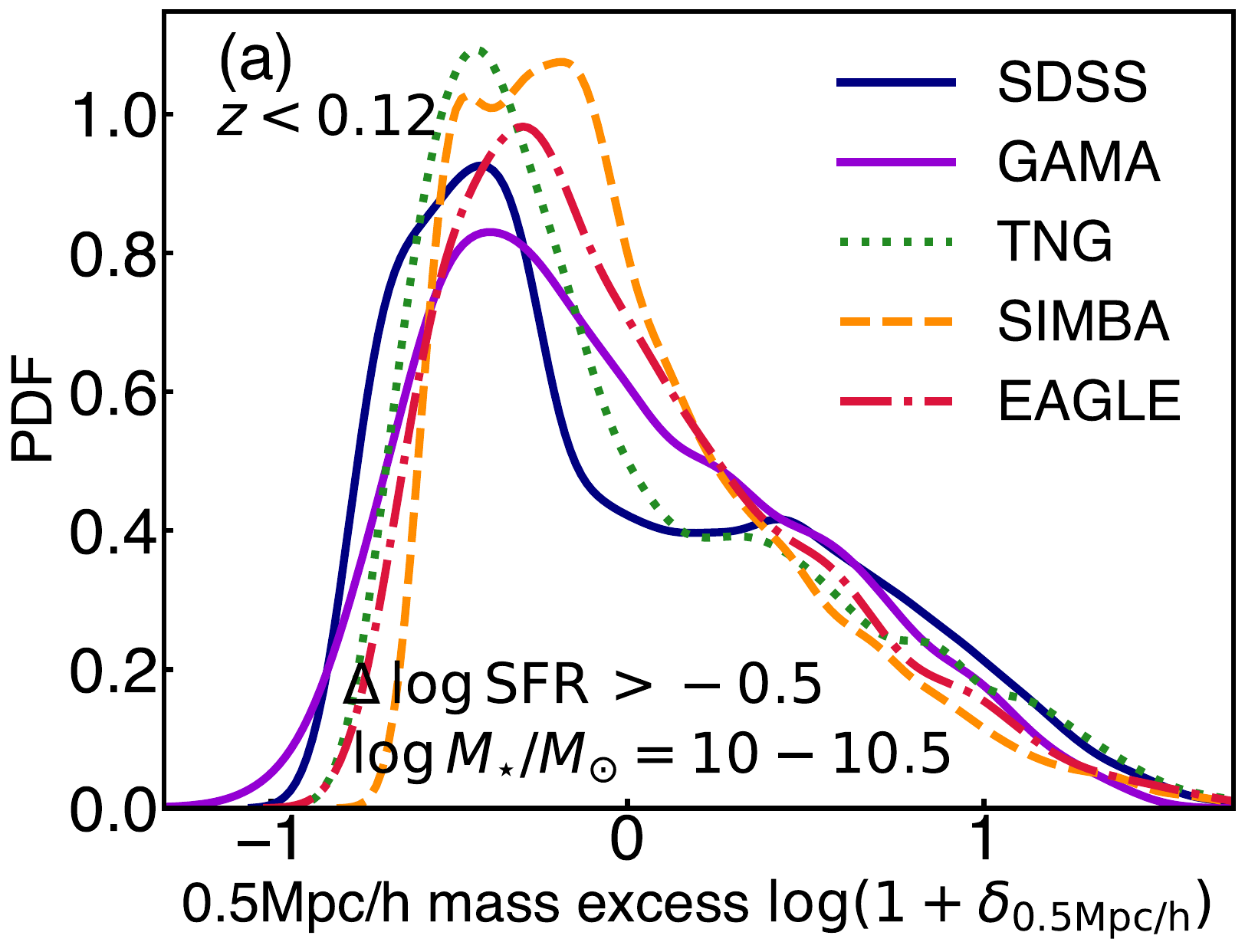}
\includegraphics[width=0.48\linewidth]
{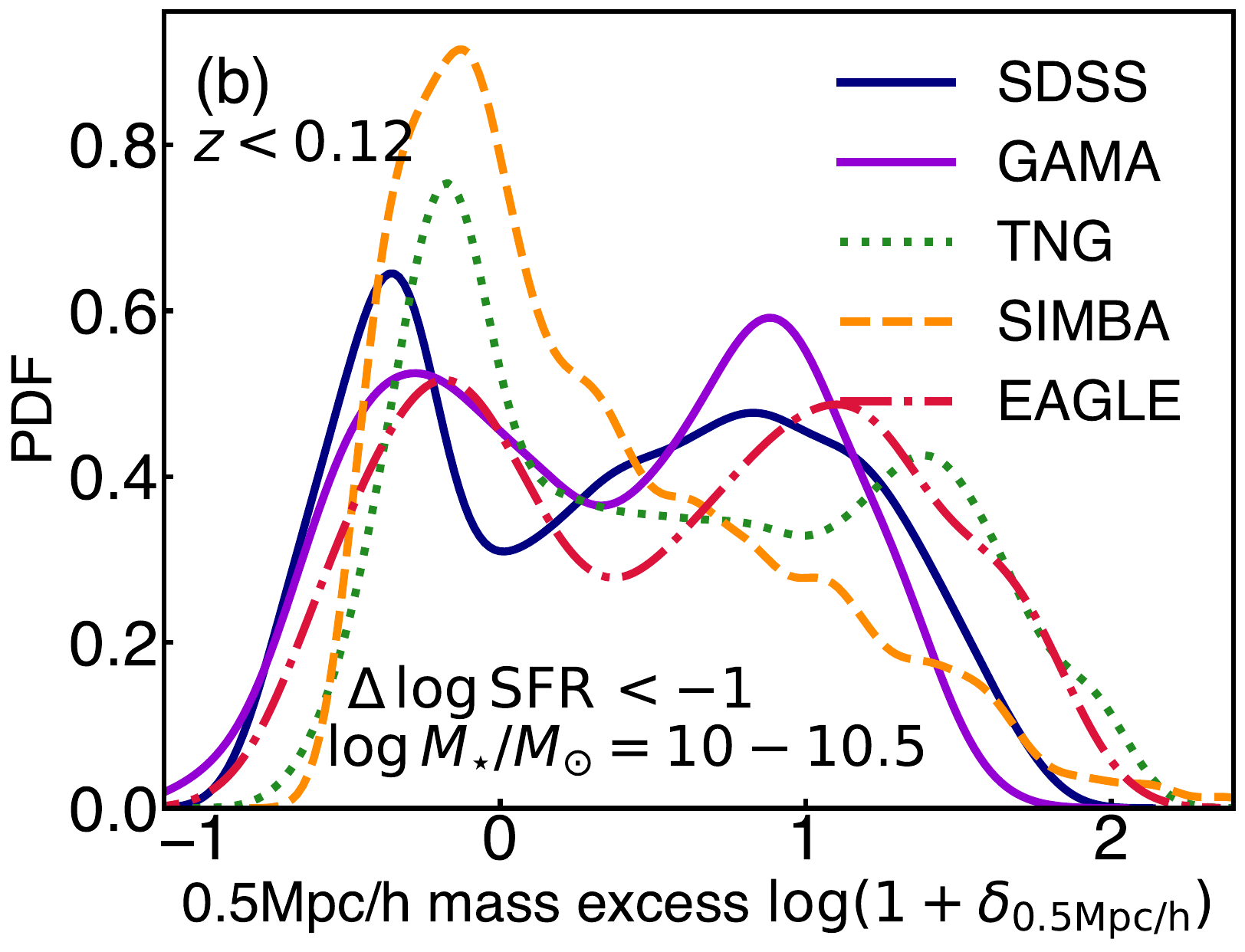}
\includegraphics[width=0.48\linewidth]
{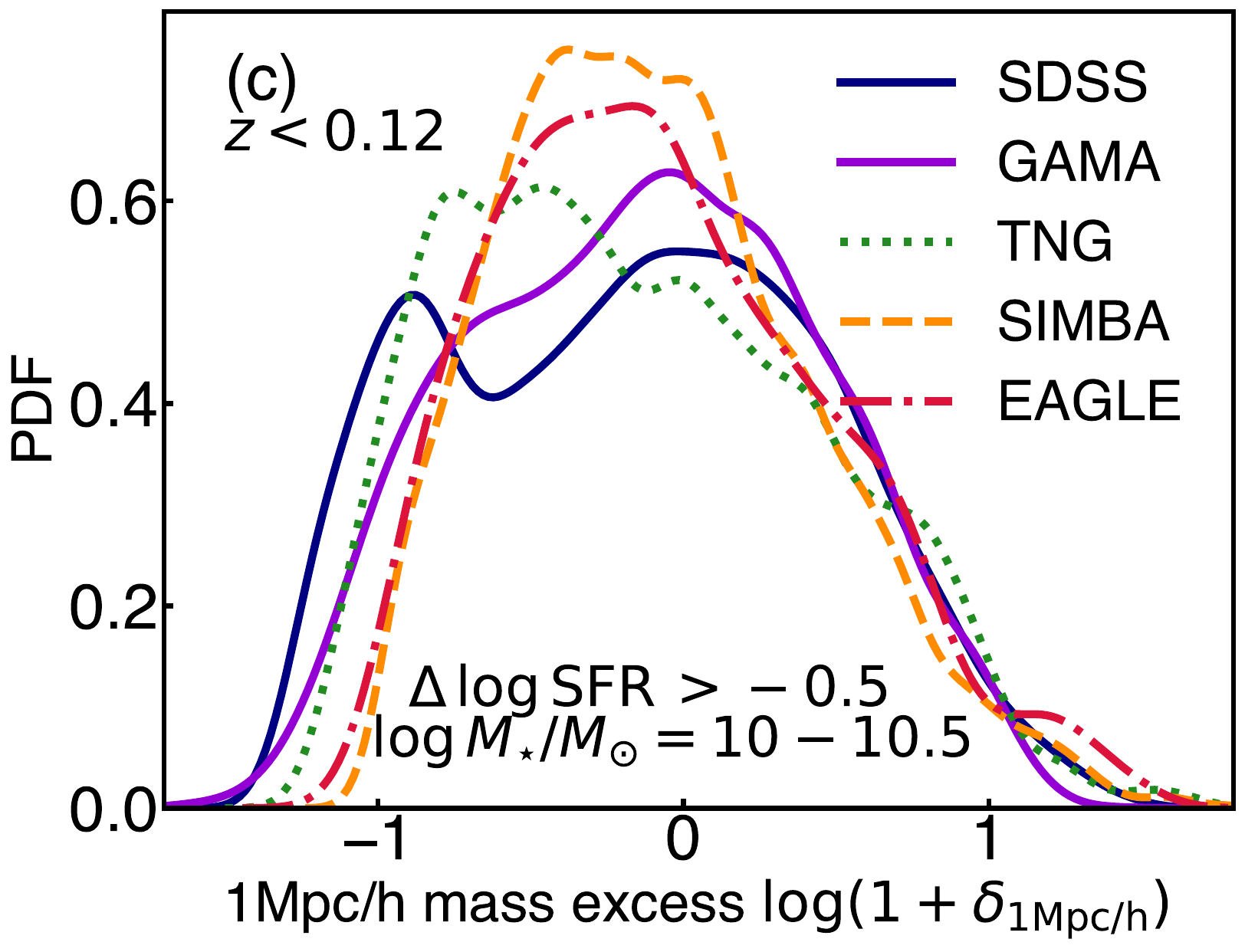}
\includegraphics[width=0.48\linewidth]
{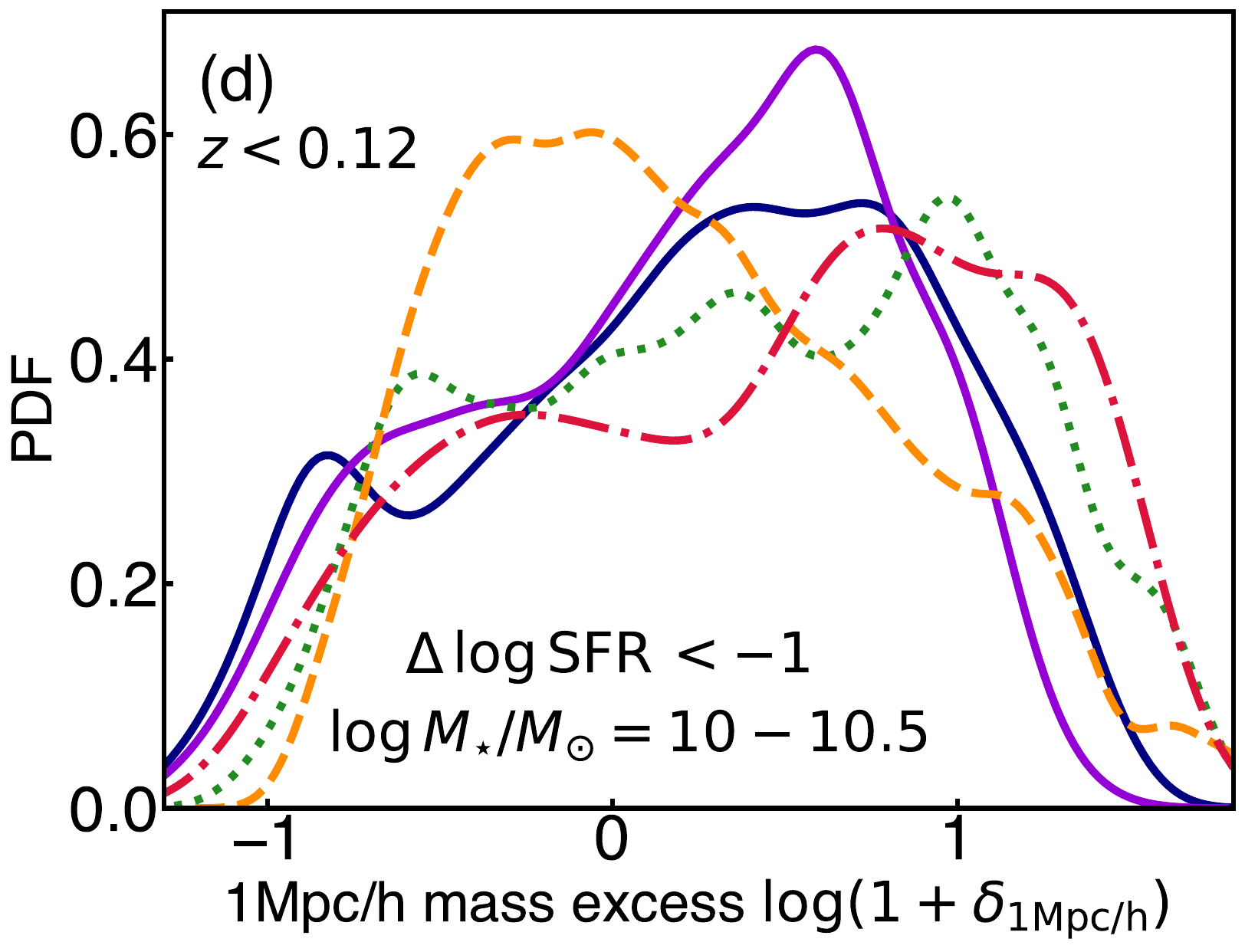}
\includegraphics[width=0.48\linewidth]
{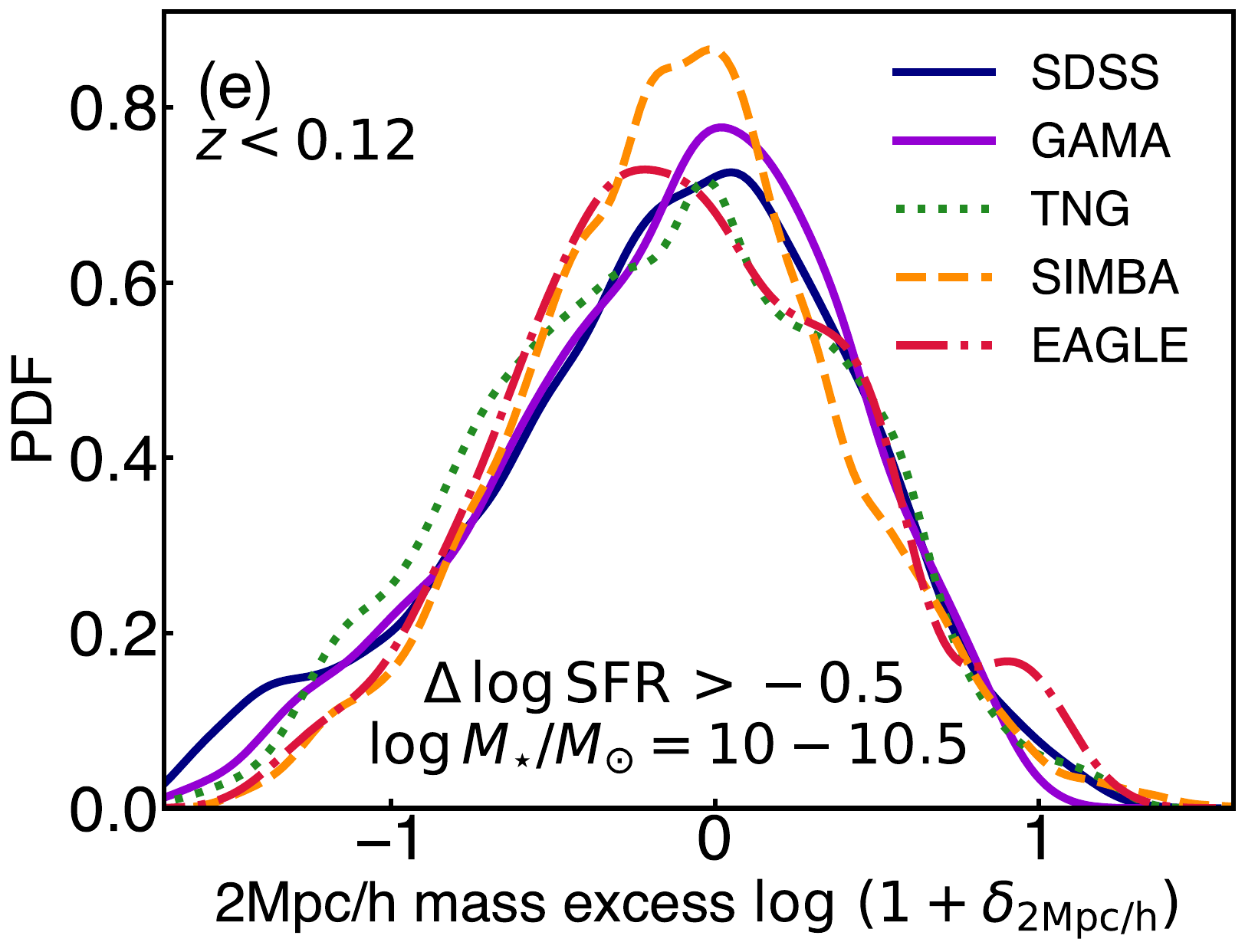}
\includegraphics[width=0.48\linewidth]
{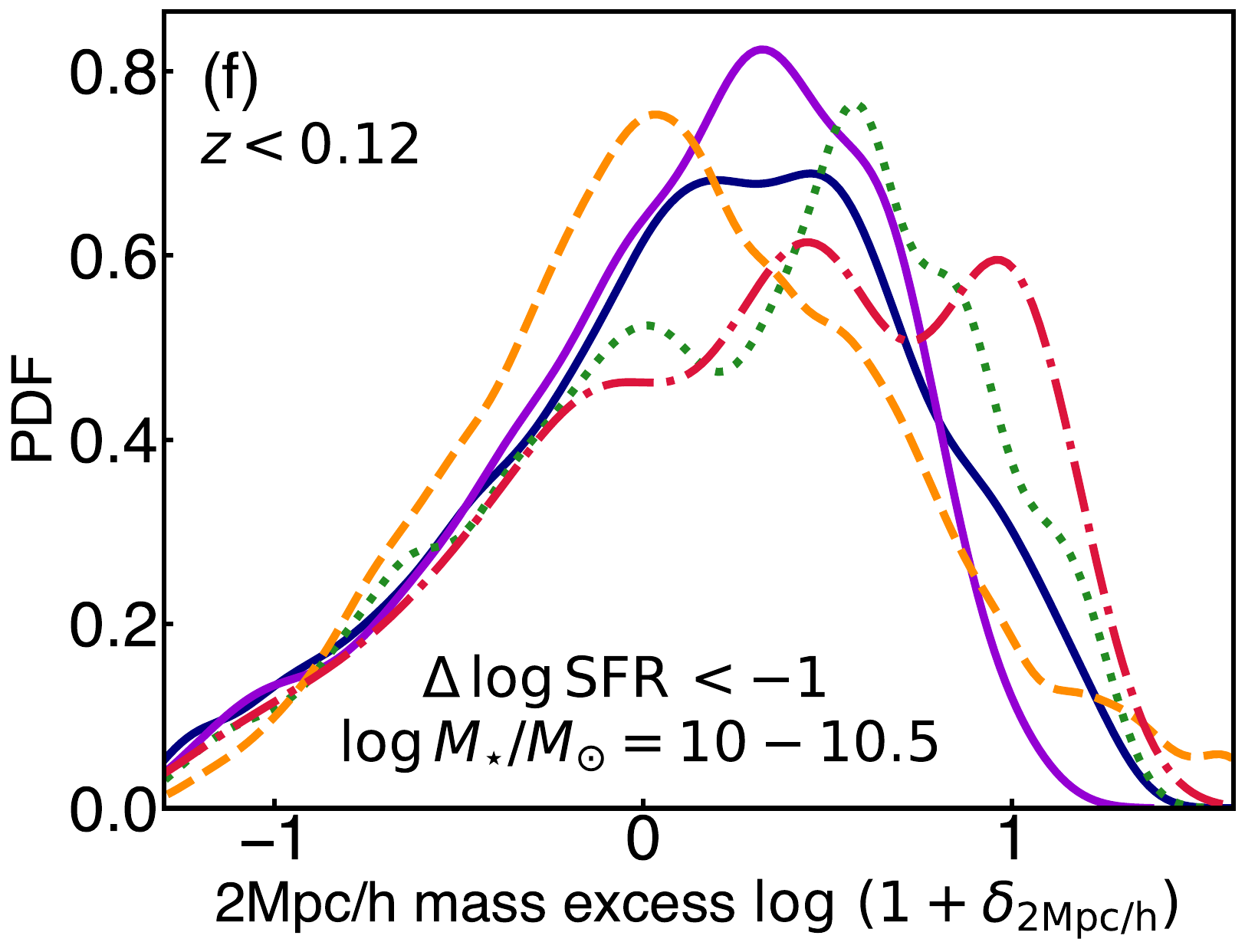}
\caption{\textbf{Stellar mass overdensity distributions for intermediate-mass galaxies.} This figure examines galaxies with $\log (M_\star/M_\odot)= 10-10.5$ at redshift $z<0.12$ in SDSS and GAMA, and compare them with predictions from the TNG, SIMBA, and EAGLE simulations. Results are shown at spatial scales of 0.5, 1, and 2\,Mpc\,$h^{-1}$ in the first, second, and third rows, respectively. The left panels correspond to star-forming galaxies, and the right panels to quiescent galaxies.\label{fig:M10del}}
\end{figure}

\begin{figure}
\includegraphics[width=0.48\linewidth]
{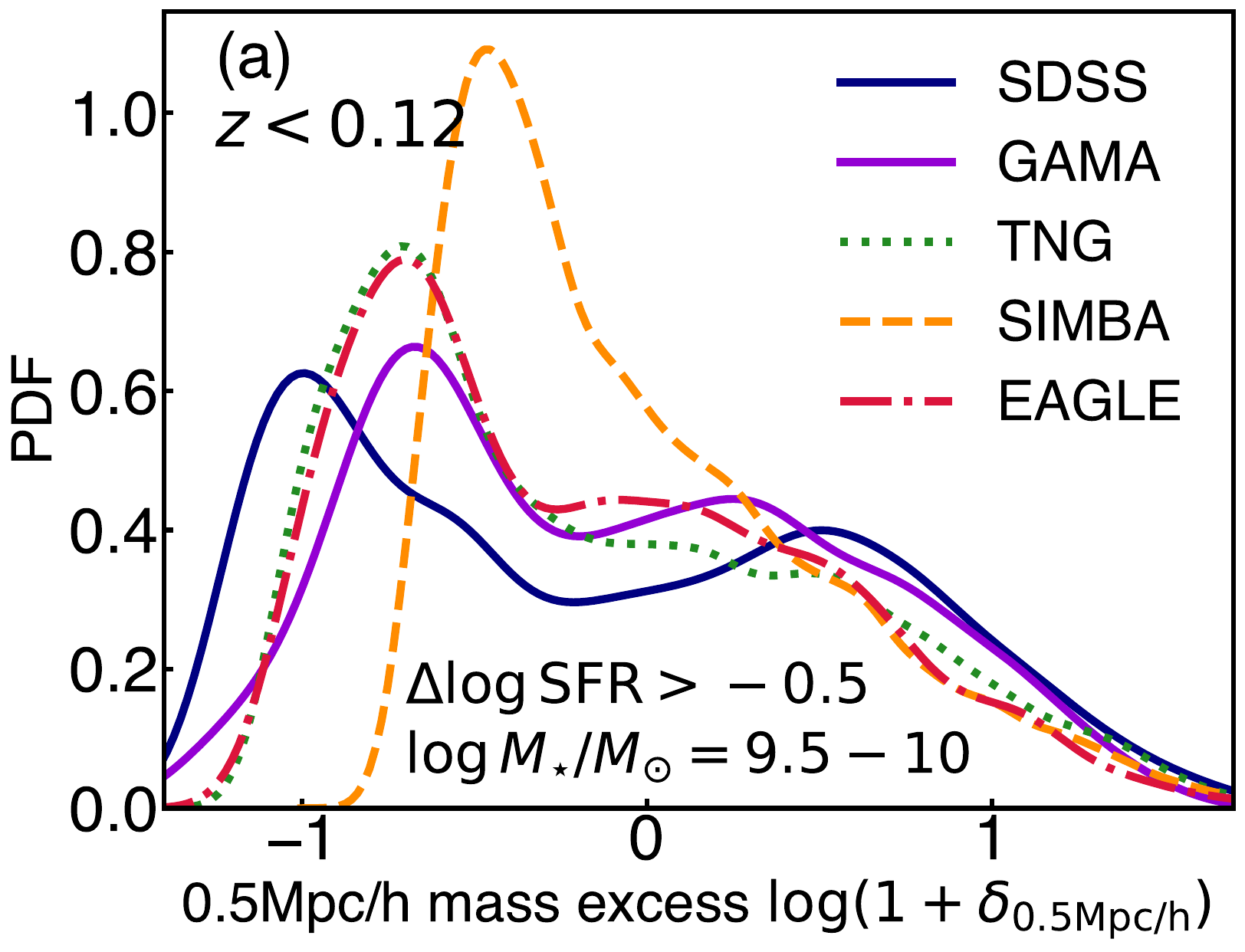}
\includegraphics[width=0.48\linewidth]
{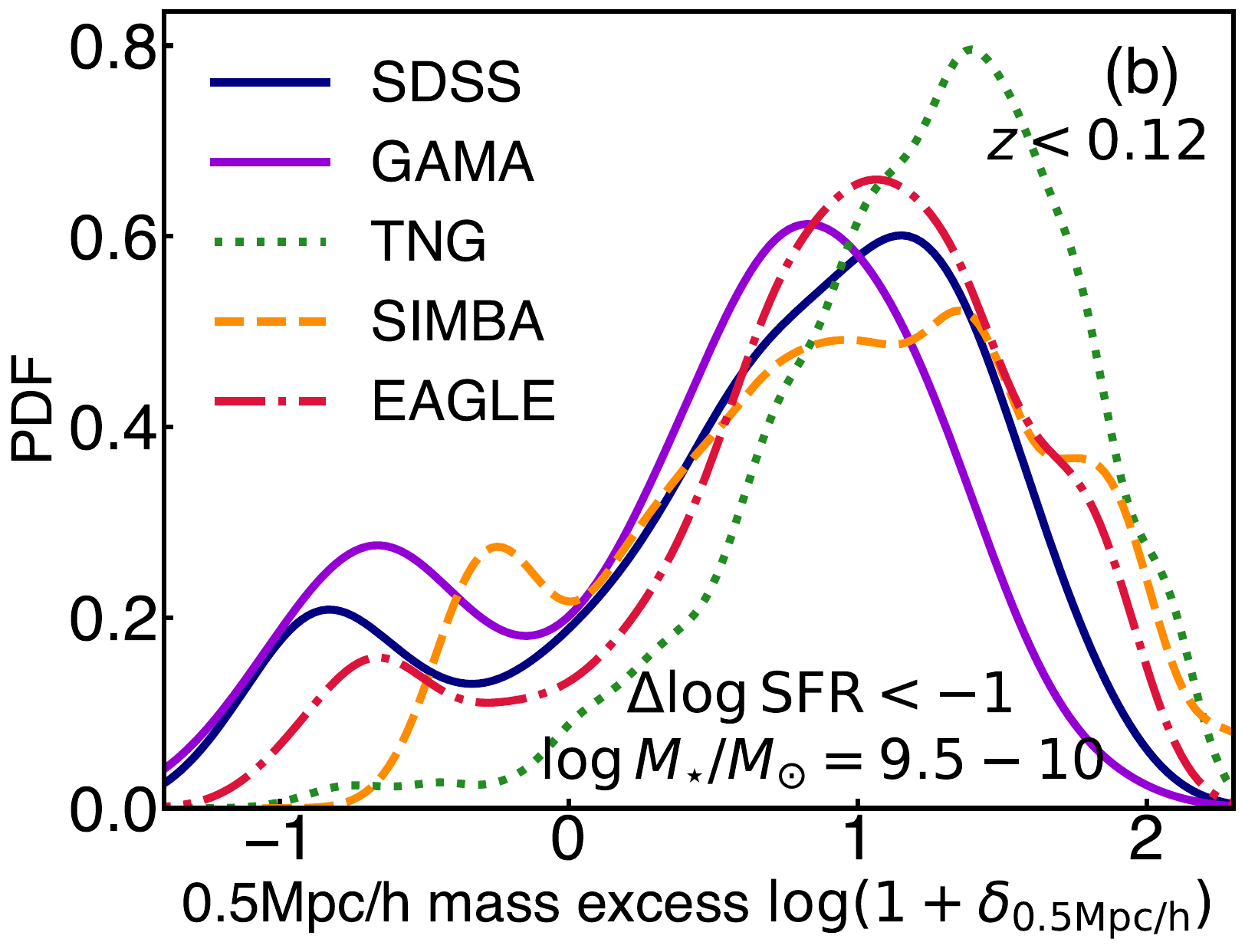}
\includegraphics[width=0.48\linewidth]
{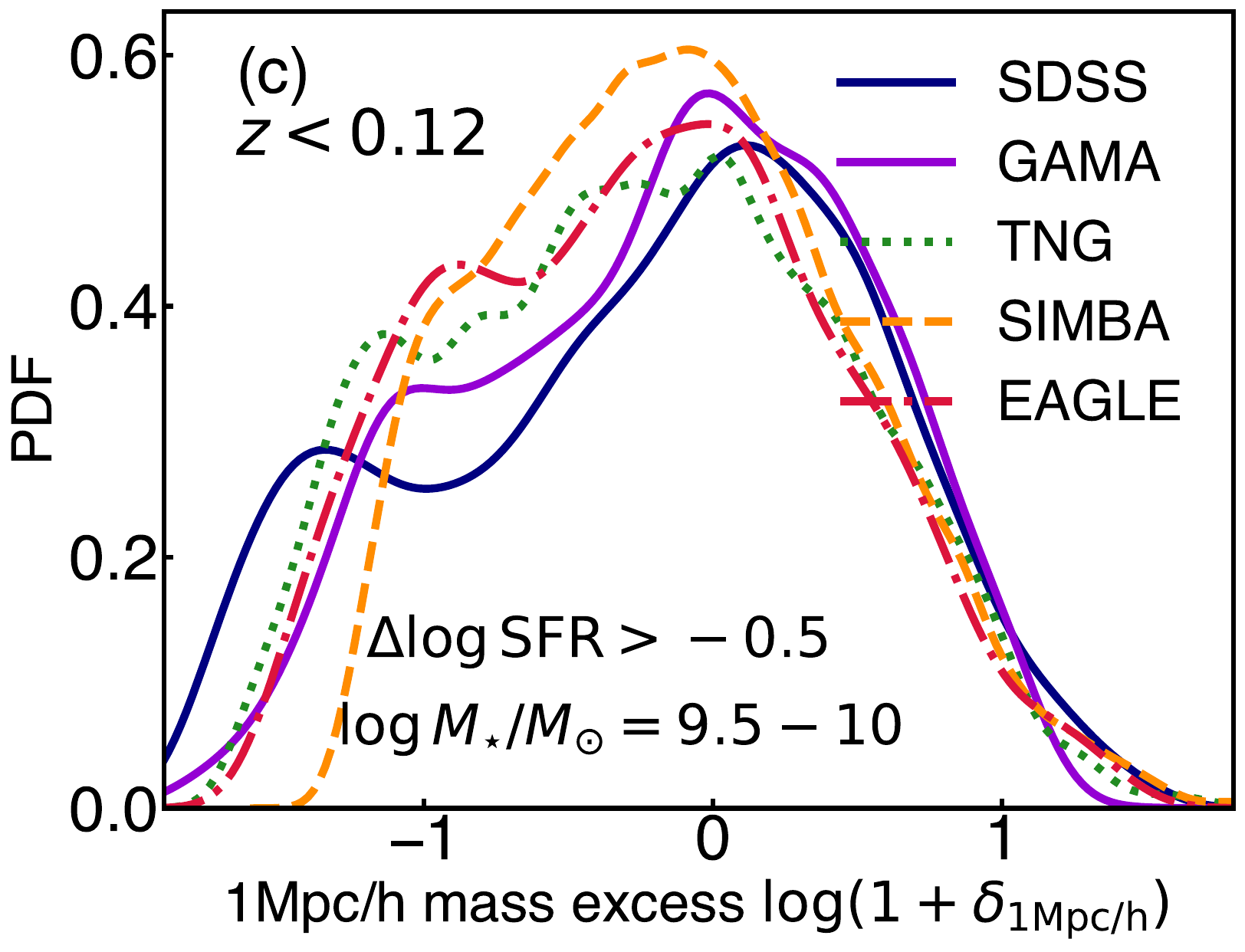}
\includegraphics[width=0.48\linewidth]
{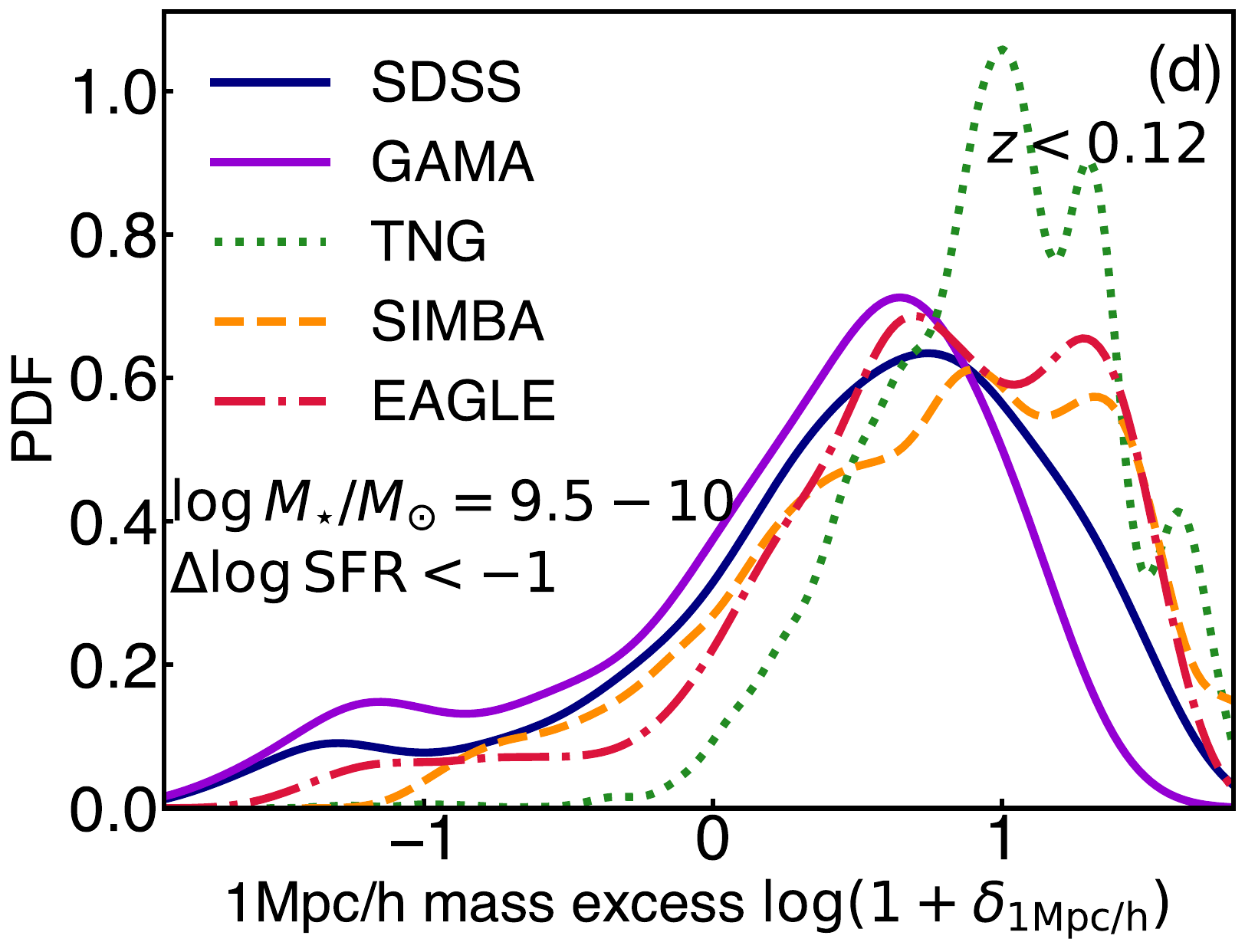}
\includegraphics[width=0.48\linewidth]
{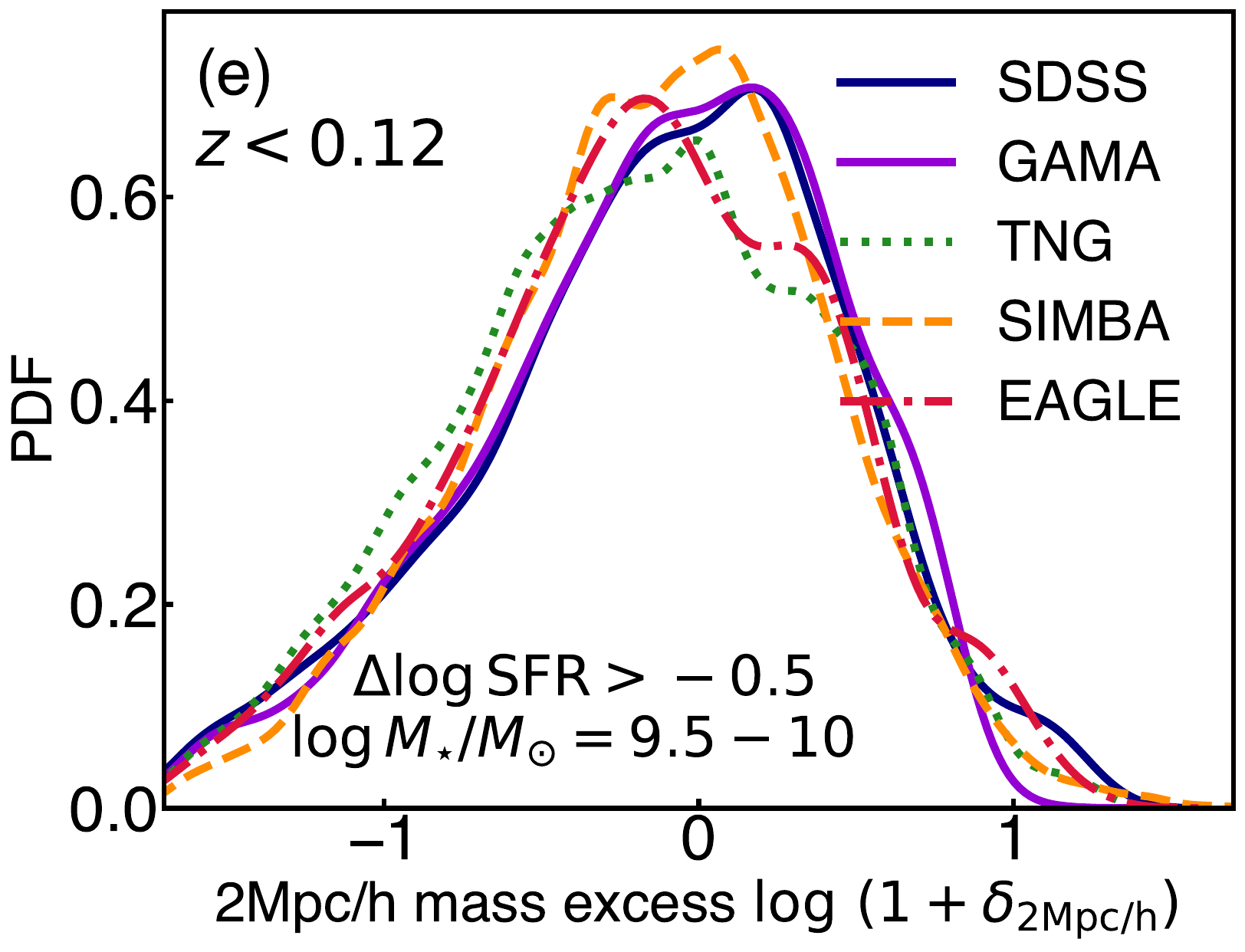}
\includegraphics[width=0.48\linewidth]
{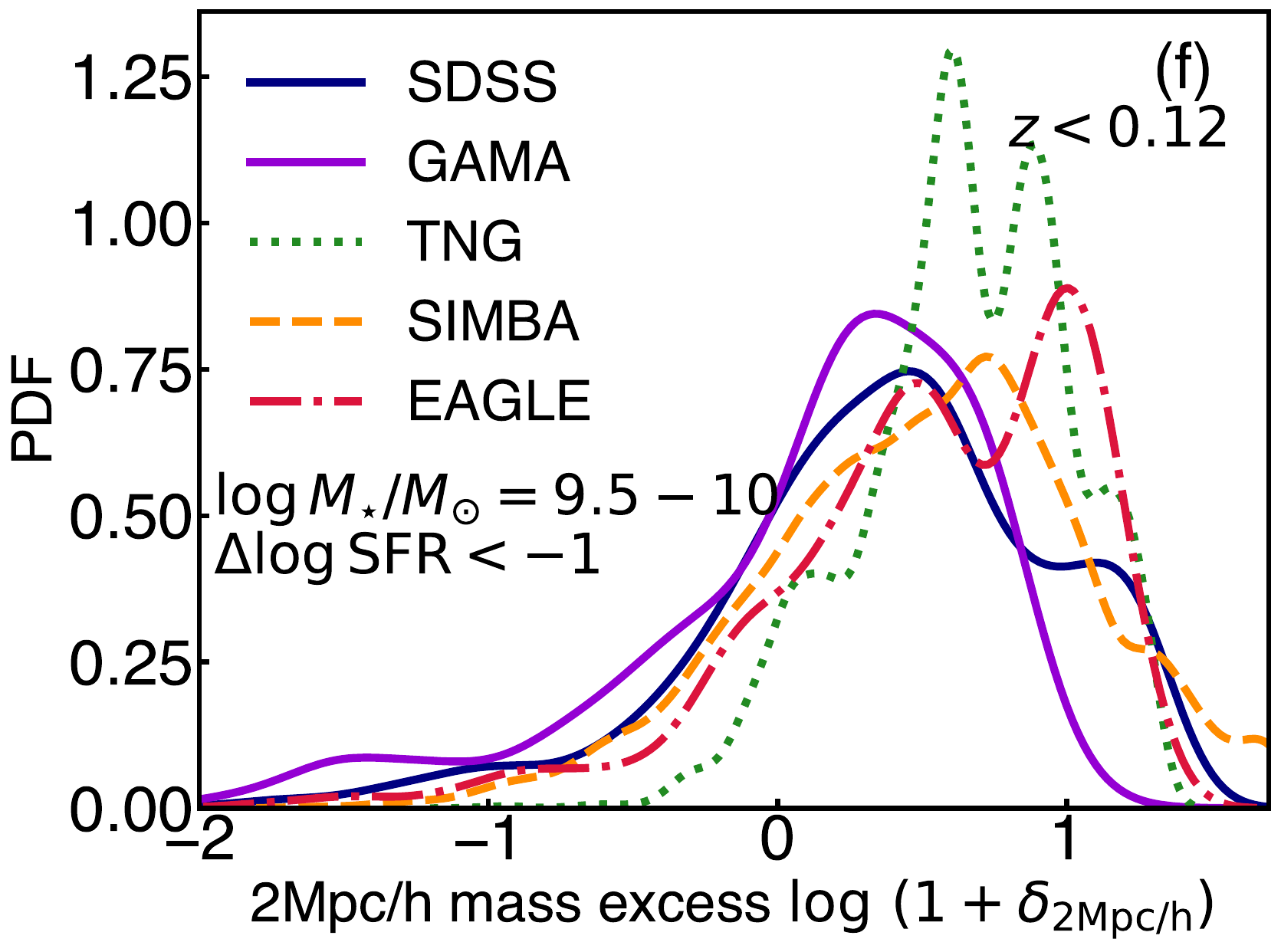}
\caption{\textbf{Stellar mass overdensity distributions for low-mass galaxies.} Galaxies with $9.5 \le \log (M_\star/M_\odot) < 10$ at redshift $z<0.12$ in SDSS and GAMA are contrasted with predictions from the TNG, SIMBA, and EAGLE simulations. The distributions are shown at spatial scales of 0.5, 1, and 2\,Mpc\,$h^{-1}$ in the first, second, and third rows, respectively. Star-forming and quiescent populations are displayed in the left and right panels.}
\label{fig:M95del}
\end{figure}

Fig.~2 compares multiscale environmental indicators between simulations and observations. On scales of $0.5-8\,\mathrm{Mpc/h}$, all three simulations effectively reproduce the mass overdensity distributions observed in GAMA and SDSS for galaxies at $z < 0.12$. However, the simulations predict different stellar mass overdensity distributions on smaller scales ($<2$\,Mpc/h), especially for galaxy subpopulations classified by $M_\star$ and $\Delta \log\,\mathrm{SFR}$, with discrepancies between the simulations and GAMA/SDSS ($p \lesssim .001$, Supplementary Figs.~\ref{fig:M105del}--\ref{fig:M95del}). 

\begin{figure}
\includegraphics[width=0.99\linewidth]{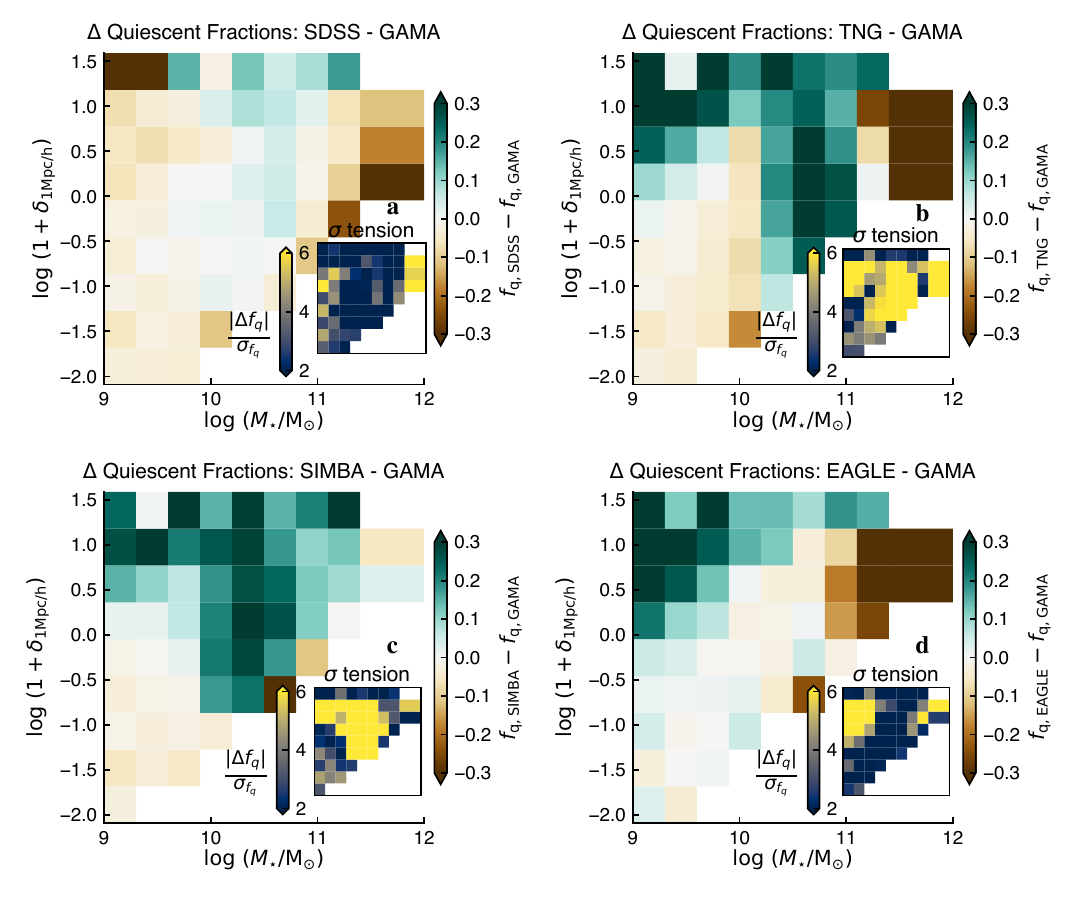}
\caption{\textbf{Differences in quiescent galaxy fractions as functions of stellar mass and environment using GAMA}. Panel (a) shows the differences of quiescent fraction measured in SDSS and GAMA as a function of stellar mass and stellar mass excess, defined within 1\,Mpc\,$h^{-1}$ of each galaxy. Panels (b)--(d) illustrate the differences between the quiescent fractions predicted by each simulation and those observed in GAMA. Bins shaded green (brown) correspond to higher (lower) quiescent fractions compared to GAMA at fixed stellar mass and environment. Insets show the absolute difference normalized by the combined Binomial uncertainties of GAMA and simulations, $| \Delta f_q | / \sigma_{f_q}$, providing a measure of the tension ($\sigma$-level). \label{fig:ME_GAMAsim}}
\end{figure}

\begin{figure}
\includegraphics[width=0.99\linewidth]{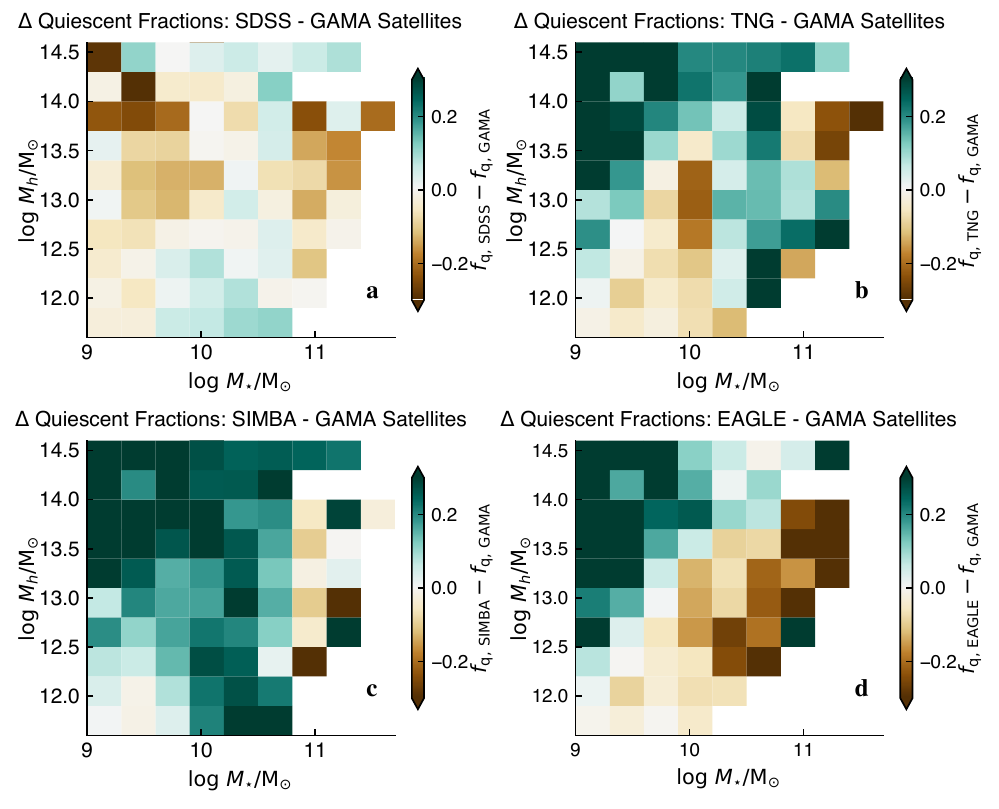}
\caption{\textbf{Differences in quiescent fractions of satellite galaxies as functions of stellar mass and halo mass in GAMA and the simulations.} Panel (a) shows the difference in quiescent fraction between SDSS and GAMA for satellite galaxies. Panels (b--d) present the residuals between the quiescent fractions predicted by each simulation and those measured in GAMA. Green (brown) indicates higher (lower) quiescent fractions relative to GAMA at fixed stellar and halo mass. For the observational data, halo masses are derived from our new measurements that combine multi-scale stellar-mass overdensities with additional observational halo-mass tracers. Panel (a) adopts the SIMBA-based halo-mass calibration, while panels (b--d) use the halo-mass calibrations corresponding to each simulation.}
\label{fig:MsMh_GAMAsim}
\end{figure}

Supplementary Figs.~\ref{fig:ME_GAMAsim} and \ref{fig:MsMh_GAMAsim} examine the relationships among stellar mass, environment, halo mass, star formation rates, and quiescent fractions, emphasizing differences between simulations and observations. These figures revisit earlier analyses with GAMA data, Both SDSS and GAMA yield similar results. However, the simulations do not accurately reproduce the dependence of $f_q$ on halo mass and stellar mass as inferred from both surveys. Discrepancies are most evident in low-mass and/or high-density regimes, with SIMBA, TNG, and EAGLE each exhibiting unique deviations. 

In addition, the predicted halo mass distributions for galaxy subpopulations differ significantly across the three simulations ($p \lesssim .001$), with offsets of $\Delta M_h \sim 0.2$--0.3\,dex in some cases (not shown in a separate figure). Future observations, particularly those separating centrals and satellites accurately and subdividing by $M_\star$ and $\Delta \log\,\mathrm{SFR}$, may help constrain these discrepancies. For massive SFGs with $\log M_\star/M_\odot = 10.5-11$, EAGLE estimates a median halo mass (with +85\%/-15\% percentiles) of $\log M_h/M_\odot = 12.66^{+0.64}_{-0.32}$, which is higher than the predictions from TNG ($\log M_h/M_\odot = 12.33^{+0.80}_{-0.22}$) and SIMBA ($\log M_h/M_\odot = 12.31^{+0.60}_{-0.21}$). EAGLE tends to predict more massive halos for both central and satellite galaxies at fixed $M_\star$, whereas SIMBA generally predicts the lowest halo masses or slightly lower than EAGLE, depending on the subpopulation.

Our results on quiescent fraction trends with environment and halo mass pertain to the study of Donnari et al.~\cite{Donnari+21fq}, which compared the quiescent fractions of central and satellite galaxies in TNG and EAGLE with similar survey data, focusing on satellites in massive halos ($M_h > 10^{13}\,M_\odot$) or considering all halos collectively. Our analysis builds on this by (1) introducing a new comparison based on an observable environmental measure, the stellar mass excess within a fixed aperture; (2) adopting a consistent, simulation-calibrated framework for estimating halo masses from observables; and (3) expanding the comparison to include SIMBA. Together, these improvements enable a more detailed assessment and tighter constraints on where and why simulations diverge from observations at fixed $M_h$ and $M_\star$. The implications of these discrepancies, along with simulation-specific behaviors and resolution-dependent effects, are presented in the Discussion and Extended Data, where we provide interpretation of these mismatches (Extended Data Fig.~2). Our versions of the trends in central and satellite quiescent fractions for the three simulations, which mirror those of Donnari et al., are given in Extended Data Fig.~6. 

At low masses ($M_\star \lesssim 10^{10}\,M_\odot$), most centrals ($\sim 80-90$\%) remain star-forming in both observations and simulations, while 40–70\% of satellites are quenched in simulations versus 20–30\% in GAMA, indicating that environmental quenching is overestimated. GAMA observations hint at a steep rise in satellite quenching below $M_\star < 10^9\,M_\odot$, consistent with theoretical expectations \citep{Fillingham+15}. Overall, GAMA and SDSS are consistent across most of the mass range, except at the extremes, contrasting with earlier studies reporting larger discrepancies \citep{Donnari+21fq}.

We note that our quiescent fraction for central galaxies is broadly consistent with the results of Dickey et al. \citep{Dickey+21}, who studied isolated quiescent systems by creating mock SDSS observations from EAGLE, TNG, and SIMBA, including forward modeling of SDSS selection effects. In their work, quiescence was defined using the 4000 {\AA} break and H$\alpha$ equivalent width. While this approach is complementary, it requires careful treatment of stellar population degeneracies and aperture effects, which our direct measurements avoid, but could be tackled by future work.

\begin{figure}
\includegraphics[width=0.99\linewidth]{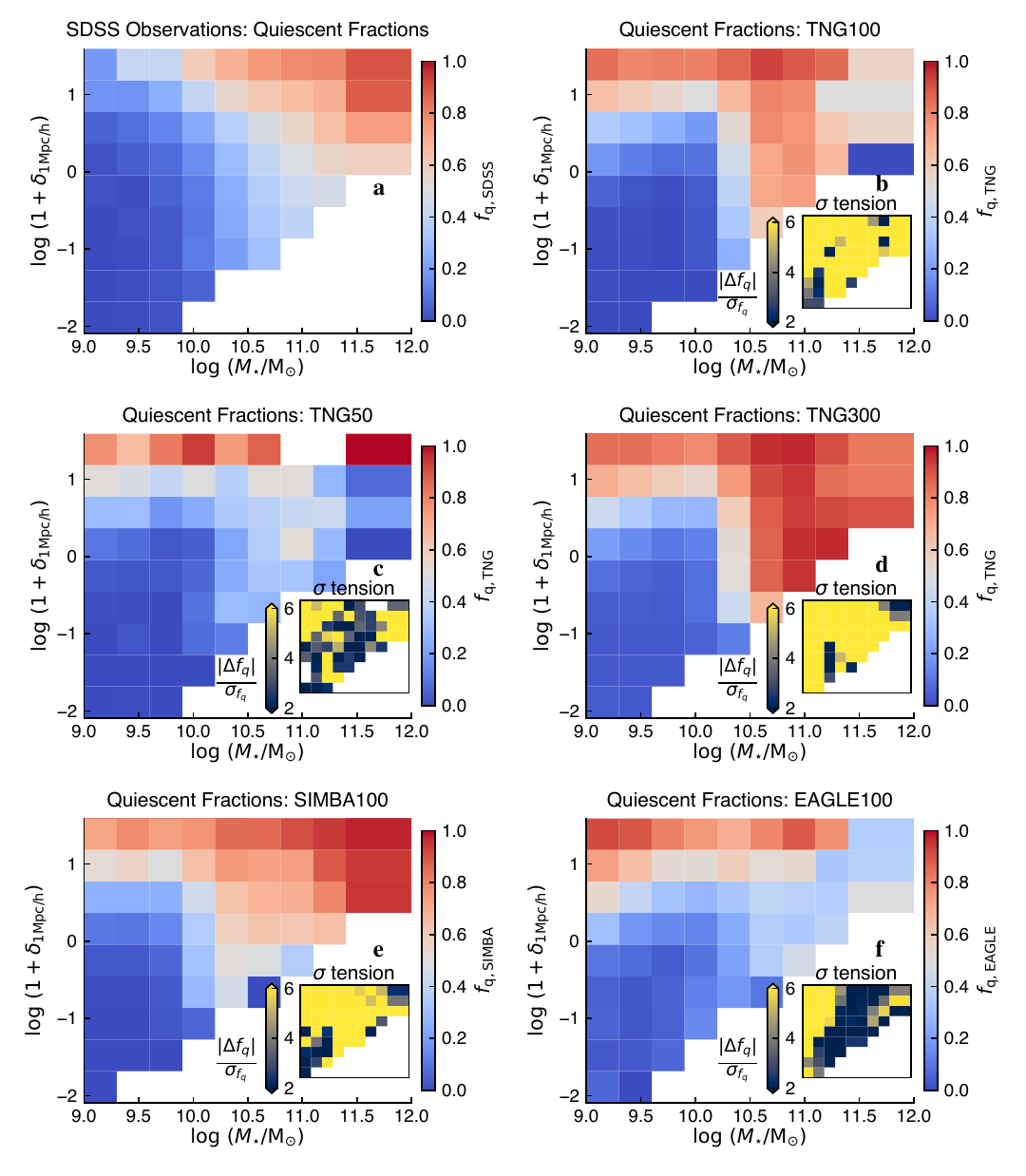}
\caption{\textbf{Trends of quiescent fractions as a function of stellar mass and environment.} Quiescent fractions from three simulations---TNG (volume variants, panels b–d), SIMBA (panel e), and EAGLE (panel f)---are compared directly with SDSS measurements (panel a). Here, the simulation values are shown rather than differences from observations as in Fig.~3. Insets show $|\Delta f_q| / \sigma_{f_q}$, where $\sigma_{f_q}$ is obtained by adding in quadrature the binomial standard errors of the SDSS and simulation quiescent fractions, providing a quantitative measure of the statistical tension ($\sigma$ level).}
\label{fig:ME_raw}
\end{figure}

Supplementary Fig.~\ref{fig:ME_raw} shows a variant of Fig.~3, presenting the quiescent fractions from the simulations directly rather than the residuals relative to observations. This version also includes results from TNG50 and TNG300. Across all simulations, the quiescent fraction of low-mass galaxies is significantly overpredicted in high-density environments, regardless of simulation volume, resolution, or feedback implementation. EAGLE predicts that many massive galaxies ($M_\star \gtrsim 10^{11}\,M_\odot$) remain star-forming, in qualitative disagreement with both observations and the TNG and SIMBA simulations. A larger-volume EAGLE variant would be useful to confirm or refine this tentative trend. However, the discrepancy also reflects the limited number of quiescent galaxies in EAGLE (Fig.~1), indicating that it arises from physical differences in the model as well as from statistical limitations. The quiescent fractions differ notably for massive galaxies among the TNG runs: TNG300 predicts higher quiescent fractions than TNG100, resulting in better agreement with SDSS at the highest masses in dense environments, but worse agreement around $M_\star \sim 10^{11}\,M_\odot$, where it overpredicts quenching by $\gtrsim 30$--$40\%$, especially in low-density environments. Overall, SIMBA provides good qualitative match to the observed quenching trends across the full range of stellar mass, whereas EAGLE shows qualitatively consistent behavior at intermediate masses or in low-density environments.

\begin{figure}
\includegraphics[width=0.99\linewidth]{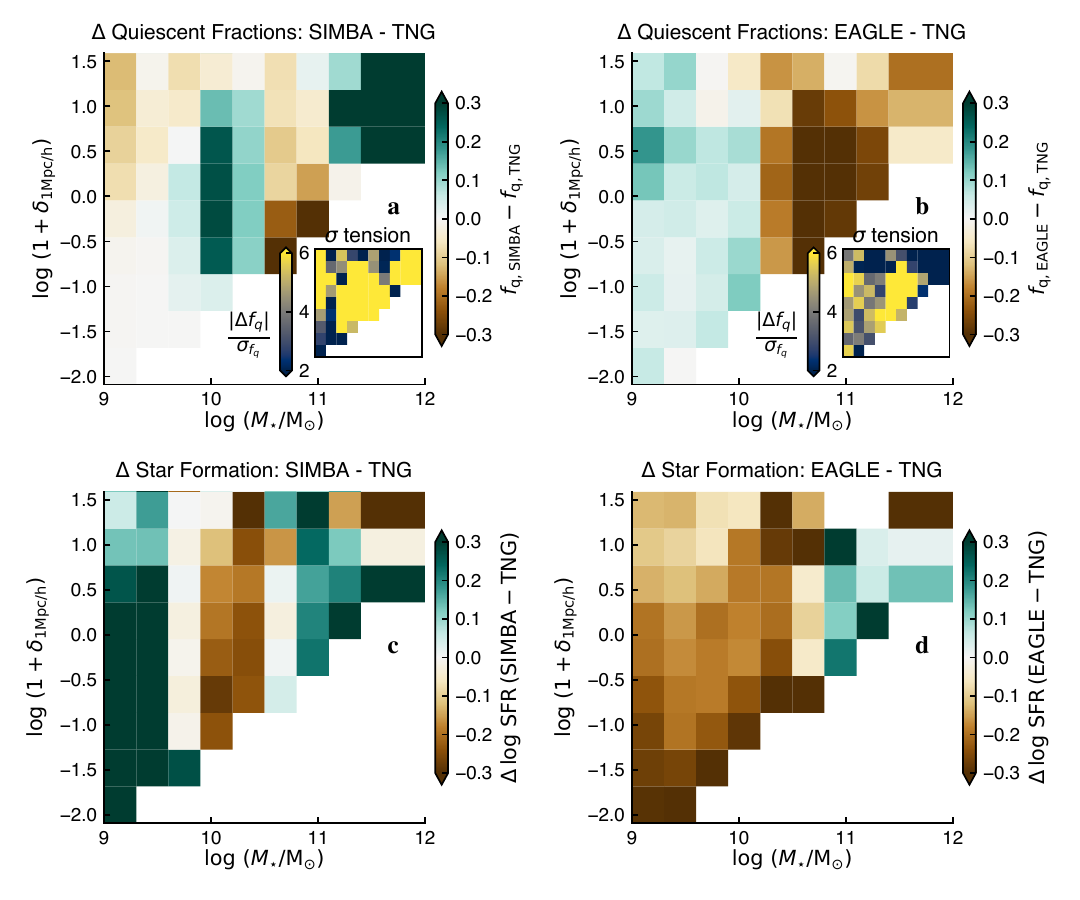}
\caption{\textbf{Differences in quiescent fractions and star formation rates of simulated galaxies as functions of stellar mass and environment.} 
Environment is quantified using the stellar mass excess within 1\,Mpc\,$h^{-1}$. Panels (a) and (b) show the differences in quiescent fraction between SIMBA and TNG, and between EAGLE and TNG, respectively, while panels (c) and (d) present the corresponding differences in star formation rates. All results are for simulations with $L \approx 100$\,Mpc boxes. Insets display the absolute difference normalized by the combined Binomial uncertainties, $| \Delta f_q | / \sigma_{f_q}$, providing a measure of the discrepancy ($\sigma$-level).}
\label{fig:ME_compare_sims_only}
\end{figure}

\begin{figure}
\includegraphics[width=0.99\linewidth]{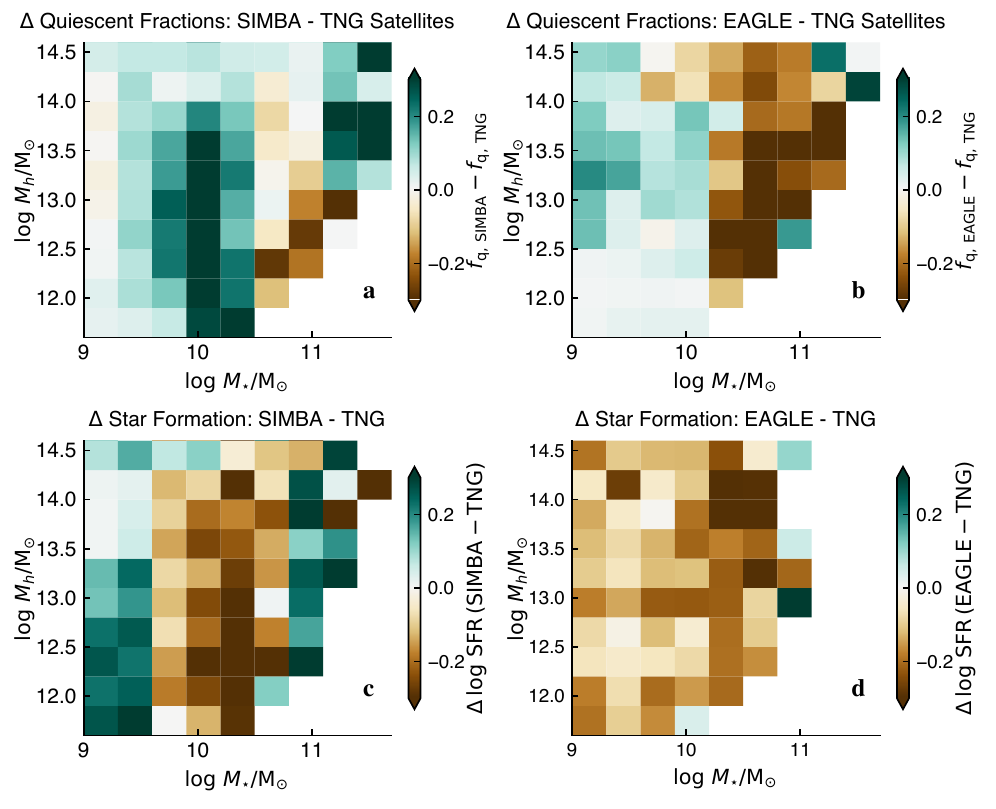}
\caption{\textbf{Differences in quiescent galaxy fractions and star formation rates as functions of stellar and halo mass among the simulations.}
Panels (a) and (b) show the differences in quiescent fraction between SIMBA and TNG, and between EAGLE and TNG, respectively, while panels (c) and (d) present the corresponding differences in star formation rates.}
\label{fig:MstarMh_fq_simsonly}
\end{figure}

Supplementary Figs.~\ref{fig:ME_compare_sims_only} and \ref{fig:MstarMh_fq_simsonly} show the differential trends of the quiescent fraction and SFR among the three simulations in the space of stellar mass versus environment or halo mass. The simulations predict markedly different trends, independent of their differences with observations (e.g., Figs.~3 and 4).

We will investigate the low-mass satellite population in more detail in future work, incorporating improved observational comparisons and refined SFR measurements. Based on GAMA data (Supplementary Figs.~\ref{fig:ME_GAMAsim} and \ref{fig:MsMh_GAMAsim}), observational limitations likely do not fully account for the discrepancy. Notably, 45--60\% of low-mass satellites in halos with $M_\mathrm{halo} > 10^{13}\,M_\odot$ are quenched or exhibit suppressed star formation ($\Delta\log\mathrm{SFR} < -0.5$\,dex). The 20--30\% offset between simulations and observations may be alleviated if models allowed satellites to retain or reaccrete gas more effectively---potentially through higher resolution or more accurate treatments of gas stripping or gas-rich galaxy interactions. These discrepancies persist even in the higher-resolution versions of the three simulations and in the larger-volume TNG300 run. Furthermore, the presence of a significant fraction ($\sim 20$--$30\%$) of low-mass satellites with enhanced star formation ($\Delta \log \mathrm{SFR} > 0$\,dex) in massive dark matter halos highlights the importance of clarifying how environmental processes---such as satellite infall, galaxy–galaxy interactions, and the time spent within the halo---regulate star formation in dense environments.

For low-mass galaxies in low-density environments, our GAMA-based estimate of the quiescent fraction ($\sim$20\%) is consistent with the trends predicted by the simulations and with measurements from nearby satellite surveys such as SAGA and ELVES \citep[e.g.,][]{Greene+23}. However, it is significantly lower than the $\sim$40\% inferred in a recent study of COSMOS galaxies at $z < 0.25$ \citep{Kaviraj+25}. This discrepancy may arise from cosmic variance or the absence of spectroscopic confirmation in the COSMOS sample. Future studies will explore additional galaxy properties beyond SFR to better understand low-mass quiescent populations across environments.

\subsubsection*{Feedback variants and resolution effects}

We utilize a suite of feedback-variant runs from the SIMBA50 simulations to isolate the impact of different AGN feedback modes. These $50\,h^{-1}\mathrm{Mpc}$-box simulations selectively disable stellar and AGN feedback (radiative, jet-mode, and/or X-ray heating AGN feedback), while keeping all other physical prescriptions fixed. In Extended Data Fig.~1, we show how the number densities of different galaxy subpopulations change across model variants compared to GAMA observations, including 16--84\% uncertainty ranges.

Without any feedback, the SIMBA simulations significantly overpredict galaxy number densities across all stellar mass and $\Delta\mathrm{SFR}$ bins, underscoring the essential role of both stellar and AGN feedback. Introducing stellar feedback alone suppresses low-mass galaxy counts, bringing number densities at the low-mass end into better agreement with observations (Extended Data Fig.~1a). However, this also leads to an overproduction of galaxies above the star-forming main sequence (SFMS) and an underproduction below it (Extended Data Fig.~1b and c), highlighting limitations in SIMBA’s current stellar feedback model. AGN feedback has limited impact at low masses, but thermal AGN feedback (without jets) lowers star formation rates in intermediate-mass galaxies ($M_\star \sim 10^{10}\,M_\odot$), modestly increasing the abundance of low-SFR or quiescent systems. The inclusion of jet-mode feedback dramatically reduces the number of massive star-forming galaxies ($M_\star \gtrsim 10^{11}\,M_\odot$). Interestingly, the abundance of massive quiescent galaxies (Extended Data Fig.~1d) is largely unaffected by AGN feedback; SIMBA reproduces their number densities even without it.

These results suggest that feedback in SIMBA not only regulates star formation but also reshapes the galaxy--halo connection \citep{Cui+21}. In the absence of feedback, SIMBA overproduces galaxies across the full stellar mass range. Incorporating feedback suppresses star formation and brings the global stellar mass function (SMF) into closer agreement with observations, as designed in the calibration strategy. Notably, even without AGN feedback, SIMBA matches the abundance of quenched galaxies while overproducing star-forming ones. This raises a key question: where do the excess galaxies go once feedback is activated? One possibility is that stellar and AGN feedback suppress star formation and/or prevents galaxy formation entirely in a subset of halos, reducing the number of halos that would otherwise host observable galaxies. Jet and X-ray AGN feedback regulate cold gas growth and drive quenching, producing inverse correlations between stellar-to-halo mass ratio and both halo formation time and galaxy quenching time \citep{Cui+21}. Thus, agreement with the global SMF may conceal a modified halo occupation function, particularly for star-forming systems. A future detailed analysis of halo assembly histories would be insightful.

Similarly, Extended Data Fig.~1e--h presents SMFs for EAGLE feedback variants. As in SIMBA, turning off AGN feedback has little effect on the low-mass end. However, the number densities of massive star-forming galaxies, especially those above the SFMS, are overpredicted without AGN feedback. Unlike SIMBA, EAGLE significantly underpredicts the abundance of quiescent galaxies in the absence of AGN feedback; their number increases when AGN feedback is reintroduced, primarily due to the quenching of high-SFR systems. The results also highlight a dependence of galaxy abundances on resolution, volume, and the thermal AGN feedback model, particularly the temperature increase per heating event.

Extended Data Fig.~1i--l further shows SMFs across the three TNG volumes. TNG300 underpredicts the abundance of massive galaxies ($M_\star \gtrsim 3 \times 10^{10}\,M_\odot$), especially star-forming and green-valley (transition) systems. The number of quiescent galaxies in TNG300, however, broadly agrees with GAMA. In contrast, TNG50 and TNG100 overproduce low-mass galaxies, particularly SFGs above the SFMS, again indicating sensitivity to resolution and volume. The global SMF shown in panel (a) has been presented previously \citep{2018MNRAS.475..648P} and used for model calibration; the new addition in our figure is the comparison of the predicted SMF split by $\Delta$SFR for galaxy subpopulations with observations.

Extended Data Fig.~2 show the satellite quiescent fractions as a function of stellar mass and halo mass for several SIMBA and EAGLE feedback variants. The stellar masses and quiescent states of satellites depend on both feedback processes and environment (halo mass). In SIMBA without any feedback, most low-mass satellites ($M_\star \lesssim 3 \times 10^{10}\,M_\odot$) are quenched in high-mass halos and star-forming in low-mass halos, highlighting the role of environmental effects or past gas consumption. The addition of stellar or AGN feedback slightly modifies this trend, making satellites more star-forming or less quiescent in the corresponding regions. Feedback, particularly AGN jet and X-ray modes, suppresses the abundance of very massive satellites ($M_\star \gtrsim 10^{11}\,M_\odot$); without feedback, simulations predict too many massive quiescent satellites in low-mass halos. AGN feedback is also essential to limit the number of massive star-forming satellites in high-mass halos, which are rarely observed.

Key insights from these two figures are: (1) the excess of quenched low-mass satellites in massive halos in the simulations is not necessarily caused by feedback;(2) AGN feedback significantly impacts the quiescent fraction of massive satellites \citep{Donnari+21fq};  (3) with large-volume, high-resolution simulations, the abundance of massive satellites can serve as a sensitive constraint on feedback models.

\subsection*{Quantifying the discrepancies between data and simulations}

Quantitative comparison metrics, comprising root-mean-square error (RMSE), mean absolute error (MAE), reduced chi-square ($\chi^2_\nu$), and the $\sigma$-level tension with respect to observations, are summarized in Table 1. All three simulations show significant discrepancies with the observations, typically exceeding $5\sigma$, with partial exceptions for EAGLE. However, their $\sim 2\text{--}3\,\times$ deviations in number density reveal systematic and informative patterns, which are discussed in detail below for each simulation. Our main results focus on simulations with box sizes of $L \approx 100$\,Mpc; results for other variants (e.g., TNG50 and TNG300) are provided in this section.

As shown in Table~1, EAGLE provides the closest match to star-forming and green-valley galaxies (GVGs; $-1 < \Delta\log\mathrm{SFR} < -0.5$), with RMSE (MAE) of $\approx 0.25\,(\lesssim 0.2)$\,dex. However, it performs worst for quiescent galaxies (QGs), with an RMSE of $\approx 0.5$\,dex. SIMBA shows the smallest discrepancy for QGs (RMSE $= 0.2$\,dex) but underperforms for lower-SFMS and green-valley galaxies, yielding an RMSE of $0.55$\,dex. TNG exhibits intermediate performance: it reasonably reproduces the abundance of massive galaxies but overpredicts low-mass highly star-forming and quiescent populations, resulting in RMSEs of 0.35--0.6\,dex across subpopulations. Its global SMF (Fig.~1a) gives the best overall fit; however, this primarily reflects its calibration rather than predictive accuracy. By comparison, the TNG300 variant underpredicts the number of massive lower-SFMS and green-valley galaxies, with an RMSE of $0.7$\,dex.

Table~2 quantifies the aggregate differences in quenched fraction or SFR for $M_\star$ vs. environment space (Fig. 3, Extended Data Fig~.3, and Supplementary Fig.~6). EAGLE best matches the observed quiescent fractions at intermediate masses ($\mathrm{RMSE} = 0.07$) but underestimates mean SFRs below $\log(M_\star/M_\odot) \lesssim 10.5$ ($\mathrm{RMSE} = 0.2$; Extended Data Fig.~3). In contrast, TNG predicts mean SFRs more accurately in low-density environments at similar masses ($\mathrm{RMSE} = 0.07$) but overpredicts the quiescent fraction near $\log(M_\star/M_\odot) \approx 10.5$ ($\mathrm{RMSE} = 0.25$). SIMBA shows the largest discrepancies in both quiescent fractions and SFRs ($\mathrm{RMSE} = 0.2$--$0.3$), with especially poor agreement near $10^{10}\,M_\odot$.

At low stellar masses and in low-density environments, all three simulations perform comparably well in reproducing quiescent fractions ($\mathrm{RMSE} \approx 0.03$). However, they diverge sharply from SDSS and GAMA in high-density environments at comparable masses ($\mathrm{RMSE} \approx 0.3$--$0.4$).

Thanks to the large SDSS sample, the $1/V_\mathrm{max}$-weighted mean quiescent fractions and SFRs are tightly constrained across the full stellar mass range, including the high-mass regime ($M_\star > 2 \times 10^{11}\,M_\odot$), with more than 100 galaxies per bin. The highest-mass bins for the fiducial TNG and SIMBA simulations contain 20--100 galaxies per bin. Fig.~1 also does not indicate undersampling. We assume binomial statistics to compute the counting uncertainties on quenched fraction and we use bootstrap sampling to take into account the uncertainties of SFRs in Figs.~3, and Extended Data Fig.~3.

Extended Data Fig.~6 presents the quiescent fraction as a function of stellar mass. Panels (a) and (b) show all satellites and centrals, respectively, while panels (c) and (d) focus on galaxies in high-density environments ($\log (1+\delta_\mathrm{1Mpc}) > 0.5$). GAMA provides more reliable constraints at low masses, whereas SDSS better samples high-mass galaxies. Quiescent fraction uncertainties are computed as $\sqrt{f_{\mathrm{q}}(1 - f_{\mathrm{q}})/N}$, where $f_{\mathrm{q}}$ is the quenched fraction and $N$ the number of centrals or satellites in each bin.  

All simulations overpredict quenching for low-mass satellites but reproduce the low fractions of low-mass centrals. At intermediate masses ($\sim 10^{10}$–$10^{11}\,M_\odot$), SIMBA and TNG overpredict quenching for both centrals and satellites, though agreement improves for high-density centrals. EAGLE reproduces high-mass satellites well but underpredicts quenching for massive centrals ($M_\star \gtrsim 10^{11}\,M_\odot$), a trend also seen in TNG100. SIMBA overpredicts quenching in massive galaxies compared to TNG and EAGLE, aligning better with GAMA but less with SDSS. TNG300 performs poorly at intermediate masses but matches high-mass quenched fractions. SIMBA predicts high-mass quenched fractions intermediate between TNG and EAGLE, though it overpredicts quenching near $M_\star \approx 10^{10}\,M_\odot$, consistent with trends in other figures. Improved constraints on massive centrals ($M_\star > 10^{11}\,M_\odot$) will aid refinement of feedback prescriptions in future large-volume simulations. Further discussion of this figure is given in the Supplementary Information.

Table~3 quantifies the discrepancies between SDSS and the simulations in the AGN luminosity function, stellar velocity-dispersion function, stellar mass function, and sSFR function, particularly for the results shown in Fig.~5. Most discrepancies have RMSE $\sim 0.4$–$1$\,dex and exceed $5\sigma$. Although different AGN subsample selections or methodological choices yield quantitatively different results (see Supplementary Information), the overall tensions remain substantial.

In addition, Extended Data Fig.~4 shows the density distributions of AGN and host properties. To include a more complete AGN sample, we relaxed the luminosity threshold to $L_\mathrm{AGN} > 10^{42}$\,erg\,s$^{-1}$ and $M_\star > 10^{9}\,M_\odot$. This relaxation is applied only in this figure; the fiducial cuts used in the main text are three times higher. The figure reveals substantial differences among the predicted AGN and host properties in the three simulations, as well as clear discrepancies with SDSS data. In Extended Data Fig.~4a, the AGN luminosity distribution in TNG peaks at $\sim 10^{43}$\,erg\,s$^{-1}$, offset by $\sim 0.2$\,dex relative to EAGLE and SIMBA. The $1/V_\mathrm{max}$-weighted median for SDSS is $4.7 \times 10^{42}$\,erg\,s$^{-1}$. The dispersions of all four distributions are comparable, $\sigma \approx 0.5$--0.6\,dex.  

In Extended Data Fig.~4b, the Eddington ratio distribution in TNG is narrower than in EAGLE and SIMBA by about 0.3\,dex and is systematically shifted to values above $10^{-3}$, with an excess near the peak. The median values for all three simulations and SDSS lie in the range $\lambda_\mathrm{Edd} = 2.5$–$4.0 \times 10^{-3}$. The SDSS distribution broadly resembles those of SIMBA and EAGLE. The Eddington-scaled mass accretion rates in TNG are $\sim 0.25$ dex lower, indicating that the offset is not driven by the post-processing conversion from mass accretion rate to luminosity.

In Extended Data Fig.~4c, the median black hole to stellar mass ratios are  $5 \times 10^{-3}$ (TNG), $2 \times 10^{-3}$ (SIMBA), and $7 \times 10^{-4}$ (EAGLE), with dispersions of 0.2, 0.4, and 0.5\,dex, respectively. EAGLE’s distribution is the closest to SDSS. The simulations also predict substantially different host stellar mass distributions (Extended Data Fig.~4d): median values of $4 \times 10^{9}\,M_\odot$ (TNG), $\sim 7\times 10^{9}\,M_\odot$ (SIMBA), $\sim 2 \times 10^{10}\,M_\odot$ (EAGLE), and $\sim 3 \times 10^{10}\,M_\odot$ (SDSS). The SIMBA distribution is narrower and more skewed, with few or no AGN hosts in dwarf galaxies, owing to its black hole seeding criterion.  

All three simulations place at least 85\% of AGNs in star-forming hosts (sSFR $> 10^{-11}$\,yr$^{-1}$). TNG contains no quiescent hosts below this threshold, resulting in a notably smaller dispersion. EAGLE’s median sSFR values are lower by about a factor of two. For SDSS, the median lies close to the threshold, such that roughly half of observed AGNs reside in star-forming hosts, while the other half are in green-valley or quiescent galaxies.

Extended Data Fig.~5 shows the black hole mass functions (BHMFs) from the simulations compared with SDSS observations. For massive host galaxies and low-luminosity AGNs, all three simulations underpredict the number densities of black holes, particularly in the range $M_{\rm BH} \sim 10^7$--$10^8\,M_\odot$. At higher AGN luminosities, TNG overpredicts while EAGLE underpredicts black holes within this mass range. By contrast, SIMBA overproduces luminous AGNs outside the $10^7$--$10^8\,M_\odot$ range. The apparent disagreement between our BHMF estimates and the Swift-BAT sample in panel (a) may reflect the BAT sample's bias against low-luminosity AGNs.

Extended Data Fig.~7 presents the relationships between halo mass and observable properties of satellite galaxies in the TNG100 simulation, including group velocity dispersion and stellar mass overdensity measured within fixed apertures of 0.5, 1, and 2 Mpc/$h$. These observables show strong correlations with halo mass, with scatter that is systematic and depends on the aperture. By combining these quantities with additional predictors using non-linear gradient boosting regression, we derive halo mass estimates that are generally consistent with those obtained by the GAMA team. Corresponding figures for EAGLE and SIMBA are provided at the end of the Supplementary Information.

\subsection*{Details on AGN demographics}

Because the density distributions of AGN and host properties differ among the simulations (e.g., Extended Data Fig.~4), it is instructive to examine how these differences affect the number density functions when restricted to various ranges of AGN and host properties. Particularly striking is the stellar–mass discrepancy between simulated and observed AGN populations, with TNG showing the most pronounced offset. However, the mismatch in host stellar mass is only one manifestation of a broader tension. Even when restricting the comparison to the same stellar-mass range, the simulated AGN populations differ substantially from observations in their multivariate properties---including AGN luminosities, host-galaxy velocity dispersions, and sSFRs.

To assess whether AGN selection effects could account for these differences, we applied multiple alternative selections (e.g., AGN luminosity cuts, host-galaxy $M_\star$ cuts, and variations in post-processing or optical AGN selection). Although these selections modify the detailed shapes of the distributions, they do not remove the discrepancies. This indicates that the divergence is not primarily driven by observational versus simulated selection choices, but instead reflects limitations in the modeling, producing a genuine tension between observations and the simulations in AGN demographics. We note that the inclusion or exclusion of weak AGNs in the observational sample affects the quantitative comparison; in particular, EAGLE provides the closest match to the observed AGN demographics when weak AGNs are excluded.

\begin{figure}
\includegraphics[width=0.99\linewidth]{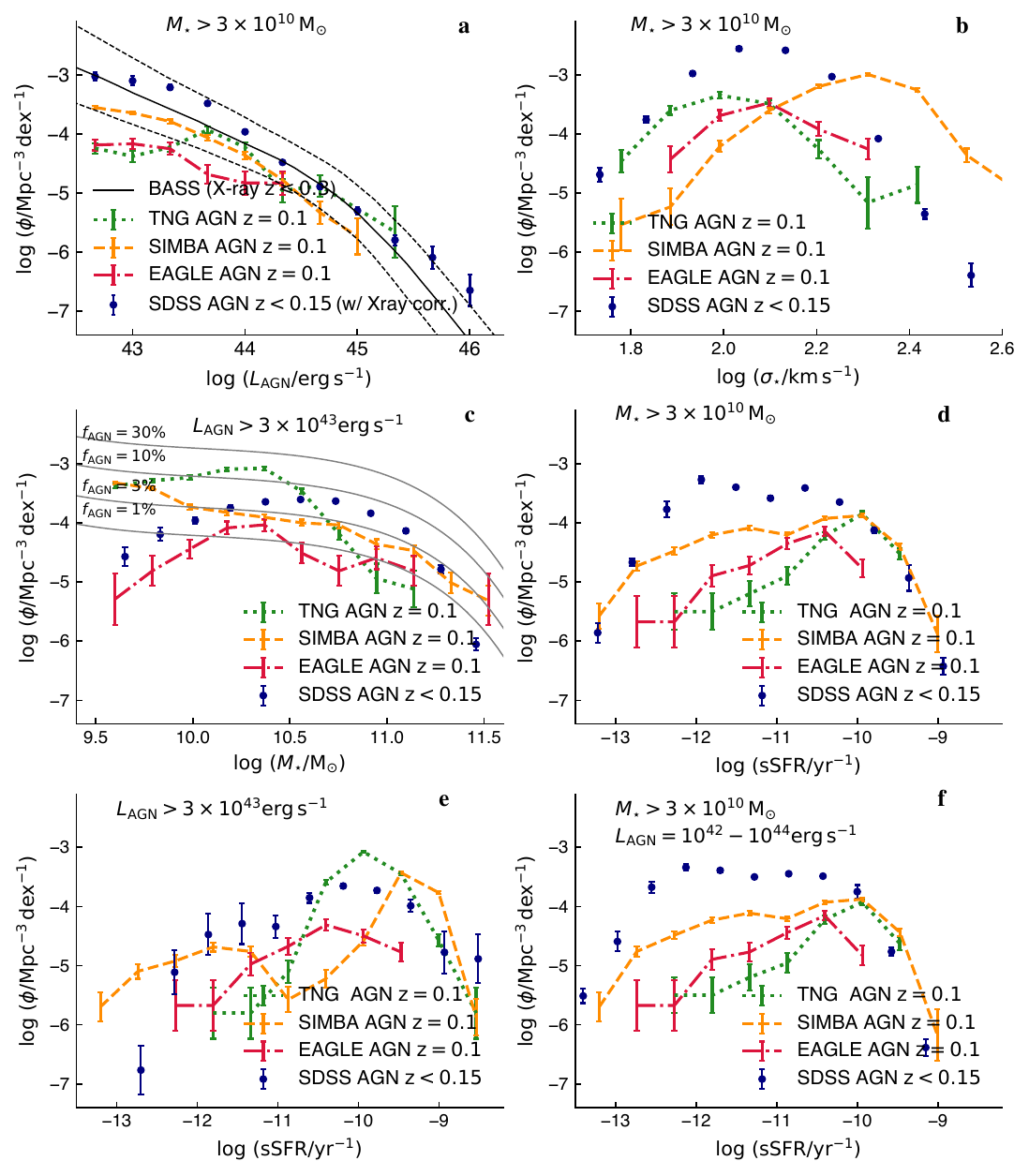}
\caption{\textbf{Effects of mass and luminosity cuts on AGN demographics.} Panels show: (a) AGN luminosity functions for massive galaxies, (b) host stellar velocity dispersion functions for massive galaxies, (c) host stellar masses of luminous AGNs, (d) host specific star formation rates (sSFR $\equiv \mathrm{SFR}/M_\star$) of massive hosts, (e) host sSFRs of luminous AGNs, and (f) host sSFRs of faint AGNs in massive hosts. Error bars indicate standard deviations assuming Poisson statistics. The median and 16th–84th percentiles of the fitted \textit{Swift}-BAT X-ray AGN luminosity function \citep{Ananna+22} are overplotted in panel (a), and a scaled-down global stellar mass function from Fig.~1a is shown in panel (c) to indicate the AGN fraction.}
\label{fig:AGNfunc_var}
\end{figure}

Fig.~\ref{fig:AGNfunc_var} shows the variants for the functions presented in the main text. Massive galaxies ($M_\star > 3 \times 10^{10}\,M_\odot$) hosting low-luminosity AGNs ($L_\mathrm{AGN} \sim 10^{42-43}$\,erg\,s$^{-1}$), with black hole masses in the range $M_\mathrm{BH} \sim 10^{7-8}\,M_\odot$, are underrepresented in all three simulations, albeit to varying degrees. These systems are predominantly quiescent or green-valley galaxies. The velocity dispersion function of massive hosts is also not well reproduced. All three simulations underpredict galaxies with $\sigma \lesssim 125$--$150$\,km\,s$^{-1}$ (with TNG showing the smallest deficit), while at higher dispersions SIMBA overpredicts and TNG and EAGLE underpredict the observed abundances.

For luminous AGNs ($L_\mathrm{AGN} \gtrsim 3 \times 10^{43}$ erg s$^{-1}$), the discrepancies are pronounced. EAGLE underpredicts their abundance, TNG overproduces them but confines them to star-forming hosts with $M_\star < 3 \times 10^{10}\,M_\odot$, and SIMBA overproduces them in low-mass star-forming galaxies ($M_\star \lesssim 10^{10}\,M_\odot$) while underproducing them in massive, low-SFR systems. Thus, none of the simulations captures the luminous AGN properties (fractions) across the full stellar mass range.  

The inferred black hole mass distributions of luminous AGNs in observations also disagree with simulations (Extended Data Figs.~4c and 5). SIMBA overproduces luminous AGNs with $M_\mathrm{BH} < 3 \times 10^{7}\,M_\odot$, while TNG overproduces them at $M_\mathrm{BH} \sim 10^{7-8}\,M_\odot$. EAGLE, by contrast, underpredicts luminous AGNs across this entire mass range. 

Our AGN black hole mass function (BHMF) broadly agrees with previous observational estimates \citep{Ananna+22}. Apparent discrepancies in the mean BHMF or AGN luminosity function are not statistically significant and can be attributed to differences in sample selection. In particular, the BASS AGN sample excludes low-luminosity AGNs with Eddington ratios below $10^{-3}$.

\begin{figure}
\includegraphics[width=0.99\linewidth]{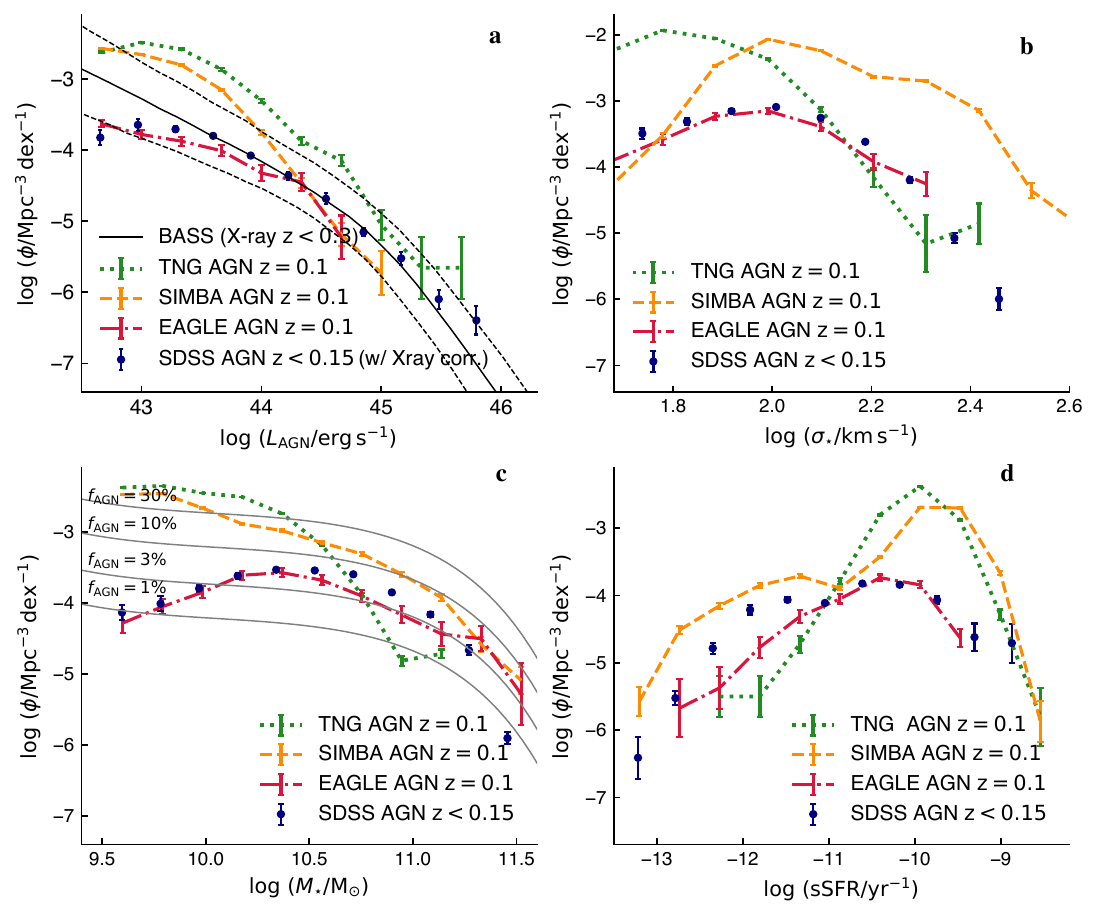}
\caption{\textbf{AGN demographics using the alternative BPT ``pure'' AGN selection.} This selection excludes weak LINERs and composite galaxies \citep{Kewley+06}. Panels show: (a) AGN luminosity functions, (b) host stellar velocity dispersion functions, (c) host stellar masses, and (d) host specific star formation rates (sSFR $\equiv \mathrm{SFR}/M_\star$). Error bars denote standard deviations assuming Poisson statistics. The median and 16th--84th percentiles of the fitted \textit{Swift}-BAT X-ray AGN luminosity function \citep{Ananna+22} are overplotted in panel (a), and a scaled-down global stellar mass function from Fig.~1a is shown in panel (c) to indicate the AGN fraction.}
\label{fig:AGNfunc_Seyf2}
\end{figure}

Figs.~\ref{fig:AGNfunc_Seyf2} show variants of Fig.~5, where the BPT AGN selection excludes weak AGNs (LINER-like or composite galaxies) \citep{Kewley+06}, respectively. Similar to Fig. 5,
significant discrepancies with observations persist for SIMBA and TNG, while EAGLE’s AGN and host properties show much better agreement with observations under this selection.

\begin{figure}
\includegraphics[width=0.99\linewidth]{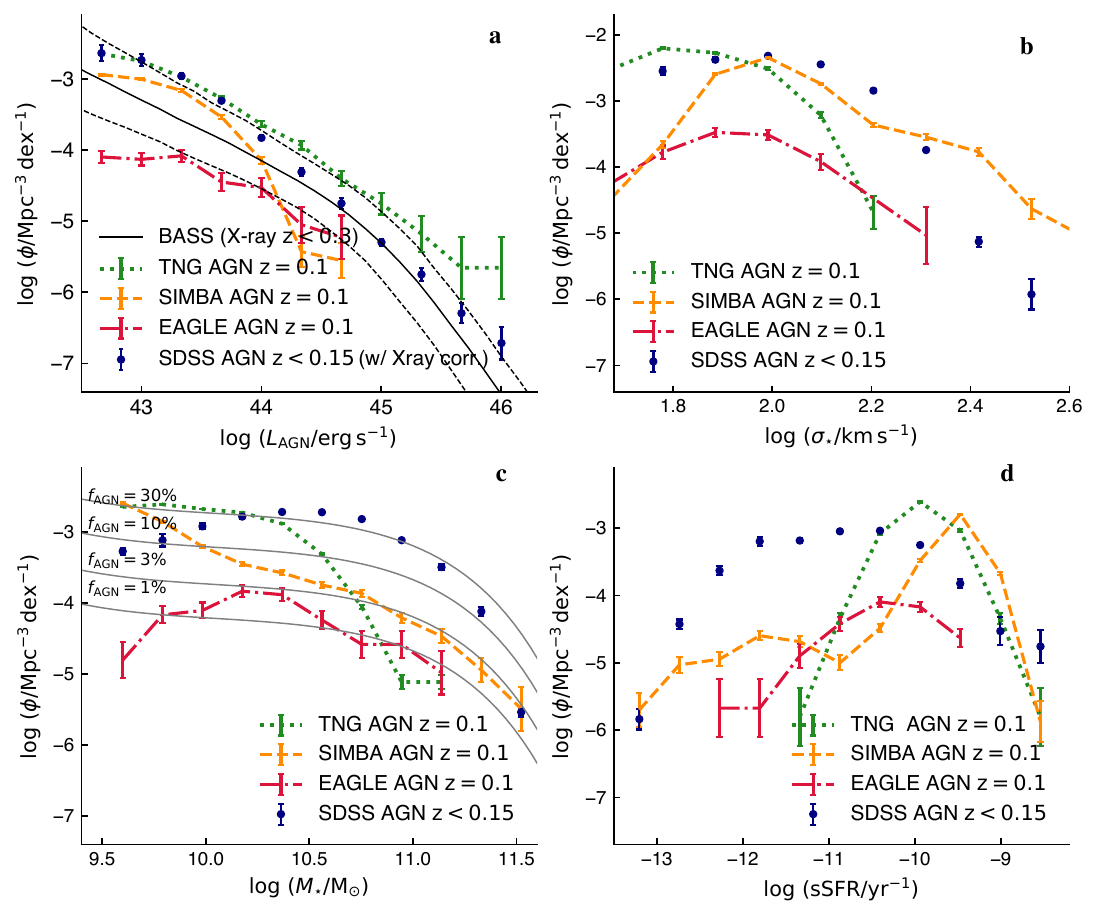}
\caption{\textbf{AGN demographic distributions using alternative simulation post-processing.} Results are shown for simulations post-processed following the methodology of ref. \citep[][]{Habouzit+21}. Panel (a) presents AGN luminosity functions; panel (b), host stellar velocity dispersion functions; panel (c), host stellar mass distributions; and panel (d), host specific star formation rates (sSFR $\equiv \mathrm{SFR}/M_\star$). Error bars represent standard deviations estimated under Poisson statistics. In panel (a), the median and 16th--84th percentiles of the \textit{Swift}-BAT X-ray AGN luminosity function \citep{Ananna+22} are overlaid, while panel (c) includes a scaled version of the global stellar mass function from Fig.~1a to indicate the AGN fraction.}

\label{fig:AGNfunc_HM}
\end{figure}

Previous AGN demographic figures adopts our fiducial post-processing \citep{Inayoshi+19} to convert BH accretion rates into observable luminosities. While an alternative post-processing \citep{Habouzit+21} yields quantitatively different results, the discrepancies among simulations and between simulations and observations persist (Fig.~\ref{fig:AGNfunc_HM}).

\subsubsection*{Details on halo mass estimation and halo-star formation trends in simulations} 

\begin{figure}
\includegraphics[width=0.49\linewidth]{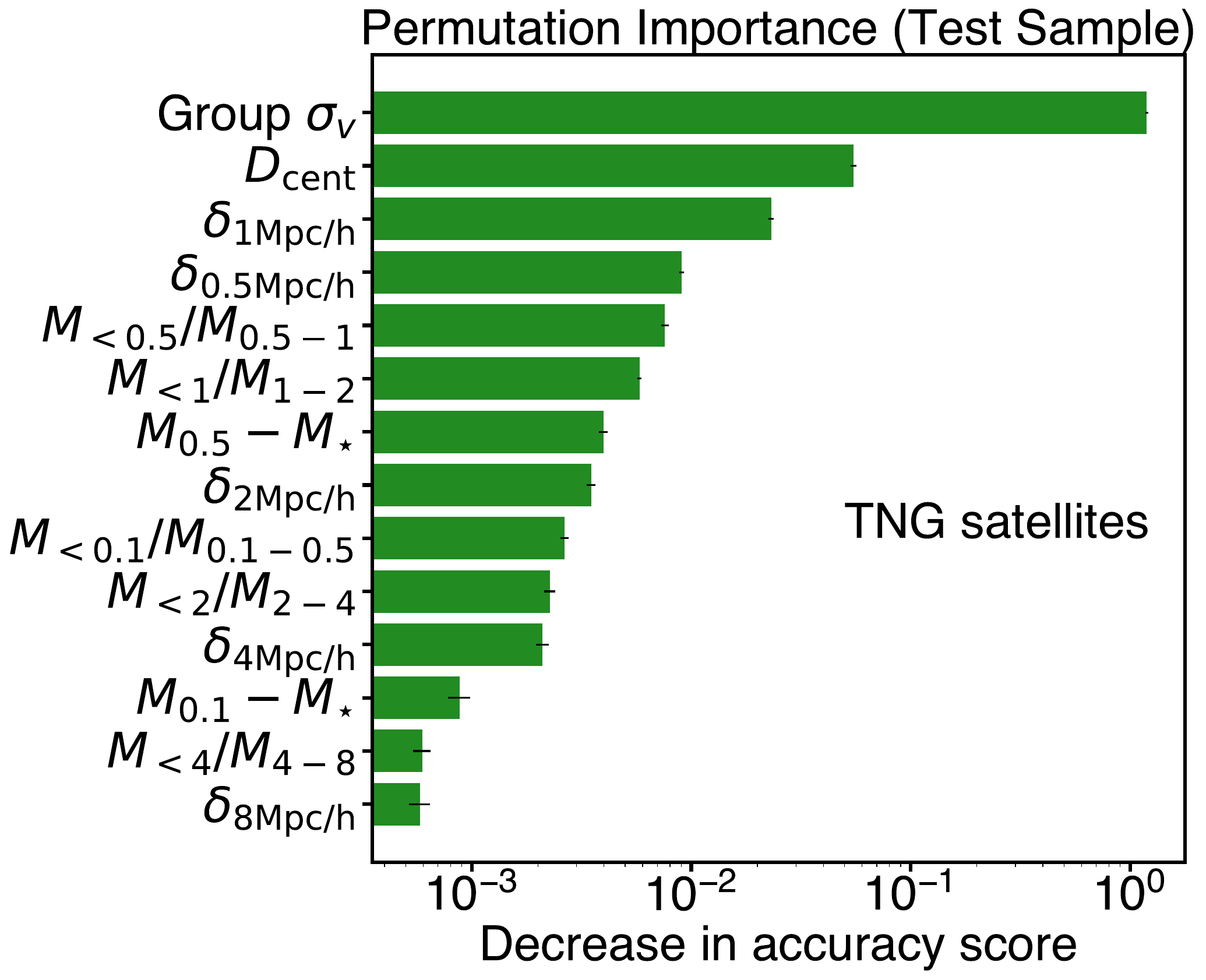}
\includegraphics[width=0.49\linewidth]{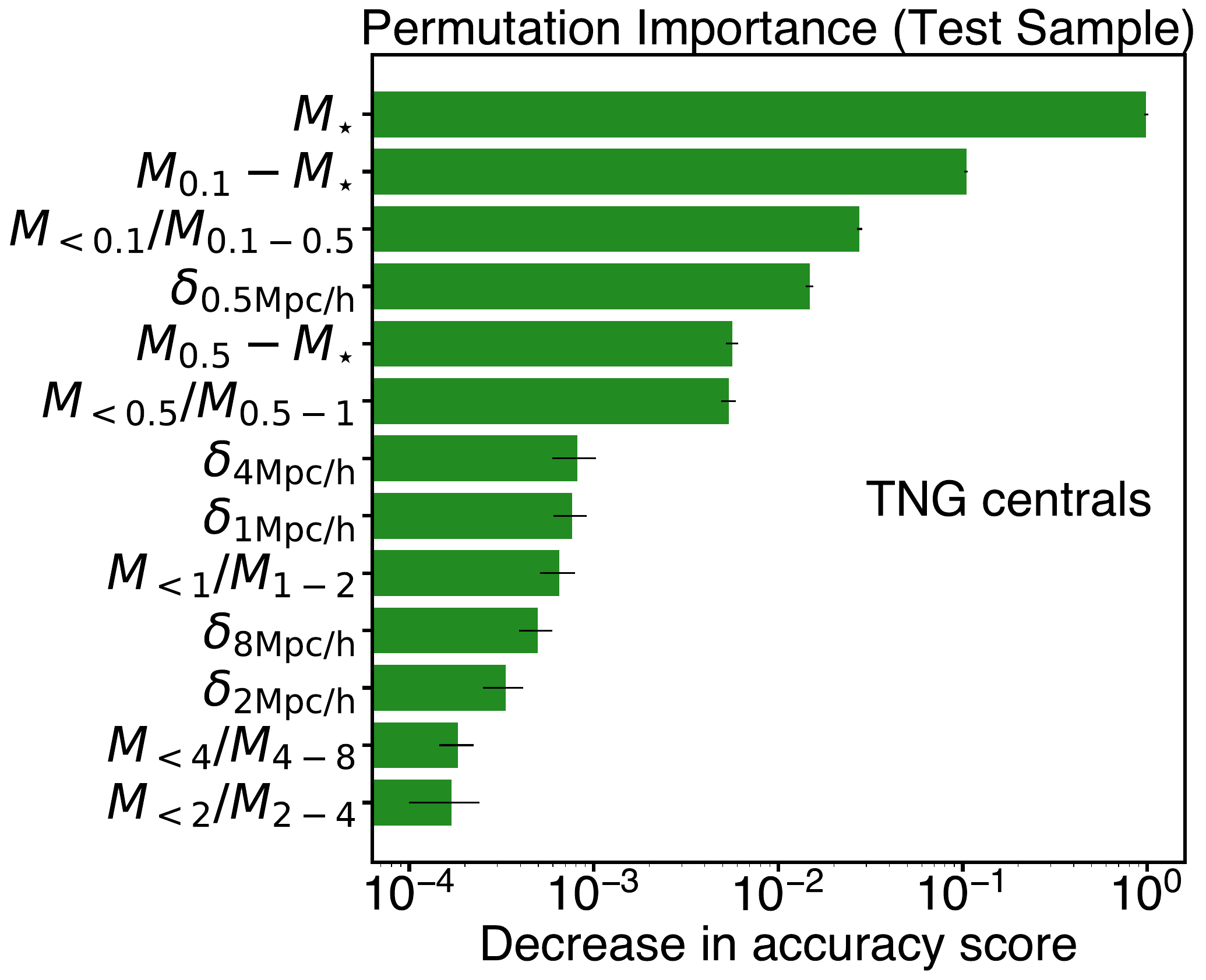}
\caption{\textbf{Variable importance for predicting halo mass of satellites and centrals}. Left panel shows the most important predictors for satellites, which include group velocity dispersion, distance to the center, and stellar mass overdensity within 1\,Mpc\,$h^{-1}$. Right panel highlights the key variables for centrals, including the stellar mass of the central galaxy, the mass of neighboring galaxies within 0.1\,Mpc\,$h^{-1}$, and the total mass overdensity within 0.5\,Mpc\,$h^{-1}$. The figures show the variable importance for TNG, but the rankings are similar for EAGLE and SIMBA.}
\label{fig:VarImp}
\end{figure}

Fig.~\ref{fig:VarImp} shows the permutation importance for satellites and centrals in the TNG simulation as an illustrative example. For satellites, the most important predictors include group velocity dispersion, distance to the group center, stellar mass overdensity within $0.5$--$1$\,Mpc/$h$, and the stellar mass ratio between $0.5$\,Mpc/$h$ and $0.5$--$1$\,Mpc/$h$, with consistent rankings across all three simulations. For centrals, the top predictors are stellar mass, stellar mass overdensity within $0.1$\,Mpc/$h$, and the stellar mass of neighbors within $0.1$\,Mpc/$h$ ($M_{0.1}-M_\star$). In the figure, $\frac{M_{< r_1}}{M_{r_1 - r_2}}$ denotes the ratio between the stellar mass enclosed within a radius $r_1$ and the stellar mass in the annular region between $r_1$ and $r_2$, for neighboring galaxies within a velocity difference $|\Delta v| \leq 1000$\,\kms, where $r_1$ and $r_2$ are in $\mathrm{Mpc}/h$. In the unified training, the most important predictors include the central’s stellar mass, group velocity dispersion, the central-to-neighbor stellar mass ratio within $\sim 1$--$2$\,Mpc, and the stellar mass excess within 1\,Mpc.

\begin{figure}
\includegraphics[width=0.99\linewidth]{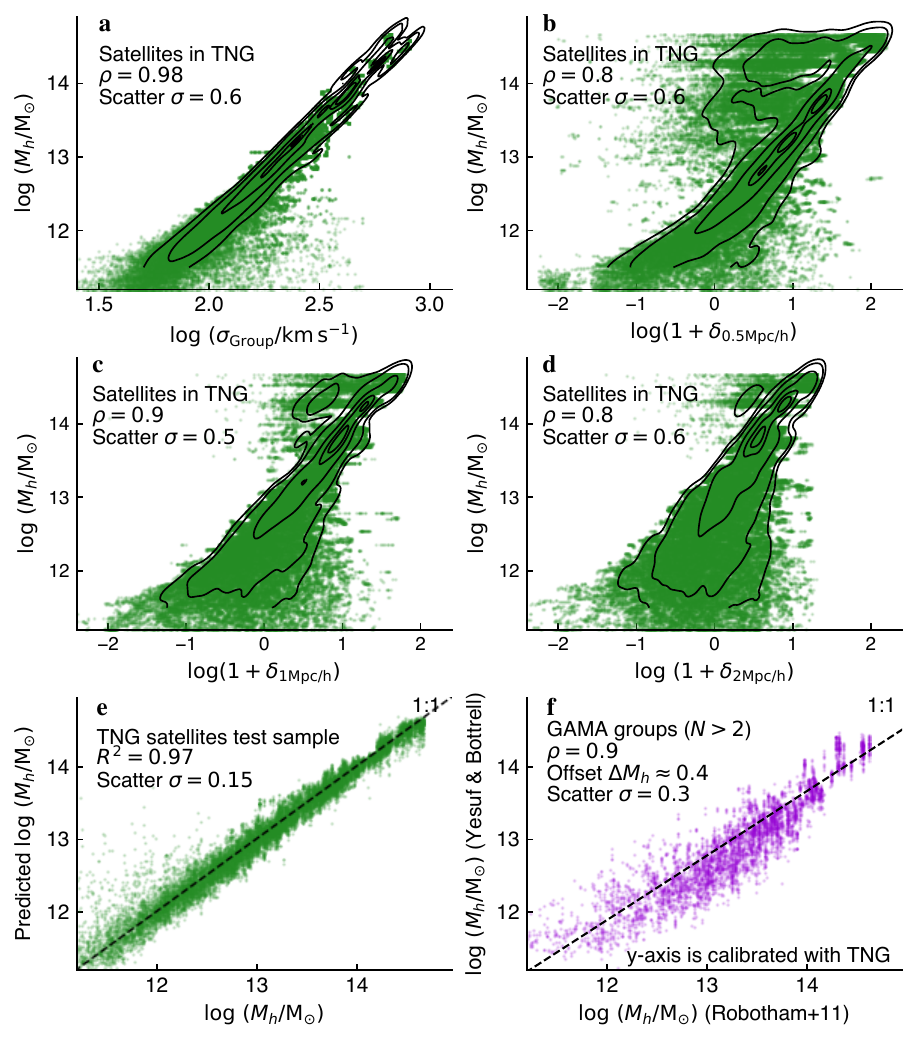}
\caption{\textbf{Satellite halo-mass estimation from observables calibrated with TNG.} Contours indicate the 5th, 15th, 50th, 85th, and 95th percentiles of the data distribution. The observables used for calibration include group velocity dispersion (panel a) and stellar mass overdensities measured within fixed apertures of 0.5, 1, and 2\,Mpc\,$h^{-1}$ (panels b–d, respectively). These quantities correlate strongly with halo mass, with scatter that varies systematically with aperture size and halo-mass sensitivity. The observables, together with additional variables, are combined non-linearly using gradient-boosting regression to predict halo masses (panel e). This calibration is then applied to galaxy groups in the observations. Panel (f) compares the resulting halo-mass estimates for GAMA satellite galaxies with previous determinations \citep{Robotham+11}.
}
\label{fig:Mh_TNGcalib}
\end{figure}

\begin{figure}
\includegraphics[width=0.99\linewidth]{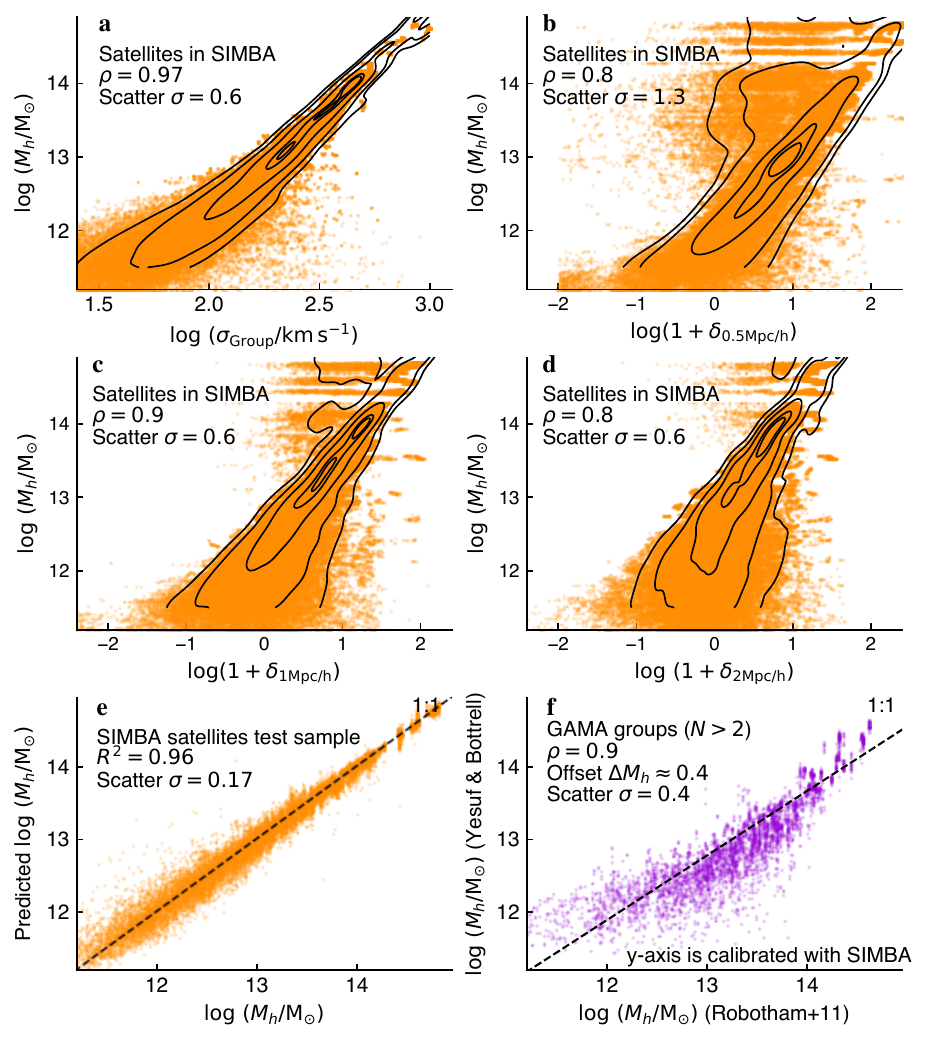}
\caption{\textbf{Satellite halo-mass estimation from observables calibrated with SIMBA.}
Group velocity dispersion (panel a) and stellar mass overdensities within apertures of 0.5, 1, and 2\,Mpc\,$h^{-1}$ (panels b--d) are used as input observables for the SIMBA-based calibration. Contours indicate the 5th, 15th, 50th, 85th, and 95th percentiles of the data distribution. These observables exhibit strong correlations with halo mass and aperture-dependent scatter. A gradient-boosting regression model combines the observables to predict halo masses (panel e), and the calibration is applied to galaxy groups in the observational data. Panel (f) compares the resulting halo-mass estimates for GAMA satellite galaxies with previous measurements \citep{Robotham+11}.}
\label{fig:Mh_SIMBAcalib}
\end{figure}

\begin{figure}
\includegraphics[width=0.99\linewidth]{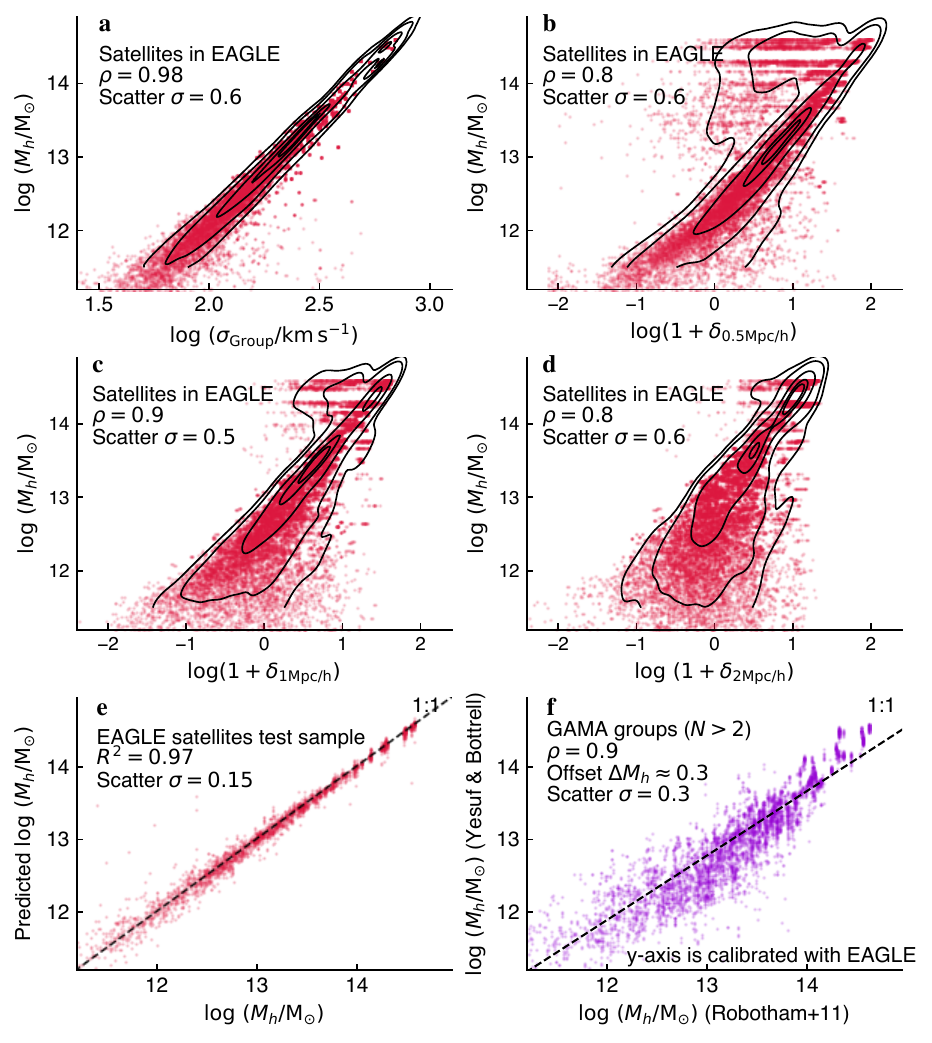}
\caption{\textbf{Satellite halo-mass estimation from observables calibrated with EAGLE.}
The EAGLE-based calibration employs group velocity dispersion (panel a) together with stellar mass overdensities measured within 0.5, 1, and 2\,Mpc\,$h^{-1}$ apertures (panels b--d). Contours indicate the 5th, 15th, 50th, 85th, and 95th percentiles of the data distribution. These observables are combined through gradient-boosting regression to infer halo masses (panel e), accounting for their strong correlations with halo mass and aperture-dependent scatter. The calibrated relation is then applied to observed galaxy groups. Panel (f) compares the inferred halo masses of GAMA satellite galaxies with previous estimates \citep{Robotham+11}.}
\label{fig:Mh_EAGLEcalib}
\end{figure}

Supplementary Figs.~\ref{fig:Mh_TNGcalib}, Figs.~\ref{fig:Mh_SIMBAcalib} and \ref{fig:Mh_EAGLEcalib} show the correlations between halo mass and observable galaxy properties of satellites in the three simulations, including group velocity dispersion and stellar mass overdensity measured within fixed apertures of 0.5, 1, and 2 Mpc/$h$. These variables correlate strongly with halo mass, though with substantial but systematic scatter. By combining these and additional predictors non-linearly with gradient boosting regression, we obtain accurate halo mass estimates that are in reasonable agreement with those derived by the GAMA team.

\subsection*{Comparison of AGN measurements with previous studies}
\begin{table}
\caption{\textbf{Parameters of the SED fitting using CIGALE code}} \label{tab:cigale}%
\begin{tabular}{@{}lll@{}}
\toprule
Parameters & parameter values\\
\midrule
 & \textbf{Star formation history}: delayed model+recent burst \\
e-folding time: $\tau_\mathrm{main}$/Myr & 500, 1000, 2000, 3000, 4000, 5000, 6000, 7000, 9000\\
Age main stellar populations in Myr & 4500, 9500, 11000 \\
Burst e-folding time $\tau_\mathrm{burst}$/Myr & 50.0, 100.0, 200.0 \\
Burst mass fraction & 0.0,0.005, 0.01, 0.05, 0.1, 0.2, 0.3 \\
\hline
 & \textbf{Simple stellar population}: Bruzual \& Charlot \citep{BruzualCharlot03}\\
Initial mass function &  Chabrier \citep{Chabrier03} \\
Metallicity & 0.02 (Solar) \\
\hline
& \textbf{Dust law}: dustatt\_modified\_starburst \\
    $E(B-V)_\mathrm{lines}$  & 0.0, 0.05, 0.1, 0.2, 0.3, 0.4, 0.5, 0.6, 0.7, 0.8, 1.0 \\
    $E(B-V)_\mathrm{star}$/$E(B-V)_\mathrm{lines}$ & 0.44 \\
    Attenuation curve power law slope & -0.5, -0.25, 0.0 \\
    Emission line extinction law & Milky Way \\
    $R_V$  &  3.1 \\
\hline
 & \textbf{AGN emission}: SKIRTOR \citep{Stalevski+16}\\
    Torus optical depth at 9.7$\mu$m  &  7 \\
    Power-law exponent of radial dust density & 1.0 \\
    Power-law exponent of angular dust density & 1.0 \\
    Angle between the equatorial plan and edge of the torus & 40 \\
    Ratio of the outer and inner radii of the torus & 20 \\
    Inclination/viewing angle &  30 (type 1),70 (type 2) \\
    AGN fraction & 0, 0.01, 0.05, 0.1, 0.2, 0.3, 0.4, 0.5, 0.7, 0.9 \\
    Extinction law of the polar dust & SMC \\
    $E(B-V)$ of polar dust & 0, 0.05, 0.1, 0.3\\
    Temperature of the polar dust in K & 100 \\
\hline
Total number model combinations & 31,434,480 per redshift\\
\botrule
\end{tabular}
\end{table}

The detailed parameter settings of our SED fitting model \citep{Boquien+19} are listed in Table~\ref{tab:cigale}. Our NLAGN measurements are in good agreement with GSWLC \citep{Salim+18}: stellar masses and SFRs show Spearman correlation coefficients of 0.86, with scatters of 0.15 and 0.4\,dex and offsets of 0.2 and 0.25\,dex, respectively.

The BLAGN sample was analyzed previously by Zhuang \& Ho 2023 \citep{ZhuangHo23}, who measured $M_\star$ and $g-r$ colors but did not report SFRs or $A_V$. Their approach combined Pan-STARRS image decomposition in five bands to remove AGN emission, followed by SED fitting with CIGALE. Our stellar masses are consistent with theirs ($\rho = 0.8$, scatter of 0.2\,dex, and offset of 0.14\,dex). However, their $g-r$ colors correlate only weakly with our SSFRs ($\rho = 0.4$). By incorporating UV and MIR data, our SED fitting likely provides more reliable SSFR estimates, capturing ongoing star formation and reducing the impact of dust attenuation on optical colors.

The SFR of AGNs can also estimated from dust-corrected [O\,II] $\lambda\lambda 3727$\,\AA\ emission, with AGN ionization corrections based on [O\,III] $\lambda5007$\,\AA\ and metallicity adjustments from $M_\star$--metallicity relations \citep{ZhuangHo19}. Our SED-based SFRs agree reasonably well with those from [O\,II] but show significant scatter ($\rho \approx 0.7$, scatter of 0.6\,dex and offset of 0.2\,dex), likely due to uncertainties in dust and metallicity corrections and aperture differences.

Our SED-based AGN luminosities broadly agree with estimates from the 5100\,\AA\ continuum and [O\,III] $\lambda5007$\,\AA\ luminosities, using bolometric corrections of 9.8 \citep{McLureDunlop04} and 600 \citep{Heckman+04}, respectively. We correct [O\,III] luminosities for dust using $2 \times A_V$ from our SED fits. Remaining discrepancies likely reflect incomplete dust correction or stellar contamination, which our SED-based approach mitigates, yielding more reliable AGN luminosity estimates.

Furthermore, our AGN selection is based on integrated BPT diagnostics, which may miss low-luminosity AGNs in dwarf galaxies where stellar emission dilutes the nuclear signal. Recent spatially resolved studies \citep{Mezcua+24} have shown that the AGN fraction in dwarfs may be significantly higher when resolved BPT diagnostics are applied, in part because many AGNs are found to be spatially offset from the galaxy centers. These AGNs are predominantly low-luminosity and not typically detected in X-rays, placing them below the luminosity regime probed in this work.

Our fiducial analysis instead focuses on moderate- to high-luminosity systems ($L_\mathrm{AGN} > 3\times10^{42}\,\mathrm{erg\,s^{-1}}$) in galaxies with $\log M_\star/M_\odot > 9.5$, where stellar dilution and spatial resolution limitations are much less severe. In this mass regime, our observed AGN fractions (rising from $\sim 1$\% to $\sim 10$\% between $\log M_\star/M_\odot =9.5$–10) broadly agree with the trend suggested by Mezcua et al. \citep{Mezcua+24}, assuming that off-centering is not significant in more massive hosts. The comparison with simulations is also illuminating: SIMBA does not seed black holes below $\log M\star/M_\odot < 9.5$, EAGLE predicts AGN fractions near 1\%, and TNG predicts $\sim 40$\%. None of these models reproduce the AGN fractions inferred in dwarfs by the aforementioned study.

Nevertheless, extending AGN selection to include spatially resolved diagnostics would improve completeness at the faint end and may reveal additional AGNs in dwarfs. While such weak AGNs may be present in up to $\sim 20\%$ of low-mass galaxies, they lie below the luminosity and mass thresholds adopted for our main comparisons. We therefore conclude that our results for the moderate- to high-luminosity AGN population remain robust, while the faint-end population in dwarfs offers an important direction for future work.

\subsubsection*{Details on SFR timescale}
The SFRs available in the catalogs of all three simulations represent instantaneous values, which do not directly correspond to the timescales probed by observational star formation tracers. However, for the TNG simulation, additional catalogs \citep{2019MNRAS.485.4817D, 2021MNRAS.506.4760D} include SFRs averaged over various timescales, allowing for comparisons with specific observational tracers. 

 The is a strong correlation between instantaneous and time-averaged SFRs (over 100–200 Myrs) for TNG galaxies, with a Spearman coefficient of 0.96 and scatter of 0.16\,dex. These timescales align closely with SFRs derived from spectral energy distribution (SED) fitting in observations. Furthermore, observations from SDSS Mapping Nearby Galaxies at Apache Point Observatory (SDSS MaNGA) confirm that SED-based SFRs are reasonably consistent with H$\alpha$-based SFRs, which trace timescales of less than 10 Myrs (Spearman correlation of 0.85 and a scatter of 0.4\,dex). Thus, while only instantaneous SFRs are available for all three simulations, their strong correlation with time-averaged SFRs, combined with the observational consistency between SED- and H$\alpha$-based SFRs, justifies their comparison with observed SFRs derived from SED fitting.

\clearpage
\FloatBarrier

\bibliographystyle{nature}

\end{document}